\documentclass[11pt]{article}

\usepackage[hyperfootnotes=false]{hyperref}
\usepackage{fullpage}
\usepackage{amsmath}
\usepackage{graphicx}

\newcommand{\etal}{\emph{et al.}}

\newcommand{\sub}[1]{\ensuremath{_{\mbox{\scriptsize \,#1}}}}
\newcommand{\supers}[1]{\ensuremath{^{\mbox{\scriptsize #1}}}}

\begin{document}


\title{On the Expectation-Maximization Unfolding with Smoothing}


\author{Igor~Volobouev \vspace{0.3cm} \\
{\it i.volobouev@ttu.edu} \vspace{0.3cm} \\
Texas Tech University, Department of Physics, Box 41051,\\
Lubbock, Texas, USA 79409
}

\maketitle

\begin{abstract}
Error propagation formulae are derived for the
expectation-maximization iterative unfolding algorithm
regularized by a smoothing step. The effective number of
parameters in the fit to the observed data is defined for
unfolding procedures. Based upon this definition, the Akaike
information criterion is proposed as a principle for
choosing the smoothing parameters in an~automatic, data-dependent manner.
The performance and the frequentist coverage of
the resulting method are investigated using
simulated samples. A number of issues of general relevance
to all unfolding techniques are discussed, including irreducible bias,
uncertainty increase due to a~data-dependent choice of
regularization strength, and presentation of results.
\end{abstract}




\section{Introduction}

Due to the finite resolution of sensors and particle
detector systems, inverse problems
involving Poisson counts received considerable attention
in particle physics~\cite{ref:phystat11, ref:reviewspano}.
In this field, reconstruction
of particle spectra that involves solving corresponding statistical
inverse problems is usually
referred to as ``unfolding''\footnote{
In the  statistical literature, the term
``deconvolution
density estimation'' is used to describe
similar problems in somewhat more restricted settings~\cite{ref:deconvolutionbook}.}.
Most widely used techniques, popularized by the code availability in
the RooUnfold software package~\cite{ref:roounfold}, include
``SVD'' unfolding~\cite{ref:svdunfold}
and iterative expectation-maximization (a.k.a. D'Agostini, or Bayesian)
unfolding~\cite{ref:dagounfold}
regularized by early stopping.

The expectation-maximization unfolding with a smoothing step
was introduced in~\cite{ref:nychka, ref:smoothedem}, where convergence of the
method was established and its relation to penalized maximum
likelihood techniques was pointed out. 
In the context of particle physics applications, this method 
became widely known after its reinvention, albeit in a less general form,
in~\cite{ref:dagounfold}. Expectation-maximization with smoothing will
be abbreviated ``EMS'' for the remainder of this paper.

This article is structured as follows.
The statistical formulation of the unfolding problem and
basic mathematical notation used throughout the manuscript
are introduced in Section~\ref{sec:problemstatement}.
In Section~\ref{sec:regularization}, the concept of regularization
is discussed and its necessity is illustrated with a simple example.
Regularization by smoothing in the EMS approach is described
in Section~\ref{sec:ems}, together with the formulae for error propagation.
A~method for determining the effective number of fitted parameters
for unfolding procedures is proposed in Section~\ref{sec:effectivenum}.
An~automatic choice of the regularization strength based on the
Akaike information criterion is advocated in Section~\ref{sec:emschoice}.
Numerical studies of the frequentist coverage of the developed
uncertainty formulae are described in Section~\ref{sec:simulation}
using two density models similar to those encountered in
particle physics applications.
Presentation and use
of the unfolded results are discussed in Section~\ref{sec:resultpresent}.
Finally, the advantages of the developed methodology are summarized in
Section~\ref{sec:conclusions}.

\section{Problem Statement}
\label{sec:problemstatement}

For the purpose of stating the unfolding problem,
it will be assumed that the detector can be described by an
operator $K$. This operator (also called {\it kernel},
{\it transfer function}, {\it observation function},
or {\it response function}, depending on the author and context)
converts probability
densities $p({\bf x})$
in the physical process space ${\bf x}$ into the densities $q({\bf y})$ in the
observation space ${\bf y}$: 
\begin{equation}
q = K p \equiv \int K({\bf y}, {\bf x}) p({\bf x}) {\bf dx}.
\label{eq:kernelaction}
\end{equation}
The response function does not have
to be fully efficient: $q$ does not have to integrate to 1 when
$p$ is normalized. In the subsequent discussion, operator $K$ will be assumed
exactly known but not necessarily invertible.

In many situations of interest, observations are described by the empirical 
density function ({\it i.e.}, there is no uncertainty associated with
each individual observation):
\begin{equation}
\rho_e({\bf y}) = \frac{1}{N} \sum_{i=1}^{N} \delta({\bf y} - {\bf y}_i).
\label{eq:edf}
\end{equation}
In this case, the probability density function
to observe a point at ${\bf y}_i$ is given by
the normalized version of $q$ called $r$: $r = q/\epsilon$.
In case $p$ is normalized,
\begin{equation}
\epsilon = \int q({\bf y}) \,{\bf dy} = \int K p \,{\bf dy}
\label{eq:efficiency}
\end{equation}
is the overall detector
acceptance for the physical process under study.

The purpose of unfolding is to learn as much as possible about $p({\bf x})$
given $\rho_e({\bf y})$ when a~parametric model for $p({\bf x})$ is lacking.
Reporting $p({\bf x})$ (rather than $q({\bf y})$) and its uncertainty
simplifies testing of theoretical models against results
obtained by multiple experiments and
allows for preservation of scientific results in
a form suitable for comparison with models that
have not been invented yet.

It should be noted that,
in case operator $K$ has a non-empty nullspace,
the problem of $p({\bf x})$ determination is
ill-posed even in the large sample limit.
The nullspace of the response function is defined
as the set of all functions $\rho({\bf x})$ satisfying
the equation $\int K({\bf y}, {\bf x}) \rho({\bf x}) {\bf dx} = 0$
which are not identical zeros. 
If $p({\bf x})$ is treated in Eq.~\ref{eq:kernelaction}
as the function to be solved for,
the general solution of this equation is $\tilde{p}({\bf x}) + \rho({\bf x})$,
where $\tilde{p}({\bf x})$ is any particular solution
and $\rho({\bf x})$ is an arbitrary
nullspace member. 
This nullspace definition is also appropriate for a~discretized
representation of ${\bf x}$ and ${\bf y}$ spaces
(discussed further in Section~\ref{sec:ems}), with $K({\bf y}, {\bf x})$
represented by a~matrix and integration over ${\bf dx}$
replaced by matrix-vector multiplication.

Nullspace forms a linear subspace in the space of functions
(for the discretized problem, in the space of vectors).  Assuming
that an inner product is postulated ({\it e.g.}, as in the $L^2$-space), an
arbitrary function can be uniquely decomposed into two additive components:
the component that belongs to the nullspace and the component that belongs
to the orthogonal complement of the nullspace. 

For numerical computations with finite precision, the effective
nullspace is enlarged. A reasonably useful definition
of this extended nullspace can be introduced procedurally,
by performing singular value
decomposition of the operator $K$ and by using the basis that consists
of right-singular vectors corresponding to ``small'' singular
values. An appropriate method for qualifying
singular values as ``small'' will, in general,
depend on the precision of floating point operations and
on algorithm implementation details.

\section{Regularization}
\label{sec:regularization}

The difficulty of the unfolding
problem can be easily appreciated from the following
argument. $K$ typically acts as low-pass filter. For measurements of
a~single scalar quantity, it can often be assumed that the detector
has resolution $\sigma$ and that $K(y, x) = {\cal N}(y - x, \sigma^2)$,
where ${\cal N}(\mu, \sigma^2)$ stands for the normal distribution
with mean $\mu$ and variance $\sigma^2$.
The detector thus simply convolves $p(x)$ with the normal distribution
which leads to the product of the corresponding Fourier coefficients.
If $q$ was exactly known, the Fourier transform of $p$ could
be obtained from $p(\omega) = q(\omega)/K(\omega)$,
where $q(\omega)$ and $K(\omega)$ are the Fourier transforms
of the observation density and the response function, respectively.
 As $q(\omega)$ is
not known, the closest available approximation is the characteristic
function of $\rho_e(y)$:
$\rho_e(\omega) = \int \rho_e(y) e^{i \omega y} dy = \frac{1}{N} \sum_{i=1}^{N} e^{i \omega y_i}$. For the normal
distribution, $K(\omega) = e^{-\sigma^2 \omega^2/2}$, so that the ratio
$\rho_e(\omega)/K(\omega)$ becomes arbitrarily large
as $\omega \rightarrow \infty$. The ``naive'' method of
estimating $p(\omega)$ as $\rho_e(\omega)/K(\omega)$ thus fails
miserably: the high frequency components of the noise contained
in the $\rho_e(\omega)$ are multiplied by an arbitrarily large factor
so that $\rho_e(\omega)/K(\omega)$ is not even square-integrable.

A number of effective
approaches to solving the pure deconvolution problem just described
are detailed in~\cite{ref:deconvolutionbook}.
These approaches invariably involve
introduction of additional smoothness assumptions about either
$p(x)$ or $q(y)$ or both. It is averred
that the high frequency components of $p(x)$
are of little interest and, therefore, can be suppressed in the
$\rho_e(\omega)/K(\omega)$ ratio so that the inverse Fourier transform
can exist. Introduction of new information by applying additional assumptions
which make an~originally ill-posed problem treatable is called
{\it regularization}.

In the problems of interest for particle physics, the action
of $K({\bf y}, {\bf x})$ on $p({\bf x})$ is usually more complicated
than the simple convolution while the necessity of regularizing the
ill-posed problem is as pressing.
In the formulation of the SVD unfolding method presented in~\cite{ref:svdunfold},
one-dimensional ${\bf x}$ is assumed, and the
regularization is performed by penalizing
the discretized second derivative of the $p(x)$
density\footnote{This regularization technique has been reinvented
many times. Depending on the problem, it is also called 
the {\it constrained linear inversion method},
the {\it Phillips-Twomey method}, {\it Tikhonov regularization}, {\it ridge regression},
or the {\it ridge-parameter approach}~\cite{ref:deconvolutionbook, ref:numrecipes}.}.
The expectation-maximization unfolding is most commonly regularized
by imposing
a~subjective limit on the number of iteration cycles. This
early stopping criterion penalizes
deviations from the distribution used to start the iterations.
However, due to the method nonlinearity, it is difficult to
supplement this statement with a concise analytic derivation
of a~penalty term that would permit
determination of the penalized likelihood maximum
as well as a~subsequent quantification of the relationship
between these deviations and the number of iterations performed.

In these unfolding approaches, it is assumed
that the ${\bf x}$ and ${\bf y}$ spaces are binned, and that
locations of the bin boundaries are beyond the control of the method.
However, it should be appreciated that
sample binning is an important part
of problem regularization. The unfolding problem
can even be fully regularized by
making very wide bins, much wider than the typical
scales associated with the detector resolution function.
This leads to a~discretized representation
of the response function by a diagonally dominant matrix
which is easily invertible. However, information about probability
density structures within each bin is now lost, as the actual
density is replaced by the uniform approximation. The
lack of knowledge about these structures gives rise to a
``systematic error'' on the unfolding result which is
subjective and difficult to formalize.

\section{Regularizing Expectation-Maximization Unfolding by Smoothing}
\label{sec:ems}

The standard expectation-maximization unfolding algorithm
iteratively updates the reconstructed values of $p({\bf x})$ according to the
formula~\cite{ref:richardson, ref:lucy, ref:dagounfold, ref:nychka, ref:smoothedem, ref:dempster, ref:vardi}
\begin{equation}
    \lambda_j^{(k+1)} = \frac{\lambda_j^{(k)}}{\epsilon_j} \sum_{i=1}^{n} \frac{K_{ij} y_i}{\sum_{\rho=1}^{m} K_{i\rho} \lambda_{\rho}^{(k)}}.
\label{eq:nosmooth}
\end{equation}
Here, $\lambda_j^{(k)}$ are the unnormalized $p({\bf x})$ values
(event counts) discretized on a grid in the physical process
space ${\bf x}$, obtained on a~$k$\supers{th} iteration. The index $j = 1, ..., m$ refers to
the~{\it linearized} cell number in this (possibly multidimensional) grid.
In this study, it will be assumed that all grid cells are small in comparison
with the typical scales associated with the response function.
All $\lambda_j^{(0)}$ values (the starting point for the iterations) can
be set to the same constant, $c = N/(\epsilon \, m)$, where $N$ is the
number of observed events and $\epsilon$ is the overall detector efficiency
for a~constant $p({\bf x})$. The number of observed events inside the
cell with linearized index $i$ ($i = 1, ..., n$) in the space
of observations ${\bf y}$ is denoted $y_i$.
Dimensionalities of the ${\bf x}$ and ${\bf y}$
spaces are arbitrary and can be different. $K_{ij}$ is the discretized response
matrix. It is the probability that
an event from the physical cell $j$ causes an observation in the cell
$i$ of the ${\bf y}$ space.
The detector efficiency for the physical cell $j$, $\epsilon_j$, is defined
by $\epsilon_j = \sum_{i=1}^{n} K_{ij}$.

These iterations are modified by introducing a smoothing step. The updating
scheme becomes
\begin{eqnarray}
    \lambda_j^{*(k+1)} & = & \frac{\lambda_j^{(k)}}{\epsilon_j} \sum_{i=1}^{n} \frac{K_{ij} y_i}{\sum_{\rho=1}^{m} K_{i\rho} \lambda_{\rho}^{(k)}}, \label{eq:semone} \\
    \lambda_r^{(k+1)} & = & \alpha^{(k+1)} \sum_{j=1}^m S_{r j} \lambda_j^{*(k+1)},
\label{eq:semscheme}
\end{eqnarray}
where $S_{r j}$ is the {\it smoothing matrix}. The smoothing step
normalization constant, $\alpha^{(k+1)}$, is
introduced\footnote{This normalization constant is not present in the
original formulation of the method intended for image analysis
applications~\cite{ref:nychka, ref:smoothedem}. Event count preservation
is substantially more desirable in the context of particle physics
data analyses, {\it e.g.}, for differential cross section measurements.} in order
to preserve the
overall event count obtained during the preceding expectation-maximization
step,
so that $\sum_{r=1}^m \lambda_r^{(k+1)} = \sum_{j=1}^m \lambda_j^{*(k+1)}$. The values $\lambda_j^{(\infty)}$ 
obtained upon iteration convergence are therefore solutions of the equation
\begin{equation}
    \lambda_r^{(\infty)} = \alpha^{(\infty)} \sum_{j=1}^m S_{r j} \frac{\lambda_j^{(\infty)}}{\epsilon_j} \sum_{i=1}^{n} \frac{K_{ij} y_i}{\sum_{\rho=1}^{m} K_{i\rho} \lambda_{\rho}^{(\infty)}},
\label{eq:semeq}
\end{equation}
where $\alpha^{(\infty)} = \sum_{r=1}^m \lambda_r^{*(\infty)} / \sum_{r=1}^m \sum_{j=1}^m S_{r j} \lambda_j^{*(\infty)}$.
Convergence  of the smoothed iterations to this
fixed point has been established in~\cite{ref:smoothedem} under
the sufficient condition that the smoothing operator (for this case, the product
$\alpha^{(k+1)} {\bf S}$) has no eigenvalues greater than unity
in absolute value.

The equation for the error propagation matrix,
$J_{rs} \equiv \frac{\partial \lambda_r^{(\infty)}}{\partial y_s}$, is obtained
by differentiating Eq.~\ref{eq:semeq} with respect to
$y_s$. In the matrix notation, this equation is
\begin{equation}
{\bf J} = (\alpha^{(\infty)} {\bf S} + {\bf A}) \left({\bf M} + {\bf B J}\right),
\label{eq:errprop}
\end{equation}
where
\begin{eqnarray}
    A_{j q} & = & \frac{\left(1 - \alpha^{(\infty)} \sum_{r=1}^m S_{rq} \right) \lambda_j^{(\infty)}}{\sum_{i=1}^{m} \lambda_i^{(\infty)}}, \label{eq:errpropmatricesa} \\
    B_{j q} & = & \frac{\delta_{j q}}{\epsilon_j} \sum_{i=1}^{n} \frac{K_{ij } y_i}{\hat{y}_i} -  \frac{\lambda_j^{(\infty)}}{\epsilon_j} \sum_{i=1}^{n} \frac{K_{iq} K_{ij} y_i}{\hat{y}_i^2},\\
    M_{j q} & = & \frac{\lambda_j^{(\infty)} K_{q j }}{\epsilon_j \, \hat{y}_q},
\end{eqnarray}
and $\hat{y}_i = \sum_{\rho=1}^m K_{i\rho} \lambda_{\rho}^{(\infty)}$ is the
fitted number of observed events in the cell with index $i$.
In practice, an equation equivalent to~\ref{eq:errprop}, $({\bf I} - (\alpha^{(\infty)} {\bf S} + {\bf A}) {\bf B}) \, {\bf J} = (\alpha^{(\infty)} {\bf S} + {\bf A}) {\bf M}$,
can be solved using the LU
factorization algorithm
(for example, as implemented in LAPACK~\cite{ref:lapack}).
Assuming that the covariance matrix of observations is denoted by ${\bf V}$,
the covariance matrix of the unfolded values, ${\bf U}$,
is estimated according to
\begin{equation}
{\bf U} = {\bf J} {\bf V} {\bf J}\supers{T}.
\label{eq:covmat}
\end{equation}

It has been suggested that, upon convergence of the
smoothed expectation-maximization iterations, one extra iteration
without the smoothing step should be performed~\cite{ref:dagostini2}.
While the utility of adding such an iteration is
questionable\footnote{Instead, the author suggests that the theory predictions
  should be processed with the smoothing matrix prior to comparison
  with unfolded results. This leads to a reduction in the unfolding
  bias introduced by regularization.
  In addition, in many situations of practical interest
  ({\it e.g.}, for QCD differential
  cross section calculations beyond the leading order in perturbation
  theory) an~algorithm for direct evaluation of $p({\bf x})$
  is not available, and only
  a~Monte Carlo procedure for sampling from $p({\bf x})$
  can be devised. Assuming that the theory sample is sufficiently copious,
  the smoothing matrix can be used to construct $p({\bf x})$ estimate
  from the empirical density function of this sample
  in the manner appropriate for
  subsequent comparison with unfolded observations.}, the error
propagation matrix for this approach, ${\bf J}^{*}$,
is given by ${\bf J}^{*} = {\bf M} + {\bf B J}$. Subsequently,
the covariance matrix of the unfolded result should be estimated as
${\bf U} = {\bf J}^{*} {\bf V}^{*} {\bf J}^{*\mbox{\scriptsize T}}$,
where the
covariance matrix of observations, ${\bf V}^{*}$, is obtained using the
$\lambda_j^{* (\infty + 1)}$ values defined by Eq.~\ref{eq:semone}
and processed according to Eq.~\ref{eq:kernelaction}.

Constructed with an appropriate method, the smoothing matrix
${\bf S}$ can have a~rather intuitive interpretation. Setting
${\bf S} = {\bf I}$ leads to the maximum likelihood estimate of
the $m$ parameters $\lambda_j^{(\infty)}$~\cite{ref:dempster, ref:vardi}.
The number of these
parameters can be large and, according to the Cramer–-Rao bound, 
the amount of information present in the observed data is insufficient
to constrain them with limited variance. The smoothing matrix introduces
a reasonable assumption that the ``nearby'' unfolded values should not
be very different. This assumption
shifts the bias-variance tradeoff of the maximum
likelihood estimate towards a strong reduction of the variance at the
cost of a slight increase in the estimator bias.

Finally, it should be noted that the error propagation formulae
are simplified for doubly stochastic smoothing
matrices\footnote{These are the matrices with non-negative real elements
satisfying the simultaneous conditions $\sum_{q=1}^m S_{rq} = 1$ for all $r$ 
and $\sum_{r=1}^m S_{rq} = 1$ for all $q$
({\it i.e.}, each
row and each column sum up to~1). Note that any reasonable smoother
should map the uniform density into itself and, therefore,
satisfy the row summation conditions.}. For
such matrices, $\alpha^{(k)} = 1$ for all~$k$ and ${\bf A} = {\bf 0}$.
This simplification is important for software implementations utilizing sparse
matrix algorithms as ${\bf A}$ is, generally, not sparse.

\section{Effective Number of Fitted Parameters}
\label{sec:effectivenum}

In various unfolding procedures, it is desirable to choose the
regularization strength by an~automatic criterion amenable to simulation
studies. A number of
possible choices are based on the comparison of the fit with
the observed data. A~least squares matching criterion is
advocated, for example,
in~\cite{ref:lindemann} and~\cite{ref:zech11}. In order to properly
estimate the $p$-value resulting from the application of
such a~criterion, one has to estimate the number of parameters
used to fit the data.

We propose to estimate the number of fitted parameters
from the effective rank of the matrix
${\bf K} {\bf J} {\bf J}\supers{T} {\bf K}\supers{T}$.
This proposal is based on the following argument\footnote{One can
also argue that the ${\bf K} {\bf J}$ matrix
plays a similar role to the hat matrix in linear regression problems.
This leads to the same conclusion about the number of model parameters.}.
The covariance matrix of the fitted folded values ({\it i.e.}, $\hat{y}_i$) is
${\bf V}_{\hat{y}}({\bf V}) = {\bf K} {\bf U} {\bf K}\supers{T} = {\bf K} {\bf J} {\bf V} {\bf J}\supers{T} {\bf K}\supers{T}$. 
If, using polynomial series ({\it e.g.}, Legendre polynomials),
one fits multiple independent
samples of random points taken from the uniform distribution, with
the number of points per sample varying according to the Poisson distribution,
the rank of the covariance matrix of the
fitted unnormalized density values calculated over these samples
equals the degree of the fitted polynomial plus one.
This is precisely the number of parameters of the fitted model.
It does not matter how many abscissae are used to construct the covariance
matrix of the fitted values as long as the number
of abscissae exceeds the degree of the polynomial and the
average number of points in the samples is ``large enough''.
While the model fitted to the observed values by various unfolding procedures
is not polynomial, some measure of the rank of 
${\bf V}_{\hat{y}}({\bf I}) = {\bf K} {\bf J} {\bf J}\supers{T} {\bf K}\supers{T}$ can still be identified with the number of model parameters.

The~effective rank of a symmetric positive-semidefinite matrix
(say, {\bf Q}) can be estimated in at least two different ways.
The first one is the exponent 
of the von Neumann entropy~\cite{ref:vonneumann}
of ${\bf Q}/\mbox{tr} ({\bf Q})$. In terms
of the ${\bf Q}$ eigenvalues, $e_i$, it is expressed
as\footnote{Naturally, $\mbox{erank}_1({\bf Q})$ is also the exponent of the Shannon entropy of the normalized eigenspectrum.}
\begin{equation}
        \mbox{erank}_1({\bf Q}) = \exp \left\{ -\sum_{i=1}^n \frac{e_i}{\|e\|} \ln \left( \frac{e_i}{\|e\|} \right) \right\},
        \ \ \ \|e\| = \sum_{i=1}^n e_i.
\label{eq:erank1}
\end{equation}
The second definition finds the ratio of the matrix trace to the largest eigenvalue~\cite{ref:vershynin}:
\begin{equation}
         \mbox{erank}_2({\bf Q}) = \frac{\mbox{tr} ({\bf Q})}{\max_{1 \le i \le n} e_i} = \frac{\|e\|}{\max_{1 \le i \le n} e_i}.
\label{eq:erank2}
\end{equation}
It is not obvious {\it a priori} which effective rank definition will lead
to better practical results for different regularization strength
selection criteria.

This method of calculating the effective number of parameters
in the fit can be applied in any unfolding procedure that supplies
the matrix for propagating errors from the observed data to the
unfolded values. It should be noted that the effective
number of fitted parameters can be overestimated in case the
bins in the observation space are only sparsely populated.
A possible adjustment of the method that can be applied in
such cases is discussed in Section~\ref{sec:adaptiveregularization}.

\section{Choosing the Smoothing Parameters for the EMS Unfolding}
\label{sec:emschoice}

For likelihood-based inference, a useful model selection principle
is provided by the Akaike information criterion (AIC)~\cite{ref:aic}.
The AIC adjusted for the finite sample size is~\cite{ref:modelsel}
\begin{equation}
AIC_c = -2 \ln L + 2 k + \frac{2 k (k + 1)}{N - k - 1},
\label{eq:aicc}
\end{equation}
where $L$ is the model likelihood,
$N$ is the sample size, and $k$ the number of parameters
in the model. Selecting a model
by minimizing $AIC_c$ avoids overfitting by reaching a compromise
between complexity of the model and goodness-of-fit.
On the use of AIC in density estimation scenarios without deconvolution
consult, for example,~\cite{ref:loader}.

With the number of model parameters estimated by the method
presented in the previous section, application of the $AIC_c$
to the EMS unfolding procedure is straightforward.
The likelihood is calculated in the
assumption of Poisson distributed counts with means
given by $\hat{y}_i$. If, for a~one-dimensional
unfolding problem, the smoothing
matrix is constructed
utilizing a~typical spatial bandwidth parameter $h$ then,
for small $h$, $k \propto h^{-1}$ represents the effective number of independent
intervals in the support of $p(x)$. In the absence of smoothing, the
likelihood tends to a constant value obtained with the Poisson means
derived by iterating Eq.~\ref{eq:nosmooth} to convergence.
Therefore, for small
values of $h$, $AIC_c$ is a~decreasing function of $h$.
At the same time, decrease in $k$ is not so pronounced for
larger values of $h$ while the likelihood is rapidly diminishing
when $h$ becomes comparable with the typical
spatial scale of the response function. Combined together,
these dependencies result in the $AIC_c$ behavior which exhibits
a minimum at some value of $h$.

The $AIC_c$ criteria that
correspond to the two different definitions of the effective
rank will be called $EAIC_c$ ($E$ in this abbreviation stands
for ``entropic'') and $TAIC_c$ ($T$ stands for ``trace'').
One of these criteria can be optimized as a~function of the
bandwidth parameter(s) used to construct the smoothing matrix.

Use of an effective rank to determine the number of model parameters
leads to the requirement that the number of discretization cells in
the observation space ${\bf y}$ should be substantially larger than
this rank. This condition should be verified once the 
unfolding is performed with the optimal smoother.

Data-dependent choice of regularization parameters introduces
an~additional uncertainty on the unfolded results due to the
variability of these parameters. While this uncertainty remains
a~subject of recent~\cite{ref:kuuselapreprint} and, potentially,
future studies, use of an automated criterion represents a~substantial
improvement upon the subjective choice of regularization strength widespread
in the current unfolding practice in particle physics.

\section{Simulation Results}
\label{sec:simulation}
\subsection{Unfolding Simulation Setup}

\begin{figure}[h!]
\begin{center}
\includegraphics[width=.49\textwidth]{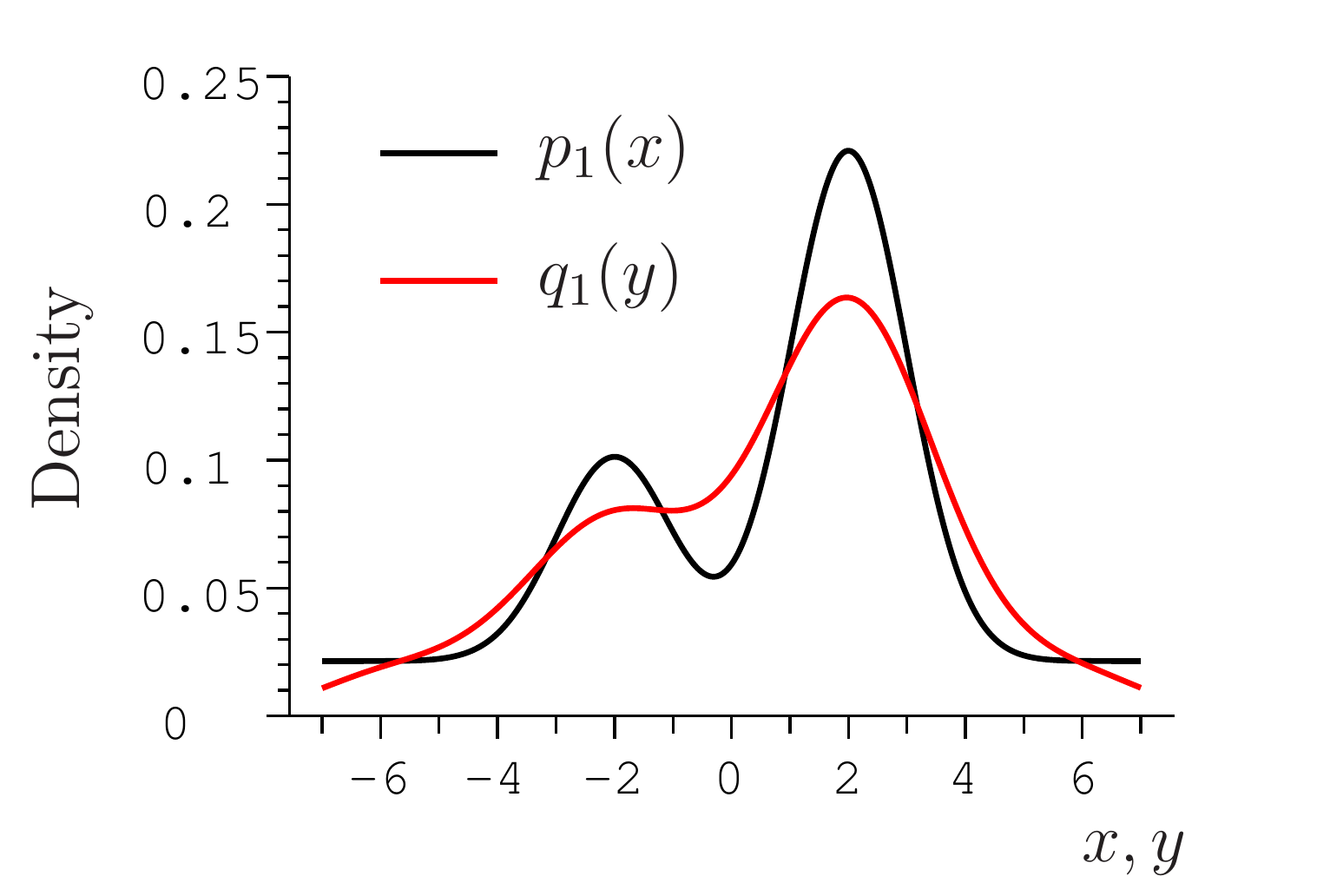} 
\caption{The first example density.}
\label{fig:density1}
\end{center}
\end{figure}

The concepts and the formulae discussed in the preceding sections
are illustrated with two examples. The first example
density consists of two Gaussian peaks
combined with the uniform distribution:
$p_1(x) = 0.2 \, {\cal N}(-2, 1) + 0.5 \, {\cal N}(2, 1) + 0.3 \, U(-7, 7)$.
$U(a, b)$ stands for the uniform distribution on the interval $[a, b]$.
The support of this distribution mixture is restricted to the interval $[-7, 7]$
in the physical process space, truncating small
fractions of the Gaussian tails outside of this interval.
The response function for this example is $K_1(y, x) = {\cal N}(y - x, 1)$.
This simulation setup was used
previously to investigate performance of
a~different unfolding technique~\cite{ref:kuuselapreprint}.
The density used in this example
is illustrated in Fig.~\ref{fig:density1}, together
with the corresponding density $q_1(y)$ for the observed data.

\begin{figure}[h!]
\begin{center}
\includegraphics[width=.49\textwidth]{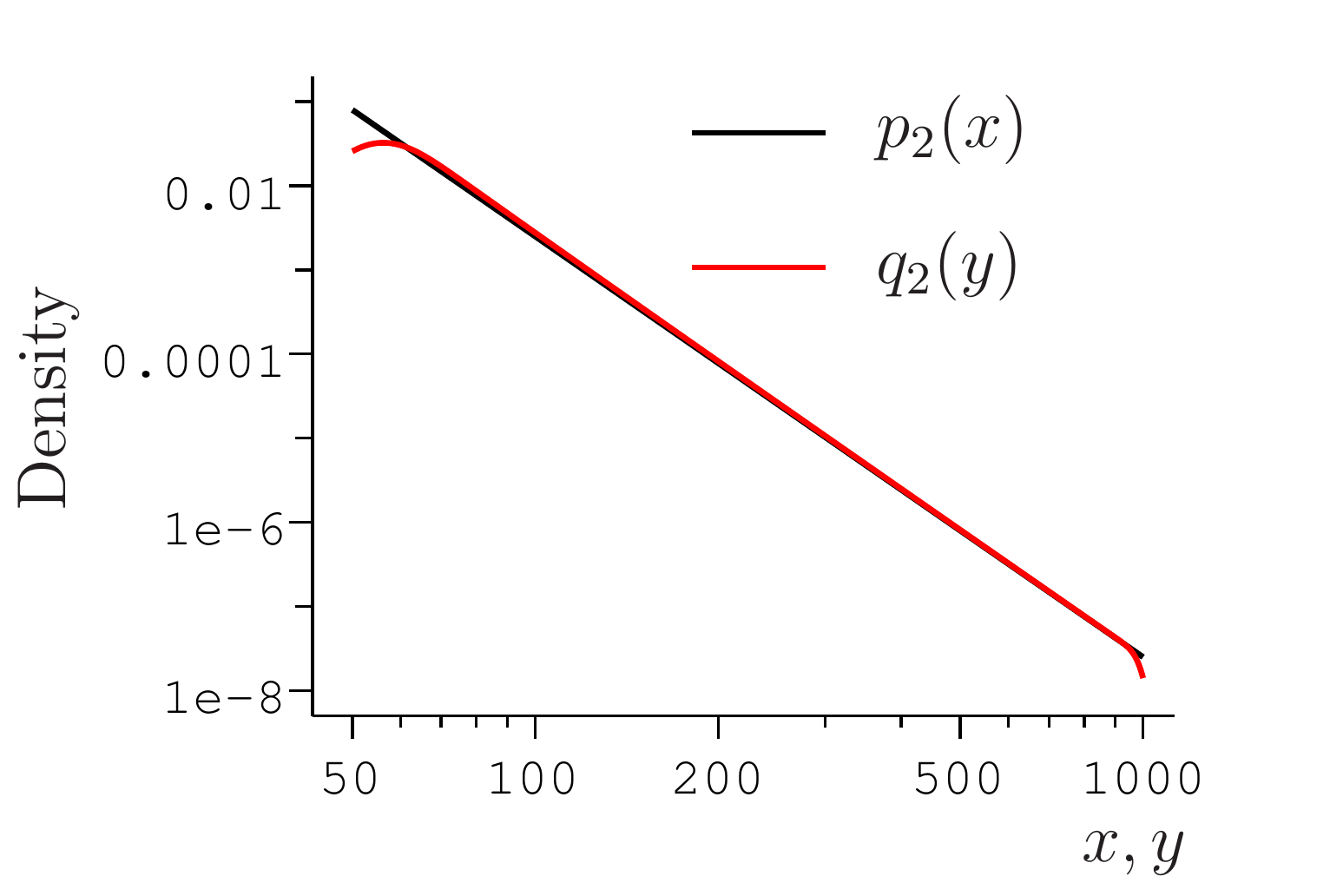} 
\includegraphics[width=.49\textwidth]{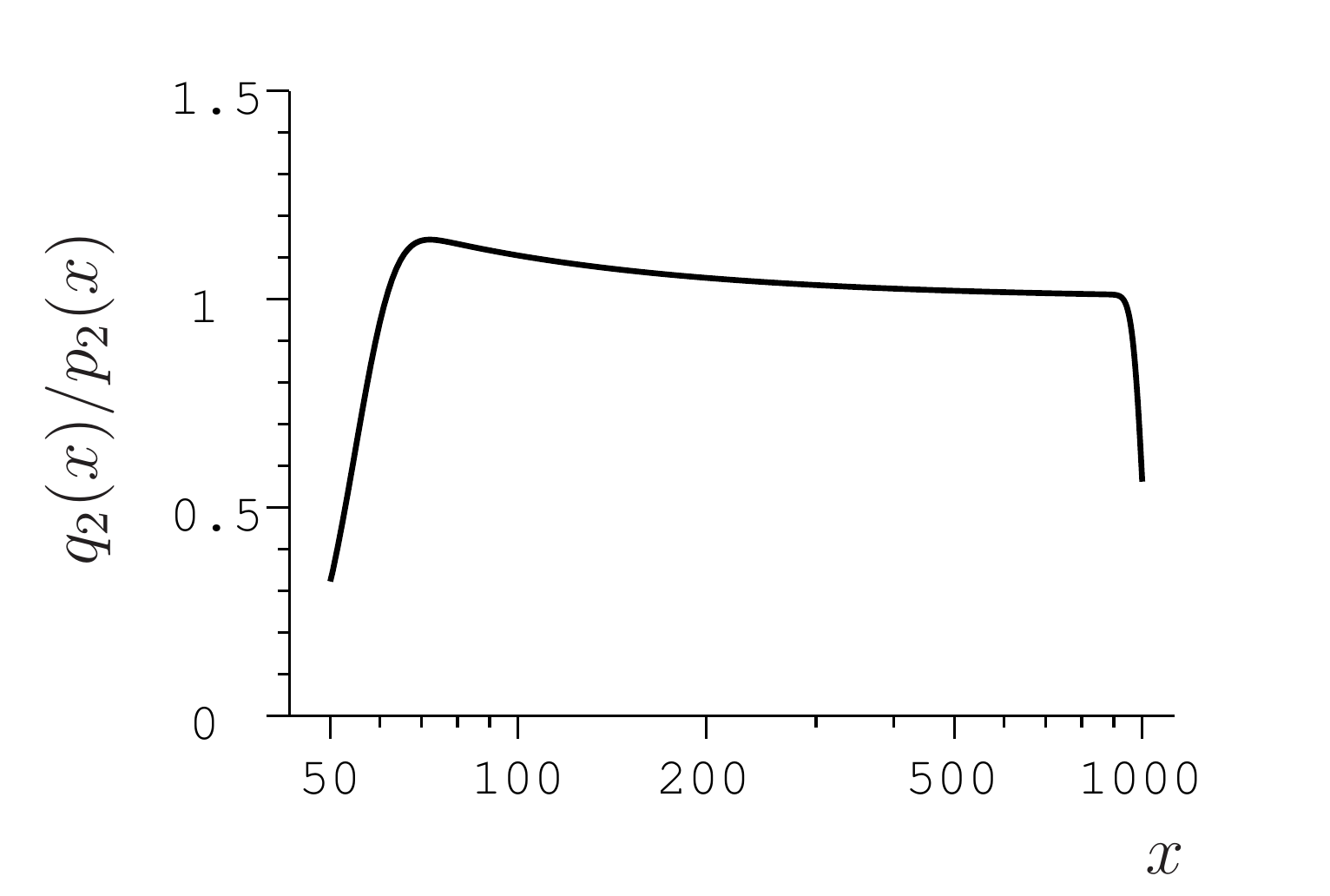}
\caption{The second example density.}
\label{fig:density2}
\end{center}
\end{figure}

The second density
is a Pareto distribution restricted to the interval $[50, 1000]$:
$p_2(x) = 4 \times 10^{12} \, x^{-5} / (16 \times 10^4 - 1)$.
In this case, the response function is chosen as
$K_2(y, x) = {\cal N}(y - x, x)$. This example emulates a
differential cross section measurement of jet production in 
proton-proton collisions, $d \sigma / d p_T$,
with jet transverse momentum resolution $\delta p_T  / p_T = 100\% / \sqrt{p_T}$.
Densities for this example are shown in Fig.~\ref{fig:density2}.
For both examples, the observed space is limited to the same interval as
the physical process space in order to 
illustrate effects of imperfect detection efficiency and to simplify plotting.

Points are drawn at random from the example densities in the
physical process space, and additional
random shifts are added to the point coordinates by sampling
the respective response functions. In order to unfold the first example
density, the space of observations is discretized into 100 uniformly spaced
bins. The physical process space is split into 420 bins, so that the standard
deviation of the response function is 30 times larger than the bin width.
For the second example, the bin width is taken to be
proportional to the width of
the response function ({\it i.e.}, the binning is uniform in the $\sqrt{x}$
variable). With 1000 bins in the physical process space, the response
is $\approx 20$~times wider than the bins.
The same binning scheme (uniform in $\sqrt{y}$)
is used for the observed data, with 200 bins.

\begin{figure}[h!]
\begin{center}
\includegraphics[width=.325\textwidth]{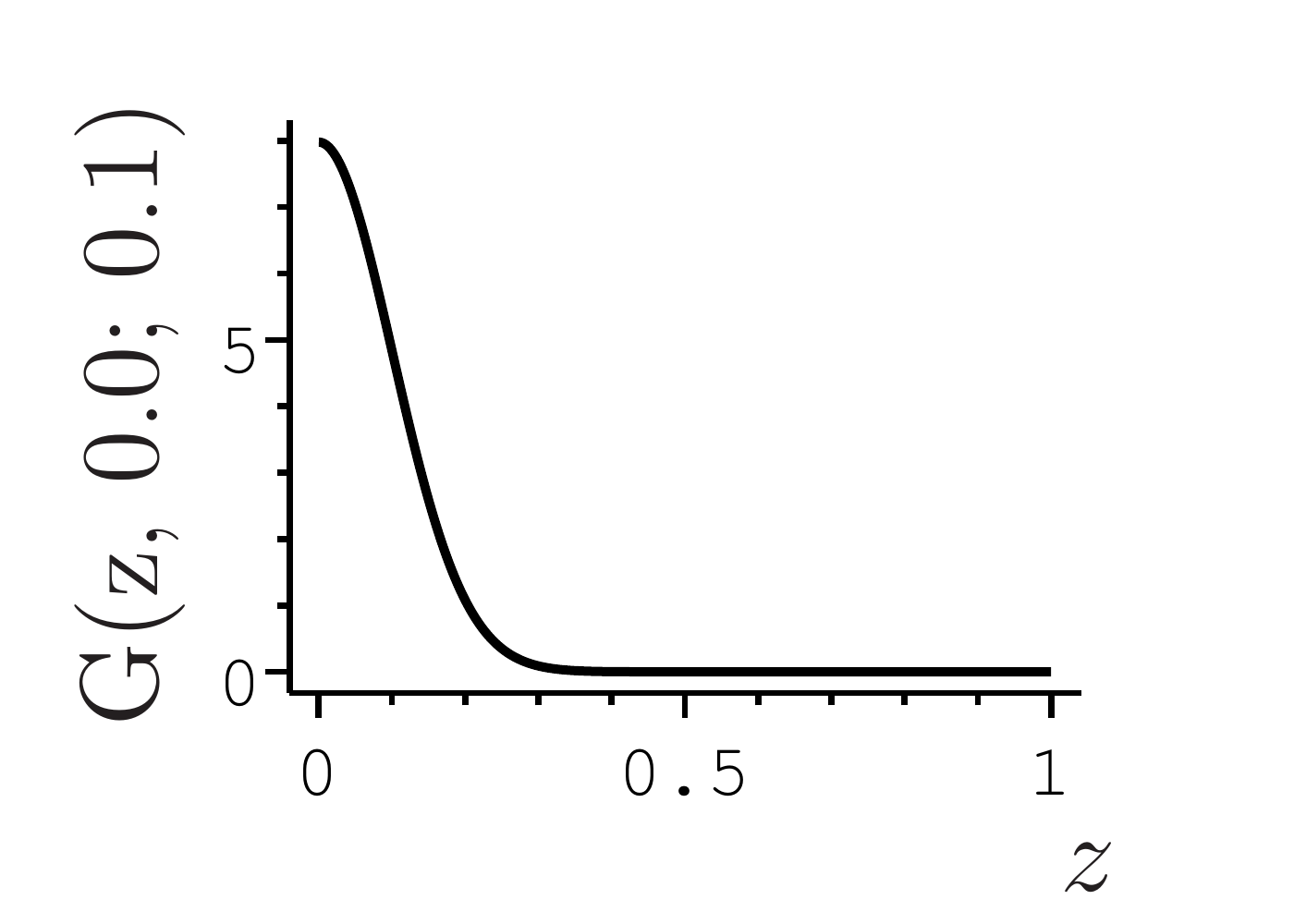} 
\includegraphics[width=.325\textwidth]{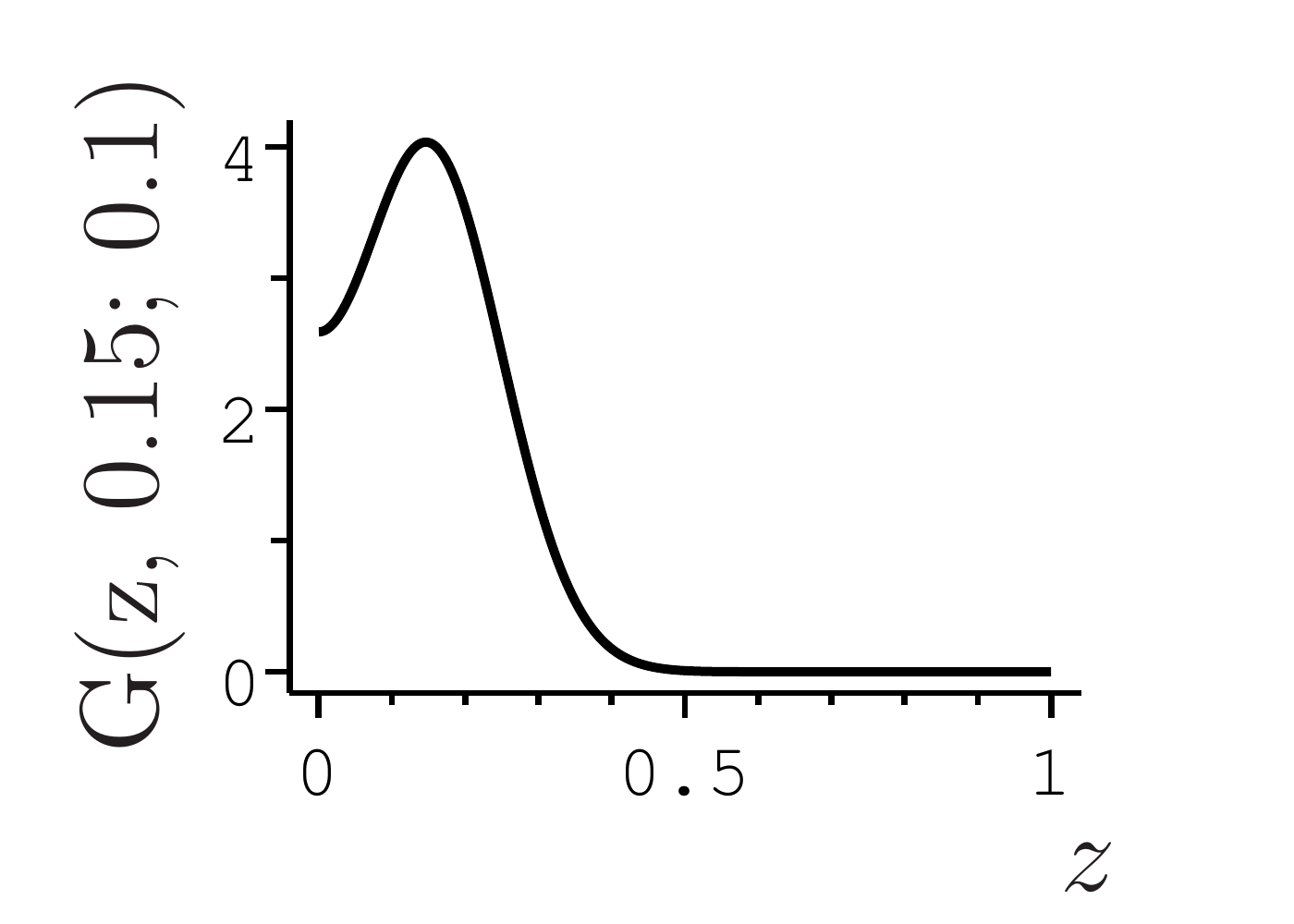} 
\includegraphics[width=.325\textwidth]{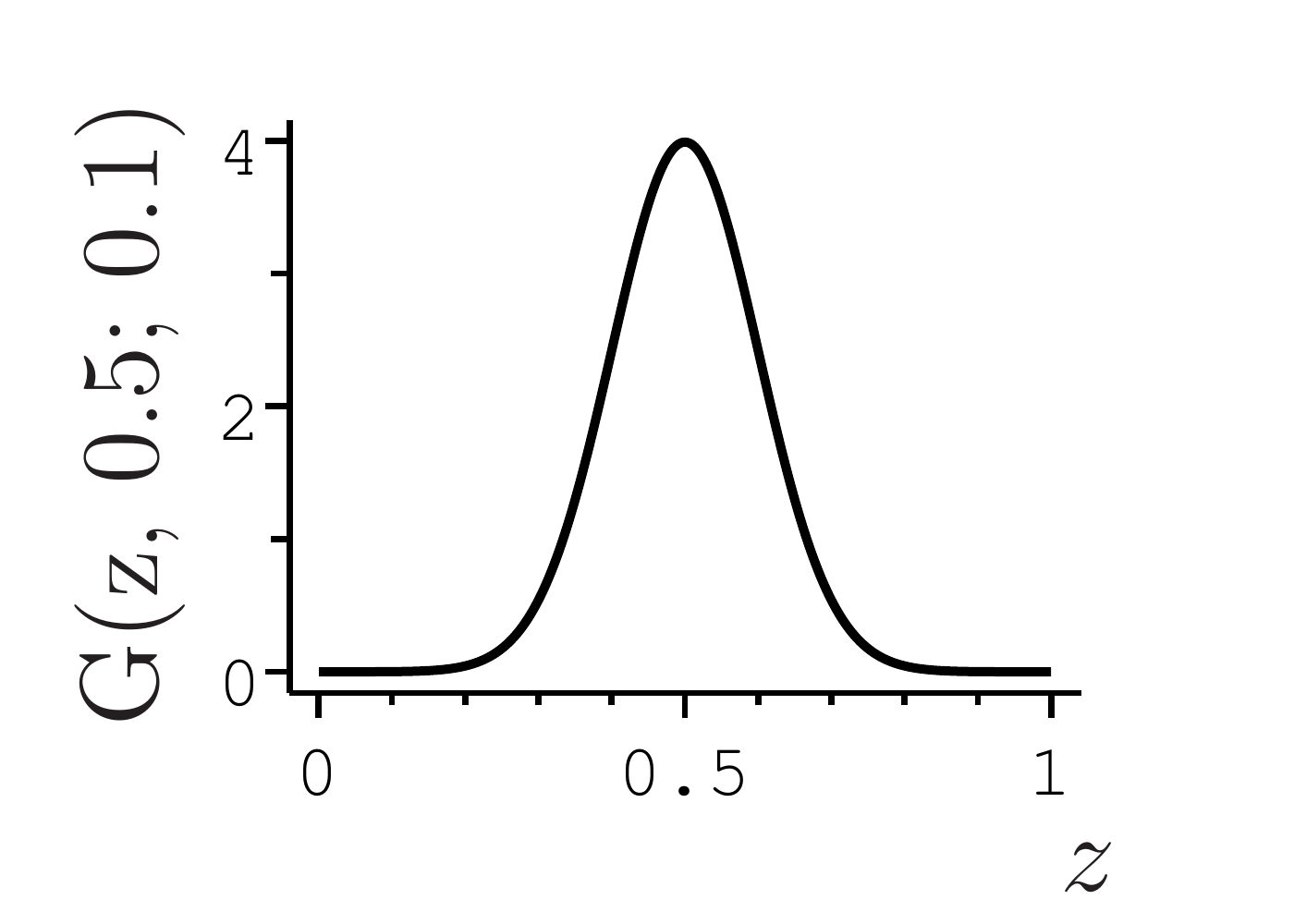} 
\caption{Example dependencies of $G(z, \xi; h)$ on $z$ for
bandwidth $h = 0.1$ and three different values of $\xi$ indicated
in the vertical axis labels ($\xi$ at the boundary, not far from the boundary,
and at the interval center). For illustration purposes,
in these plots
the ratio of $h$ to the length of the $z, \xi$ interval is significantly
larger than the typical ratios employed in the unfolding examples.
Individual
elements of the smoothing matrix are obtained by integrating $G(z, \xi; h)$
over the corresponding discretization cells $\Delta z \times \Delta \xi$.}
\label{fig:gf}
\end{center}
\end{figure}

For the purpose of this study, the smoothing matrix is generated by
discretizing the Green's function of the homogeneous heat equation in
one dimension with Neumann boundary conditions ({\it i.e.}, with
boundaries not permeable to the heat transfer). This function is
fully determined by a~single parameter: product of time, $t$, and
thermal diffusivity, $\alpha$. For the convenience of presentation, the bandwidth
parameter $h$ used in this study is the standard deviation of the Green's
function far away from the density support boundaries: $h = \sqrt{2 \alpha t}$.
On the $[0, 1]$ interval, this function is given
by\footnote{This representation of the Green's
function is easily obtained by the method of
images~\cite{ref:duffy}.
Due to very fast convergence for small values of $h$,
these series are better suited for numerical
evaluation than the mathematically equivalent
trigonometric series derived by separation
of variables.}
$G(z, \xi; h) = \sum_{i=-\infty}^{\infty} \left[{\cal N}(z - \xi + 2\,i, h^2) + {\cal N}(z + \xi + 2\,i , h^2)\right]$ with $z, \xi \in [0, 1]$. This function is illustrated in Fig.~\ref{fig:gf}.
For an arbitrary interval $[a, b]$,
the Green's function values can be obtained by appropriate
shifting and scaling of its arguments, including the bandwidth.
The $z$ variable maps to the row number of the smoothing matrix while
$\xi$ maps to the column number.
Multiplication of the smoothing matrix by a~vector in the
discretized physical process space thus approximates the integral
$p\sub{smoothed}(x) = \int_0^1 G(x, \xi; h) \, p(\xi) \, d \xi$.

As $\int_{0}^1 G(z, \xi; h) \, dz = \int_{0}^1 G(z, \xi; h) \, d\xi = 1$,
the smoothing matrix constructed in this manner is doubly stochastic.
All of its eigenvalues belong to the $[0, 1]$ interval
which guarantees convergence of the EMS
unfolding iterations. For the second example density, $\sqrt{x}$
is used in place of the Green's function arguments $z$ and $\xi$, so that
the characteristic bandwidth of the smoothing matrix is increasing
concordantly with the response function resolution.
Both the smoothing matrix construction and the subsequent EMS
unfolding are performed with the aid of the NPStat software
package~\cite{ref:npstat}.

\begin{figure}[h!]
\begin{center}
\includegraphics[width=.49\textwidth]{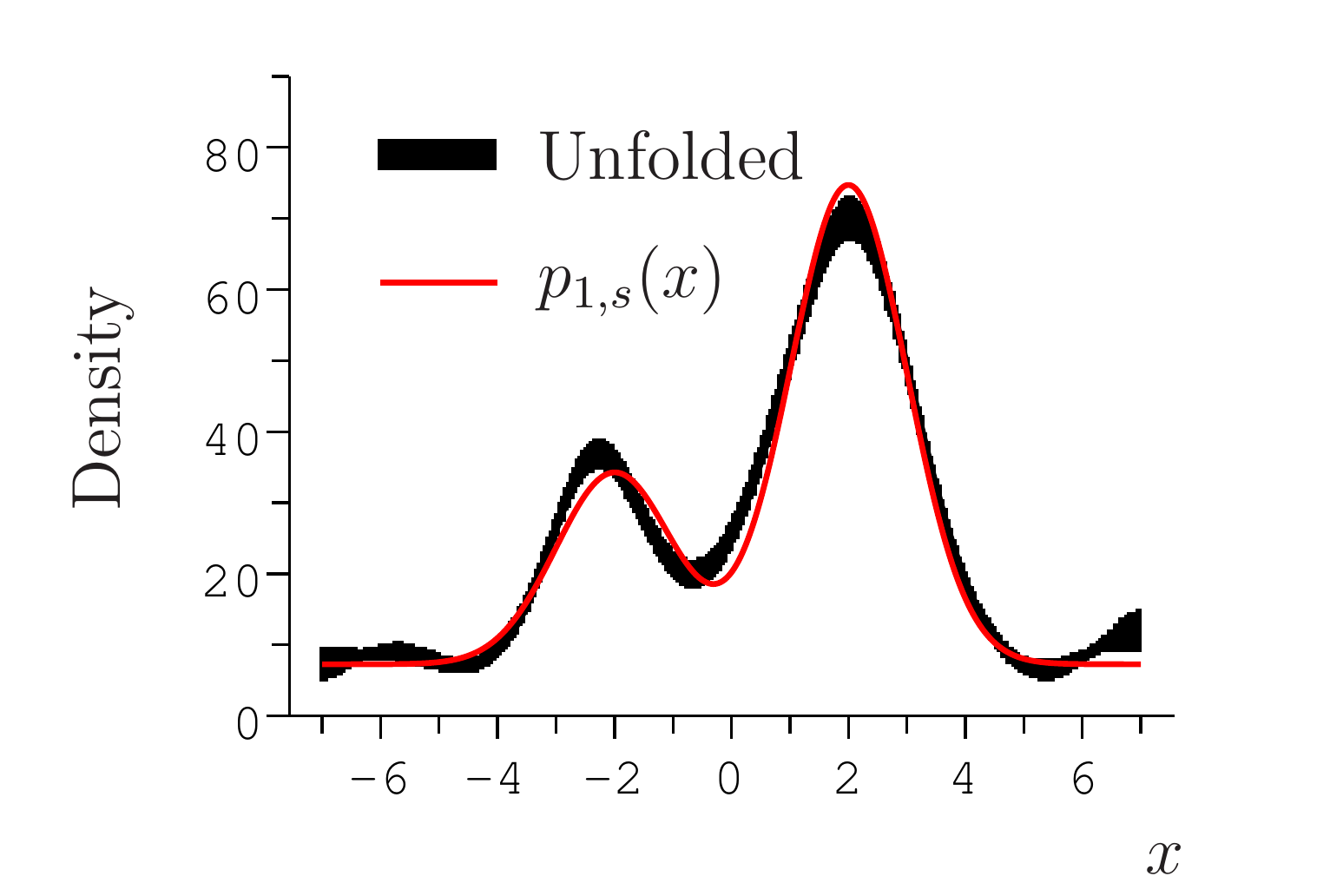} 
\includegraphics[width=.49\textwidth]{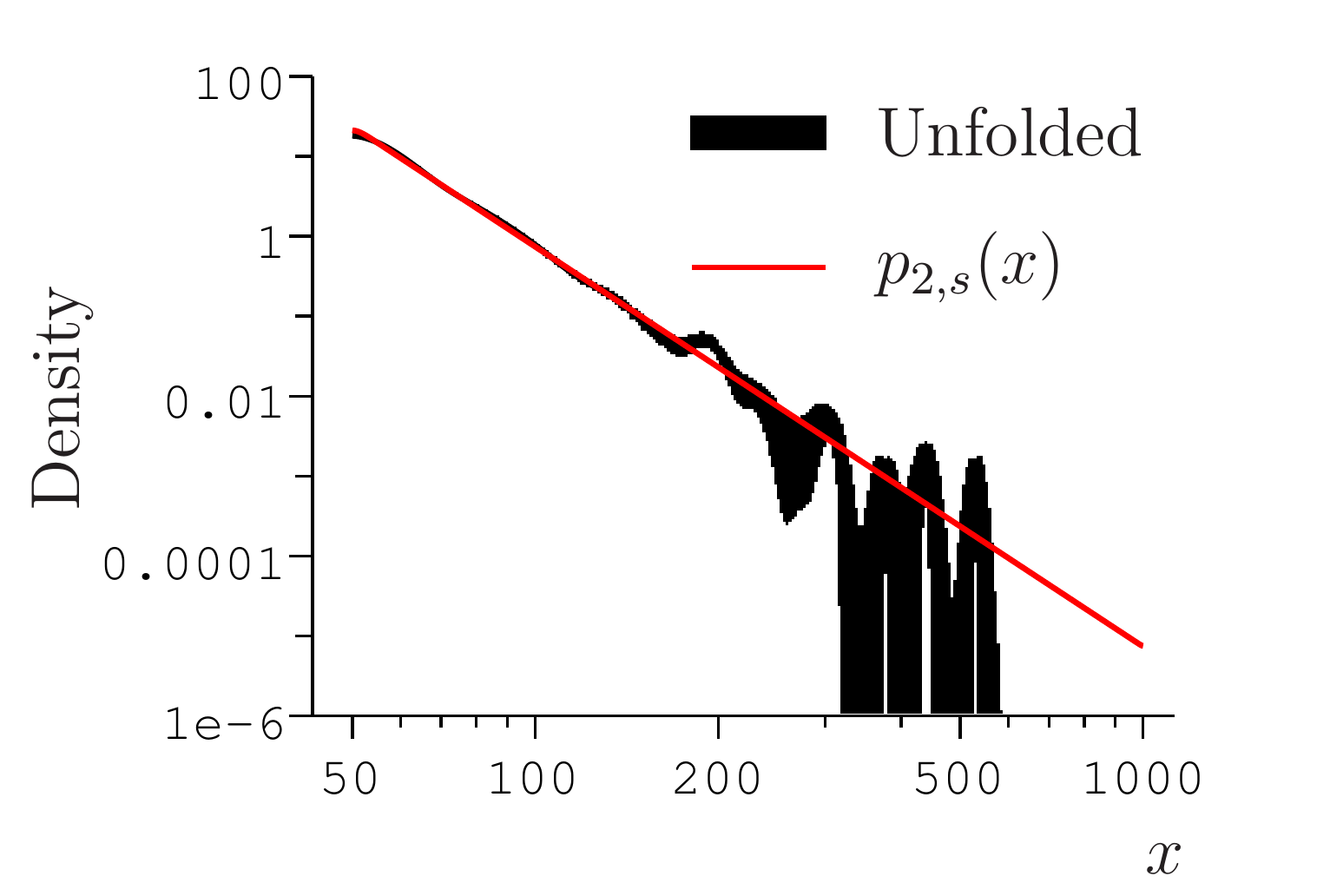}
\caption{Example unfolding results. Subscript $s$ stands for ``smoothed''.
         The smoothing matrix bandwidth is $h = 0.08$ for the plot on the
         left (the first model)
         and $h = 0.11$ in the $\sqrt{x}$ space for the plot
         on the right (the second model).
         The filled black regions correspond to $\pm 1$ standard
         deviation interval estimates.
         Poisson uncertainties on the observed event counts
         are propagated to the unfolded results
         according to Eqs.~\ref{eq:errprop} and~\ref{eq:covmat}.
         It is apparent that, for the second model, the event sample size
         is insufficient to reconstruct the density reliably
         at large values of $x$.}
\label{fig:unfexample}
\end{center}
\end{figure}

Example unfolded results for random samples containing 10,000
simulated observations each are shown in Fig.~\ref{fig:unfexample}.
In this figure, the unfolded densities are compared with the
corresponding physical process densities processed by the respective
smoothing matrices.

\begin{figure}[h!]
\begin{center}
\includegraphics[width=.49\textwidth]{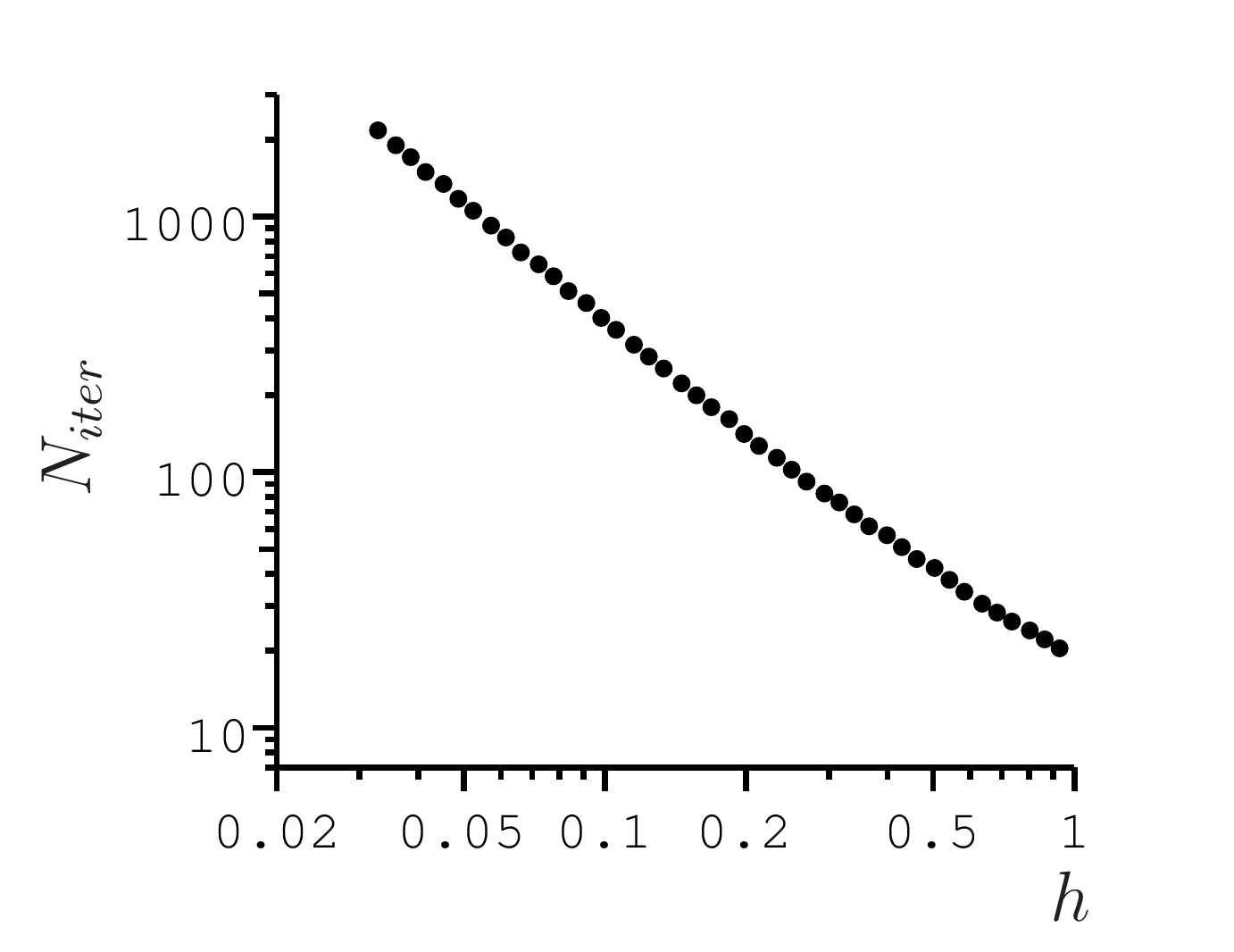} 
\includegraphics[width=.49\textwidth]{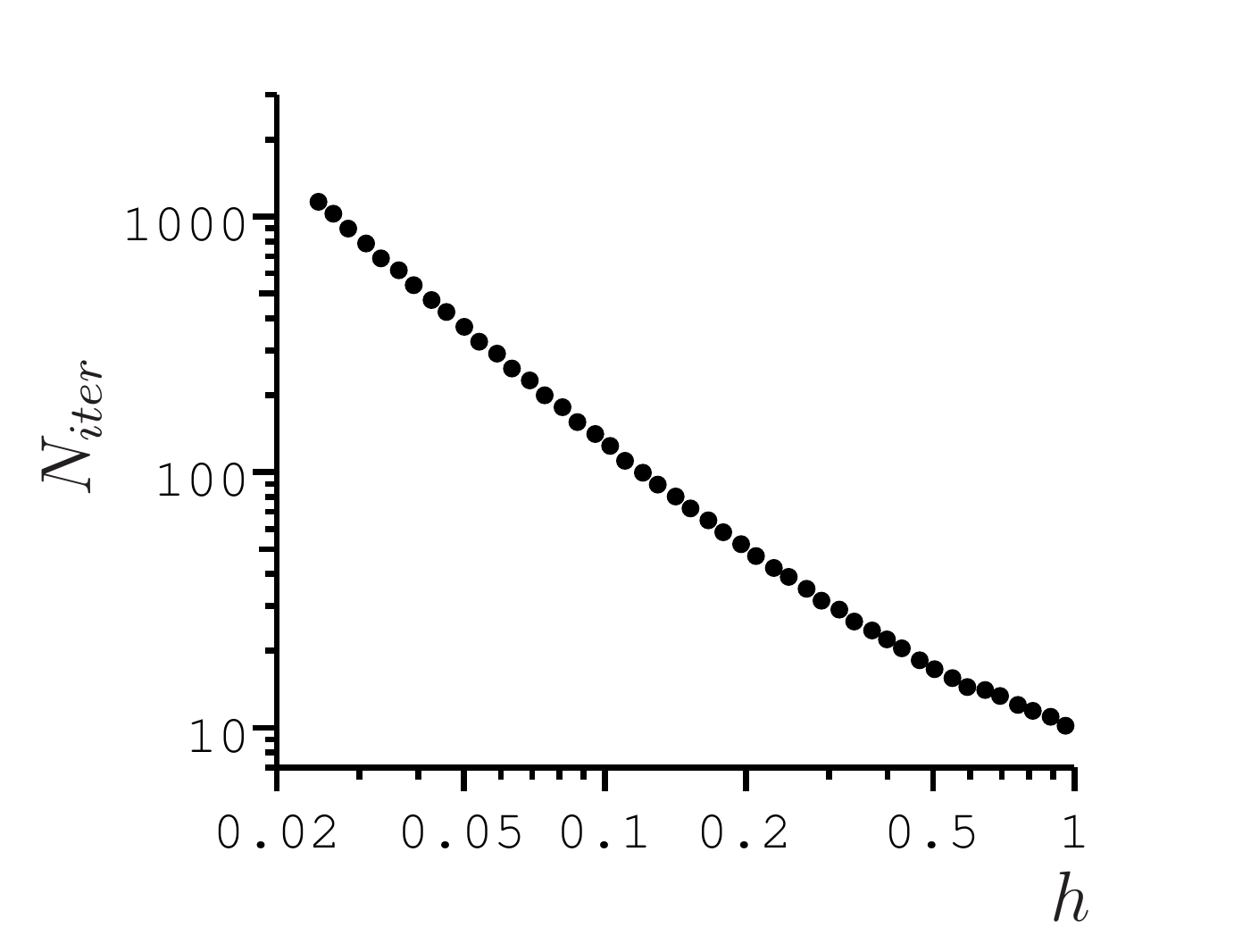}
\caption{Left: dependence of the average number of iterations needed
         for convergence, $N_{iter}$, on the
         smoothing matrix bandwidth for the first example distribution.
         1,000 points per sample were generated on average. Right: the same dependence
         for the second example distribution, with 10,000 points per sample on average.
         The number of iterations does not change
         appreciably from one simulated sample to another, so
         standard deviations would not be visible in these plots.
         The decrease in $N_{iter}$ with increasing $h$ is in qualitative
         agreement with the discussions of EMS convergence
         rates in~\cite{ref:nychka, ref:smoothedem}.}
\label{fig:niter}
\end{center}
\end{figure}

In the simulation studies presented in this manuscript,
EMS unfolding iterations are declared converged upon reaching
the iteration number $k > 0$ satisfying the condition
$\frac{\sum_{j=1}^{m}\left|\lambda_j^{(k)} - \lambda_j^{(k-1)}\right|}{\sum_{j=1}^{m} \left(\lambda_j^{(k)} + \lambda_j^{(k-1)}\right)} \le \frac{\varepsilon}{2}$, where $m$ is the
number of cells in the discretization of $x$ space, and
$\varepsilon = 10^{-9}$. This condition results in a~typical relative difference
between $\lambda_j^{(k)}$ and $\lambda_j^{(\infty)}$ of the order $\varepsilon$.
This value of $\varepsilon$ is chosen so that the relative precision of
$\hat{y}$ evaluation (which is on the order of $\varepsilon \sqrt{m}$)
is substantially better than $1/N$ for all sample sizes considered.
The dependence of the number of iterations needed for
convergence on the smoothing matrix bandwidth is illustrated in Fig.~\ref{fig:niter}.
Note that, for an~implementation of the EMS unfolding
algorithm utilizing dense matrices, the computational complexity
of each iteration is ${\cal O}(m^2)$ while
the computational complexity of solving Eq.~\ref{eq:errprop} is ${\cal O}(m^3)$.
Therefore, the iterative stage dominates the computation time only for
very small bandwidth values.

\subsection{Fixed Regularization Strength}

For the coverage studies presented in this article, the total
sample counts are allowed to fluctuate according to the Poisson
distribution with averages described in the text or in the figure
captions. The number of points actually generated is larger
(on average, by factor $1/\epsilon$) because
only the points that end up inside the observation intervals
after the addition of random shifts are counted.
Example densities processed by the smoothing matrix are used
as references in coverage calculations without bias
correction (after bias correction
the reference no longer matters).

\begin{figure}[h!]
\begin{center}
\includegraphics[width=.49\textwidth]{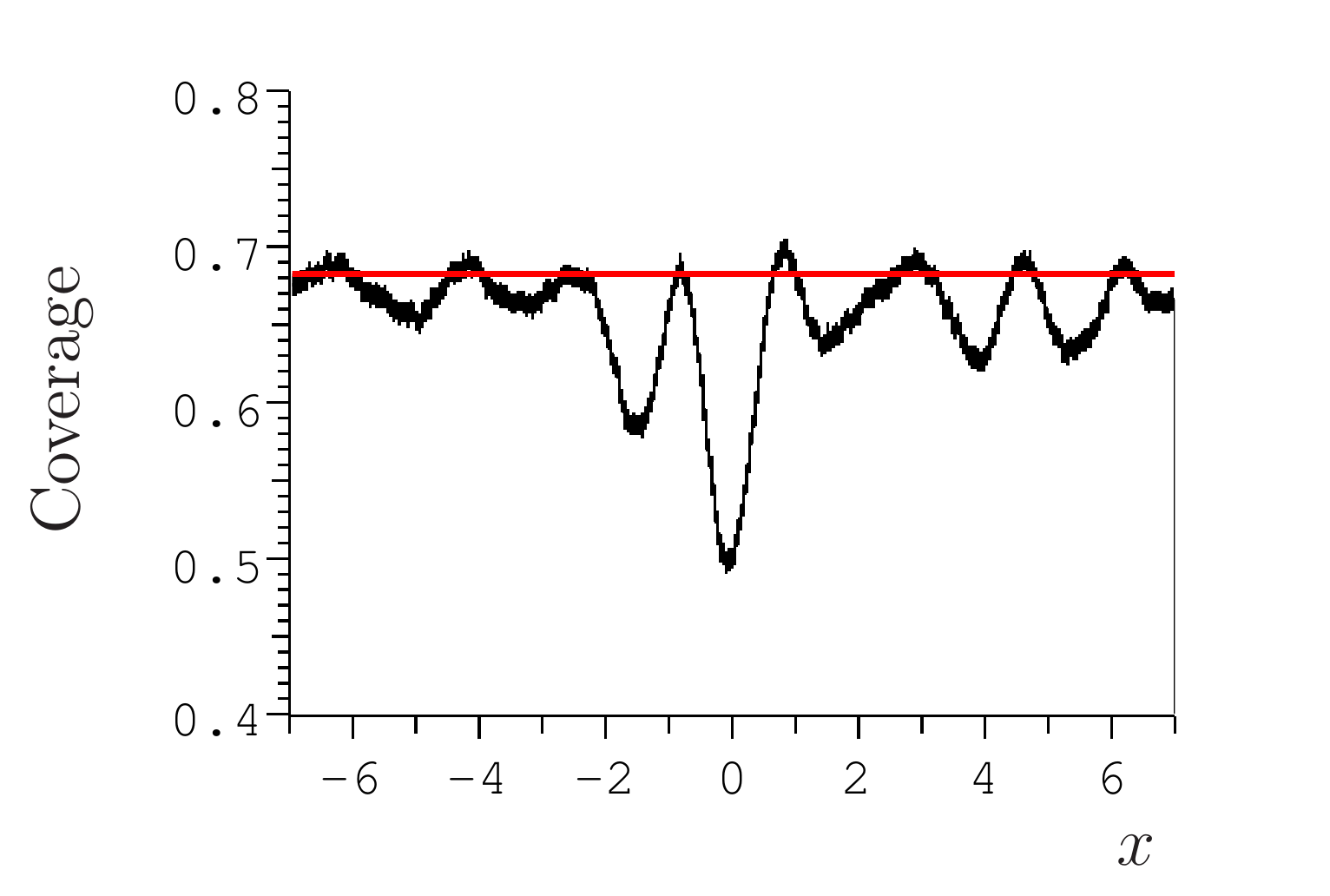} 
\includegraphics[width=.49\textwidth]{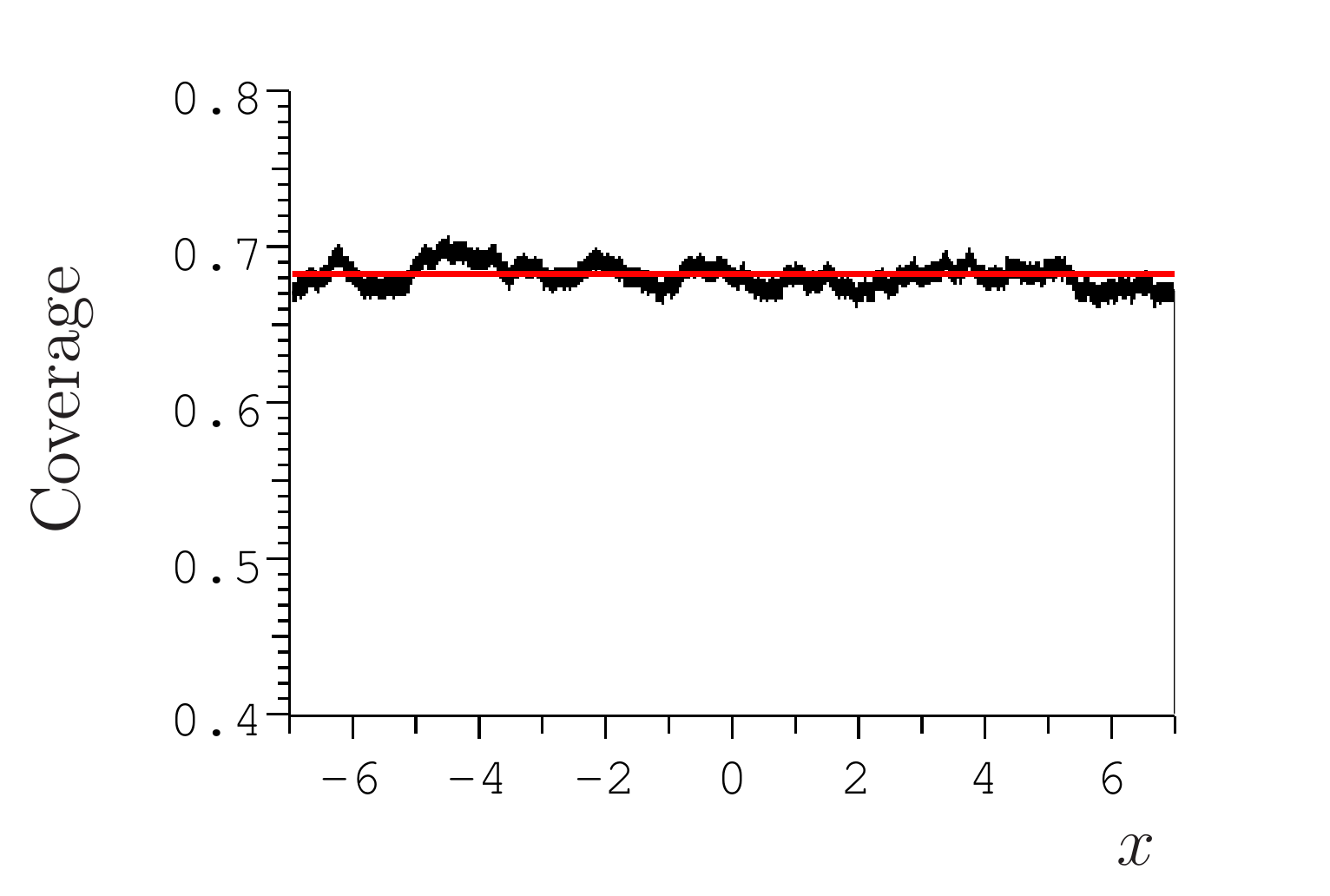}
\caption{Pointwise 
         frequentist coverage of the unfolding result for the first example
         density. 5,000 simulated samples are used, with 10,000 points
         on average per sample. The black vertical bars are drawn at each
         $x$ according to the binomial statistical uncertainty of the coverage
         determination method.
         For the plot on the left, the bias correction is not
         performed. The smoothing matrix bandwidth is fixed at $h = 0.08$.
         The red line is drawn at 68.3\%.}
\label{fig:coverage1}
\end{center}
\end{figure}

\begin{figure}[h!]
\begin{center}
\includegraphics[width=.49\textwidth]{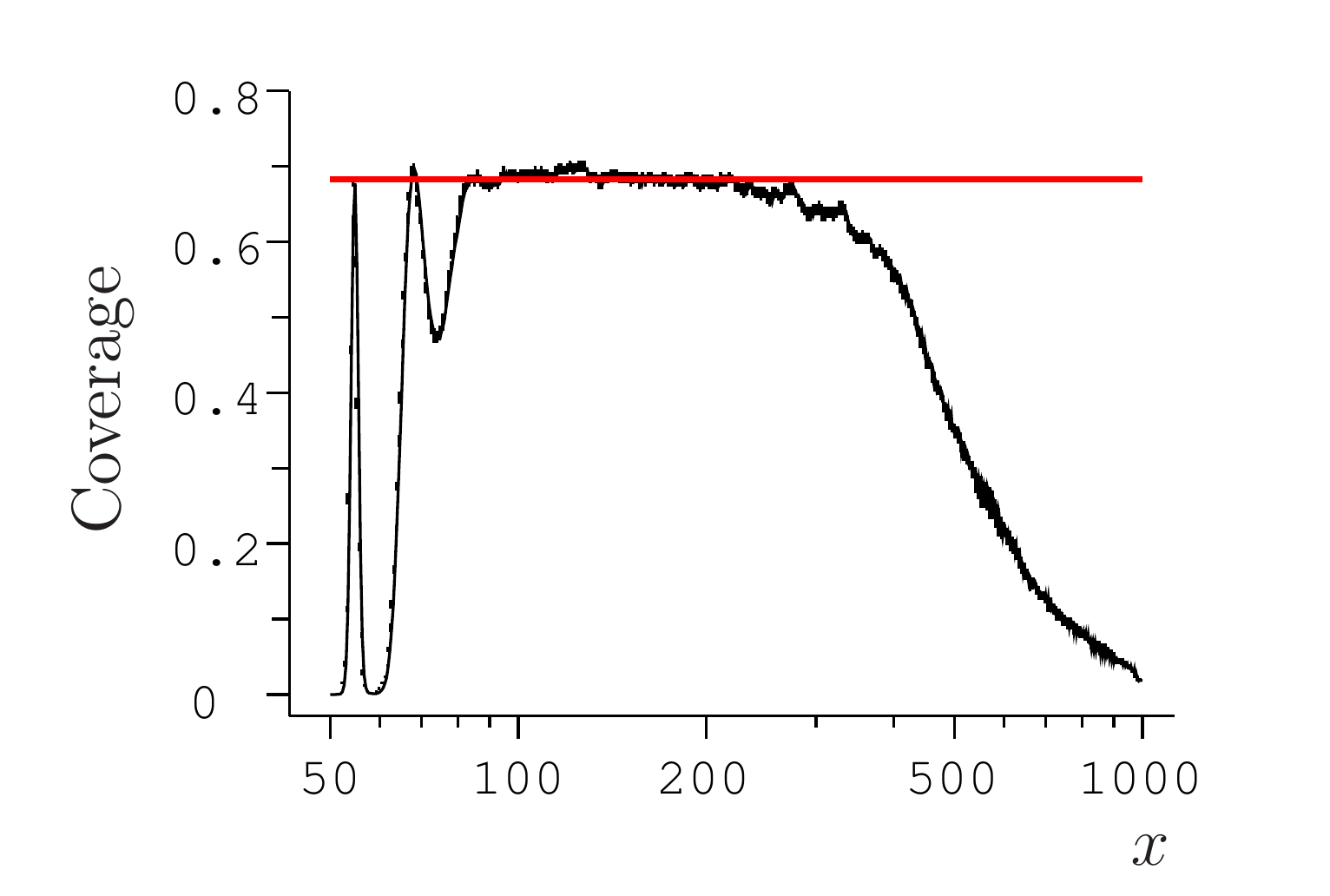} 
\includegraphics[width=.49\textwidth]{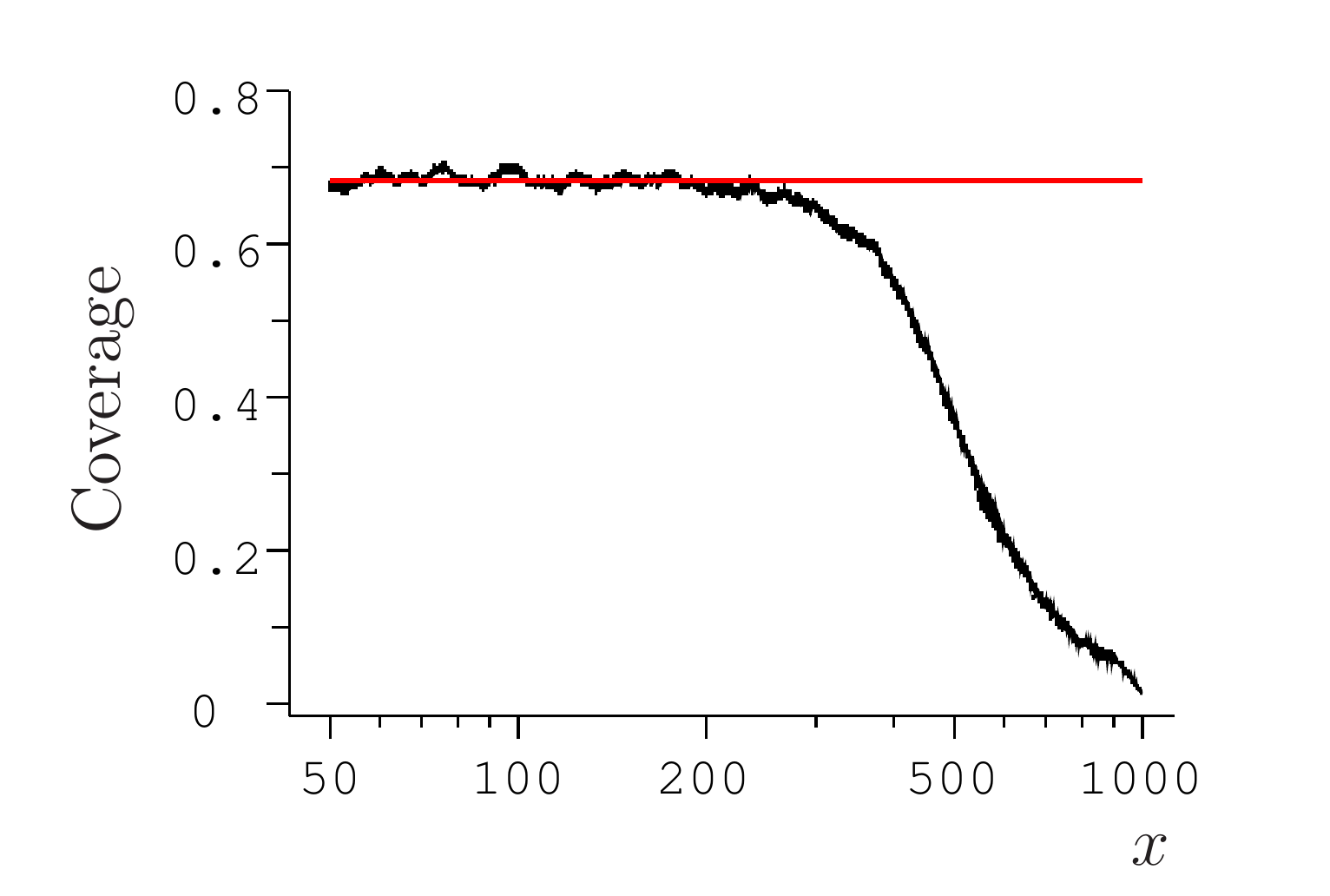}
\caption{Pointwise 
         frequentist coverage of the unfolding result for the second example
         density. 5,000 simulated samples are used, with 10,000 points
         on average per sample.
         For the plot on the left, the bias correction is not
         performed. The smoothing matrix bandwidth is fixed at $h = 0.11$
         in the $\sqrt{x}$ space. Degradation of frequentist coverage above
         $x = 250$ is due to the breakdown of the linear error propagation
         approximation for small samples. The fraction of the second example
         distribution in the $x > 250$ region is 0.16\%.}
\label{fig:coverage2}
\end{center}
\end{figure}

\begin{figure}[h!]
\begin{center}
\includegraphics[width=.49\textwidth]{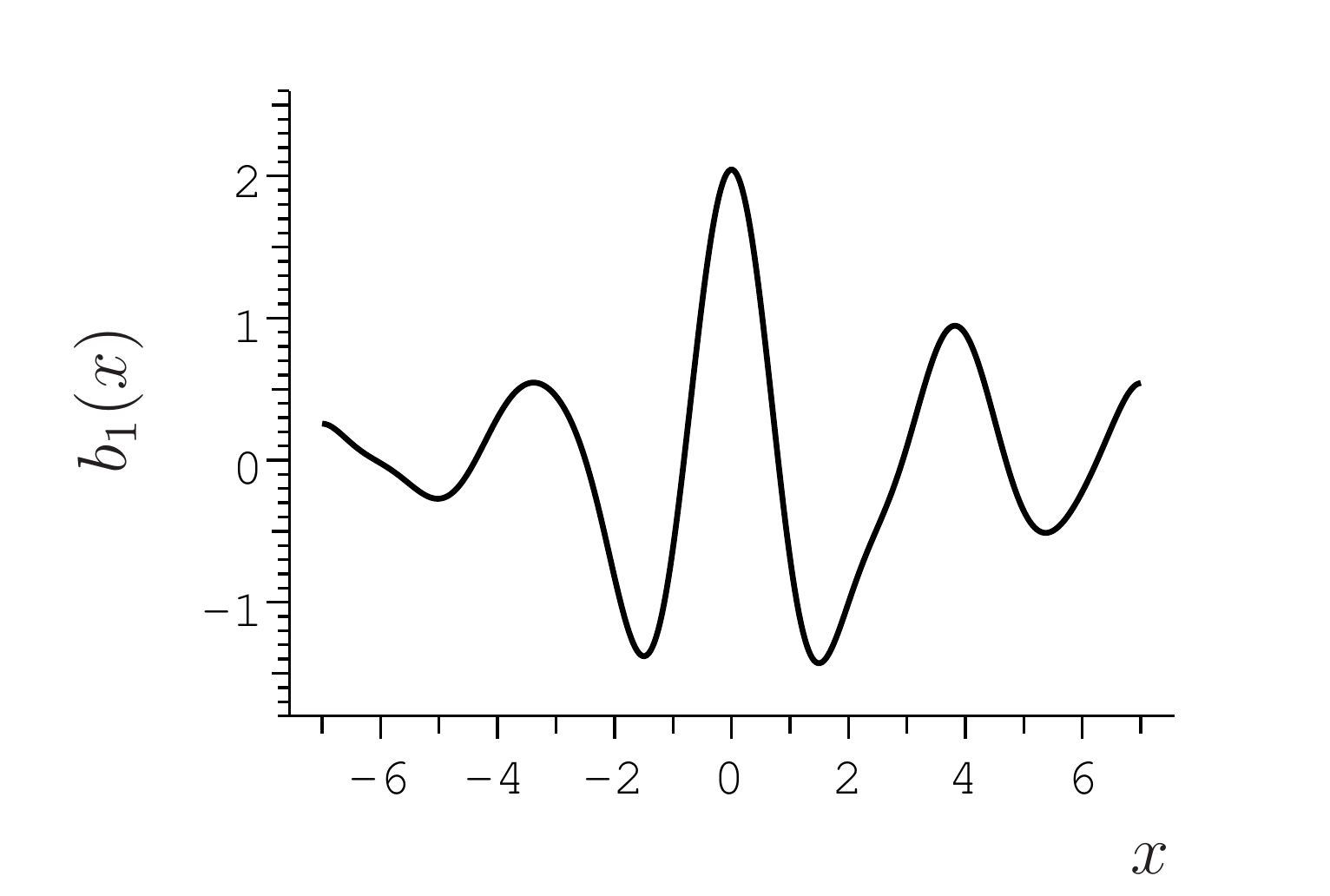} 
\includegraphics[width=.49\textwidth]{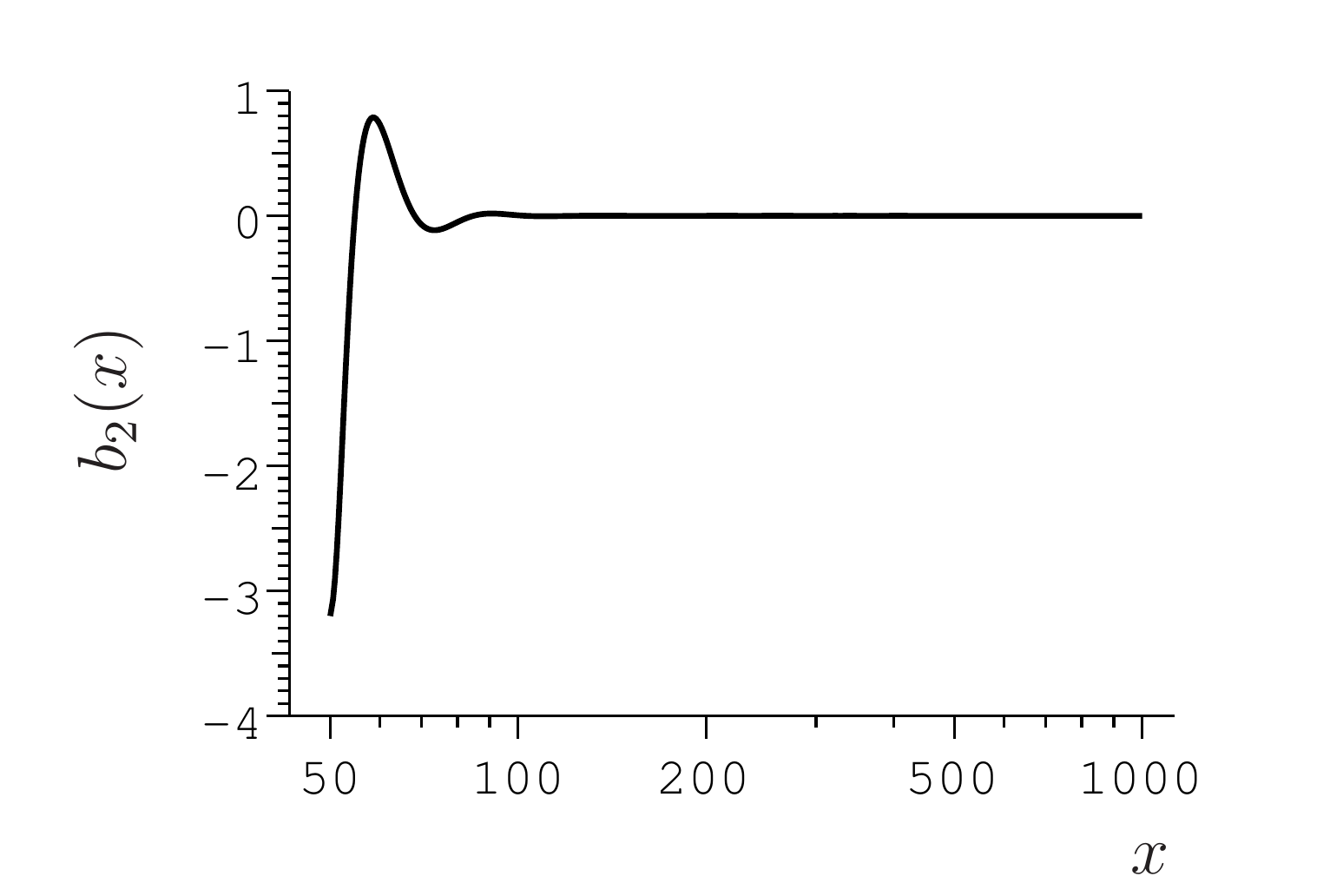}
\caption{Unfolding bias for the example densities, with 10,000 points on average per sample.}
\label{fig:bias}
\end{center}
\end{figure}

The pointwise frequentist coverage of the EMS unfolding method is illustrated
in Figs.~\ref{fig:coverage1} and~\ref{fig:coverage2} for density estimation
performed with fixed regularization strength.
The expected frequentist
coverage of 68.3\% by $\pm 1$ standard deviation intervals is recovered
after correcting for the unfolding bias, $b(x)$.
Bias is defined as the average difference
between the unfolded result and the unnormalized physical
process density filtered by the smoothing matrix. Its dependence on $x$
is illustrated in Fig.~\ref{fig:bias} for the example densities.
It is worth emphasizing that bias correction of this kind
can be used to validate covariance matrix estimates in simulation studies
but cannot be performed when the process density is not known in advance.

\begin{figure}[h!]
\begin{center}
\includegraphics[width=.49\textwidth]{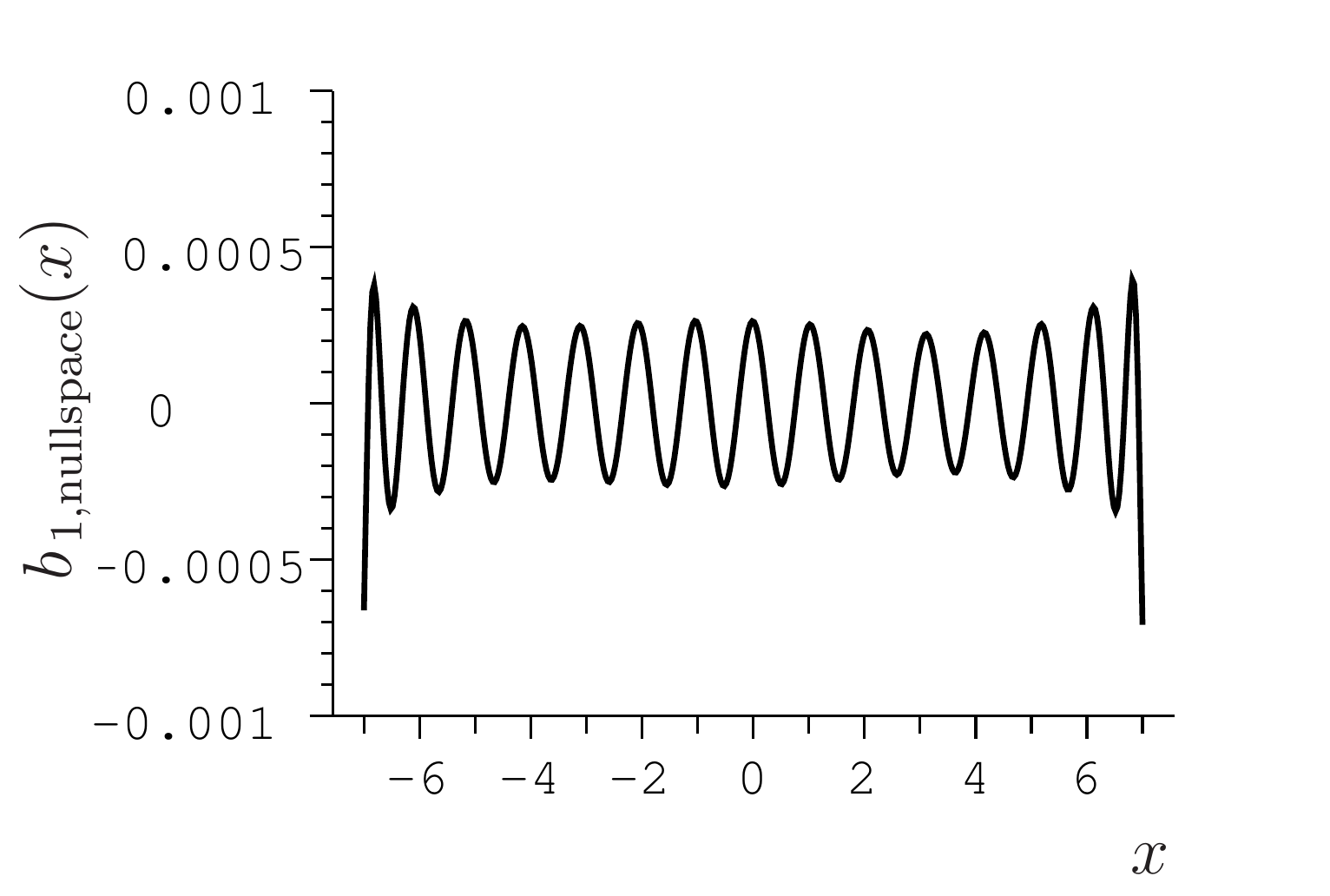} 
\includegraphics[width=.49\textwidth]{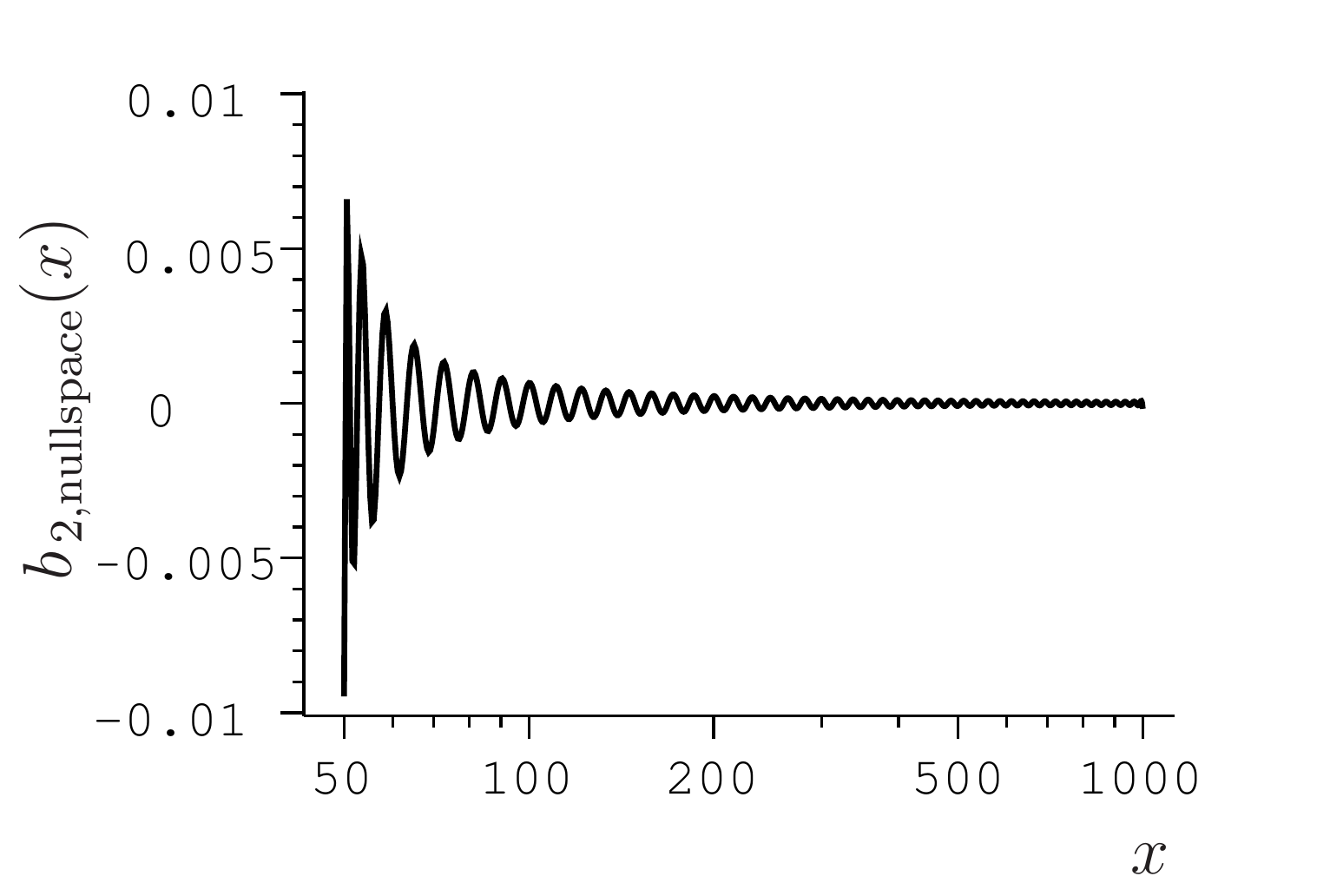}
\caption{Part of the unfolding bias
that belongs to the
effective nullspace of the corresponding response function.
Definition of this nullspace was given in Section~\ref{sec:problemstatement}.
To make these plots, the response function singular
values were considered ``small'' if their ratios to the largest singular value
did not exceed $3 \times 10^{-8}$.
This approximate cutoff is of the order $\varepsilon \sqrt{m}$.}
\label{fig:nullspacebias}
\end{center}
\end{figure}

For the example densities under study, the component of the unfolding bias
that belongs to the effective nullspace of the response function
is illustrated in Fig.~\ref{fig:nullspacebias}.
This component is, in some sense, irreducible --- information about it is
destroyed and can not be recovered without
introducing further assumptions beyond those needed to regularize the problem.
The irreducible bias is inherent in all unfolding
procedures and can be viewed as an argument against using
unfolding methods in the data analysis practice.
Comparison of theoretical predictions with experimental
results in the observation space avoids this issue altogether. However,
due to the complexity and uniqueness of large particle physics
experiments, realistic response functions are not exactly known\footnote{This
lack of knowledge introduces systematic uncertainties on the unfolded
results. Treatment of systematic uncertainties is a complex subject
which is beyond the scope of this article.}, and standards for developing
and publishing
response functions together with their uncertainties have not been
established yet.

\begin{figure}[h!]
\begin{center}
\includegraphics[width=.49\textwidth]{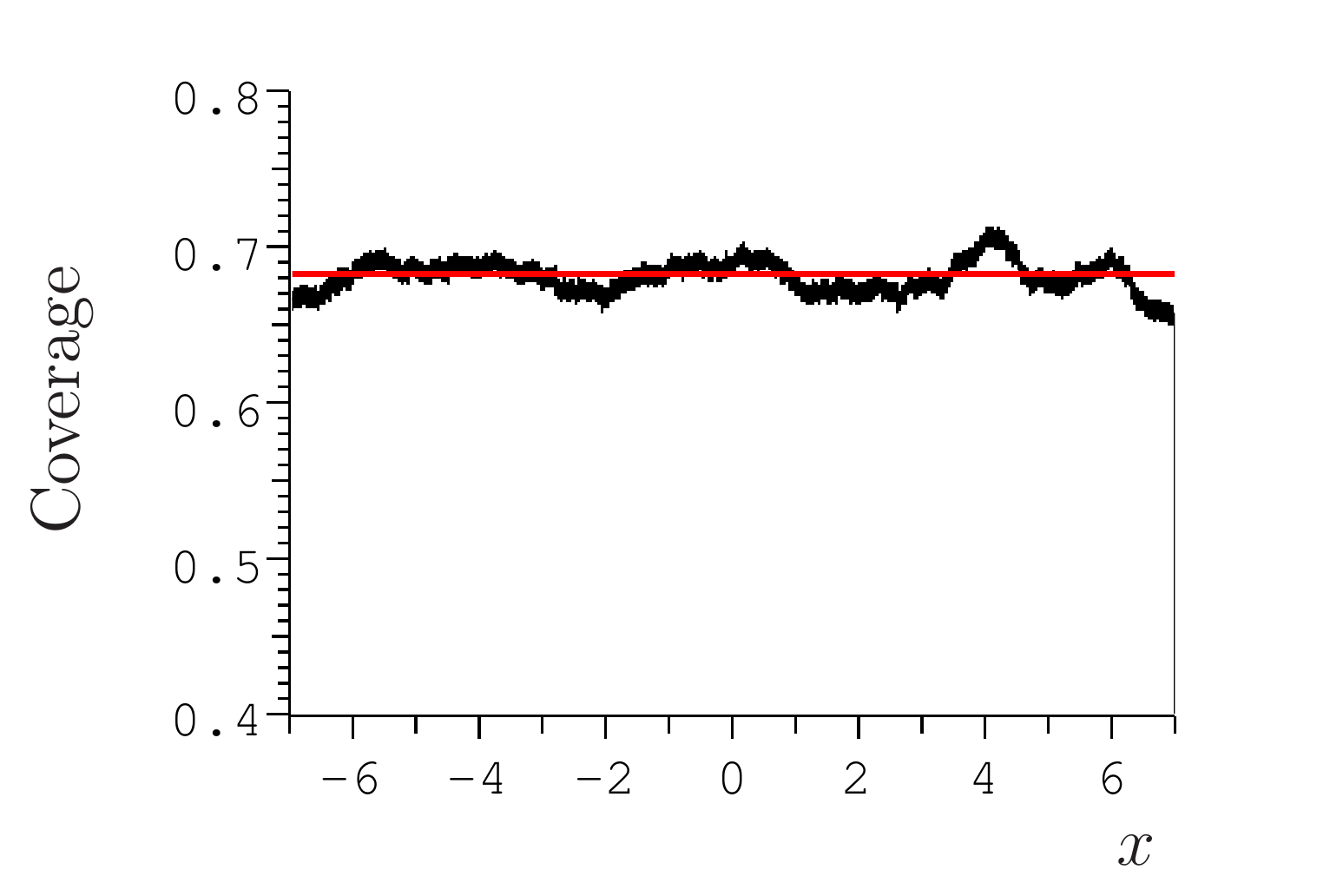} 
\includegraphics[width=.49\textwidth]{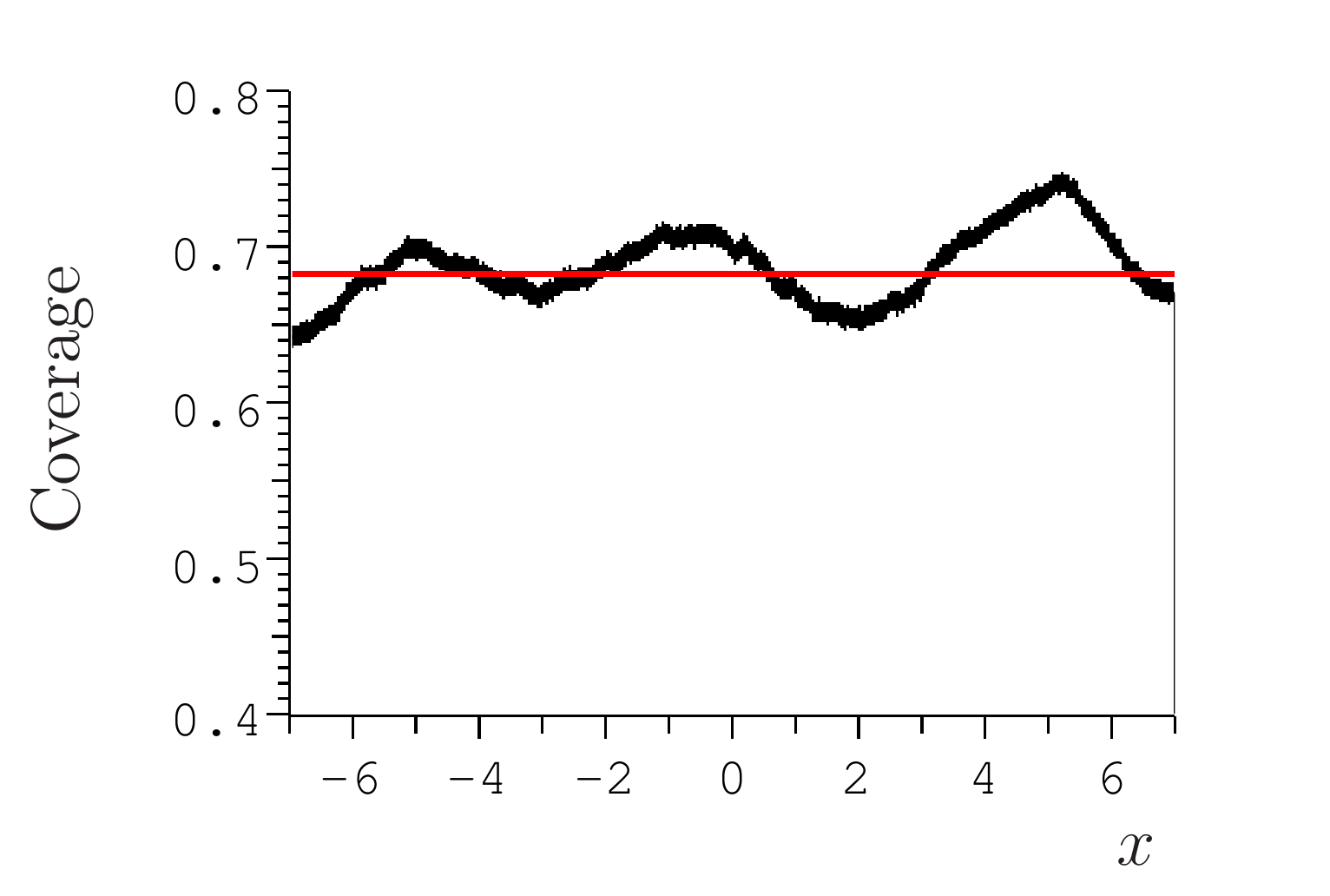}
\caption{Bias-corrected pointwise 
         frequentist coverage of the unfolding results for the first
         example density with $N = 1,000$ points on average per sample (left)
         and with $N = 100$ points (right). The smoothing matrix bandwidth is set to 0.2
         for the left plot and to 0.8 for the right plot.}
\label{fig:covdegrade}
\end{center}
\end{figure}

As the expectation-maximization unfolding procedure is nonlinear,
the frequentist uncertainty coverage is also affected by the fidelity
of the linear error propagation approximation. The breakdown of this
approximation at high values of $x$ for the second model
is apparent from Fig.~\ref{fig:coverage2}.
The effect of the sample size, $N$,  on the first model is
illustrated in Fig.~\ref{fig:covdegrade}.
The difference between Gaussian and Poisson distributions
becomes more pronounced for smaller values of $N$, and the influence of
nonlinearities in the unfolding procedure is increased for smaller
samples due to the increase in the relative statistical uncertainty
of the observation counts.

\subsection{Adaptive Regularization Strength}
\label{sec:adaptiveregularization}

For the EMS unfolding examples considered in this section,
the regularization strength is defined by the smoothing matrix
bandwidth. As the bandwidth is increasing, the variance of
the unfolded result is decreasing at the cost of increase
in the bias with respect to the true density. The optimal tradeoff between
the bias and the variance will,
in general, depend on the manner in which the unfolded results
are to be used. If the results are to be
compared with model
distributions\footnote{It is assumed that for all such comparisons
model densities are processed by the same smoothing matrix.}
lacking high frequency components
then larger bandwidth values may be preferred, as the variance is
reduced without loosing
relevant information. On the other hand, a~reduction in bandwidth
may be beneficial if the results are to be searched for sharp peaks.
It should also be noted that, although the width of the smoothing
matrix is proportional to the width of the response function
everywhere in $x$ for the examples considered in this section,
other schemes of constructing the smoothing matrix can be envisioned,
possibly along the lines of variable-bandwidth kernel density
estimation techniques~\cite{ref:kde}.

\begin{figure}[h!]
\begin{center}
\includegraphics[width=.49\textwidth]{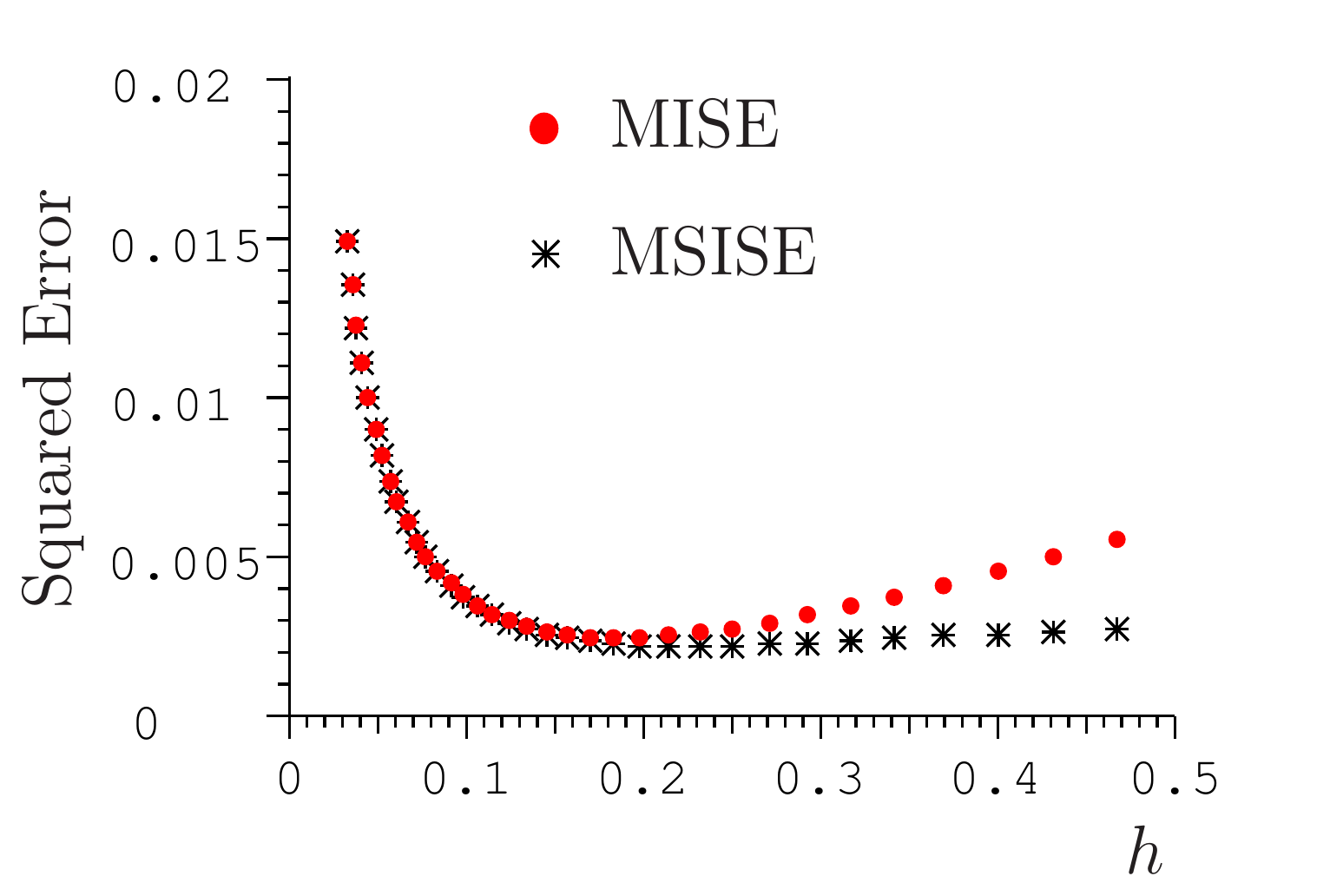} 
\includegraphics[width=.49\textwidth]{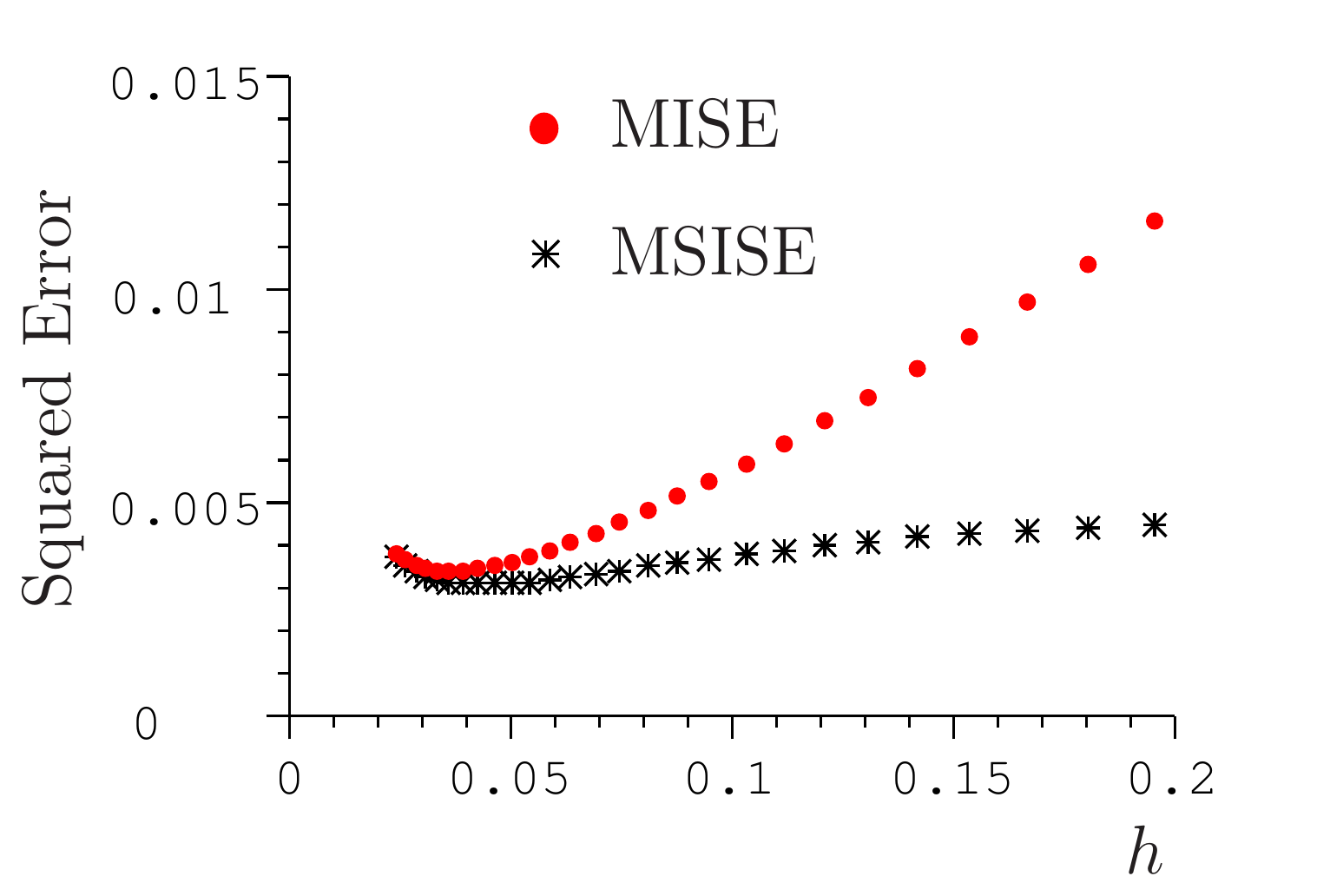}
\caption{Left: dependence of MISE and MSISE on the
         smoothing matrix bandwidth for the first example distribution.
         1,000 points per sample were generated on average. Right: the same dependence
         for the second example distribution, with 10,000 points per sample.}
\label{fig:mise}
\end{center}
\end{figure}

While the most relevant distance measure between a normalized density
function, $p(x)$, and its estimate, $\hat{p}(x)$, will be application-dependent,
a simple and convenient distance is given by the integrated squared error:
$\mbox{ISE}(p, \hat{p}) = \int (p(x) - \hat{p}(x))^2 dx$. It may also be
interesting to consider the smoothed ISE (SISE)
in which the original density is processed by the smoothing matrix
before comparison with the estimate. 
The mean integrated squared error (MISE) and the mean smoothed ISE (MSISE)
are, respectively, the ISE and the SISE averaged over many simulated
samples. The dependence of MISE and MSISE on the smoothing matrix
bandwidth is illustrated in Fig.~\ref{fig:mise} for the example densities
considered.

\begin{figure}[h!]
\begin{center}
\includegraphics[width=.49\textwidth]{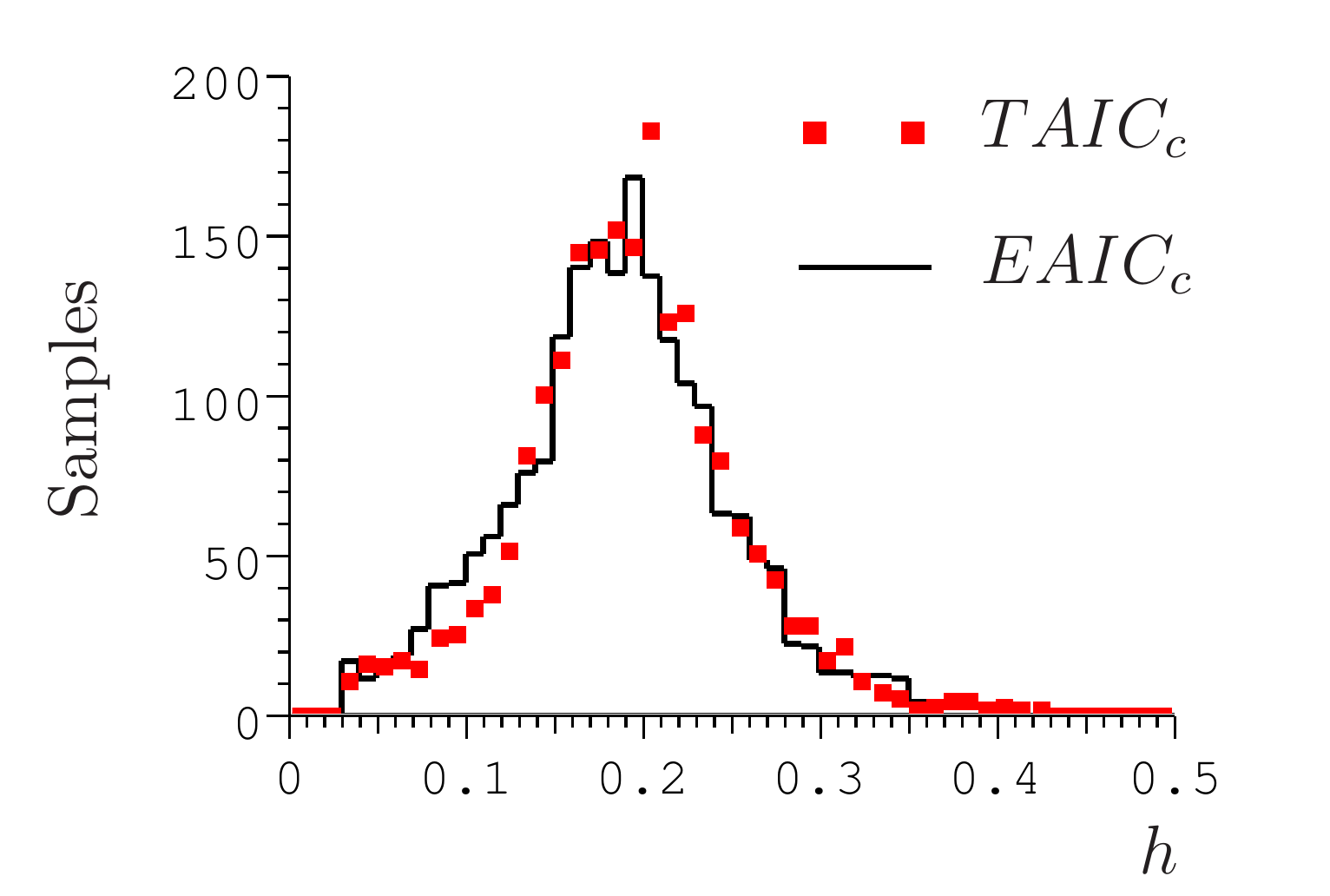} 
\includegraphics[width=.49\textwidth]{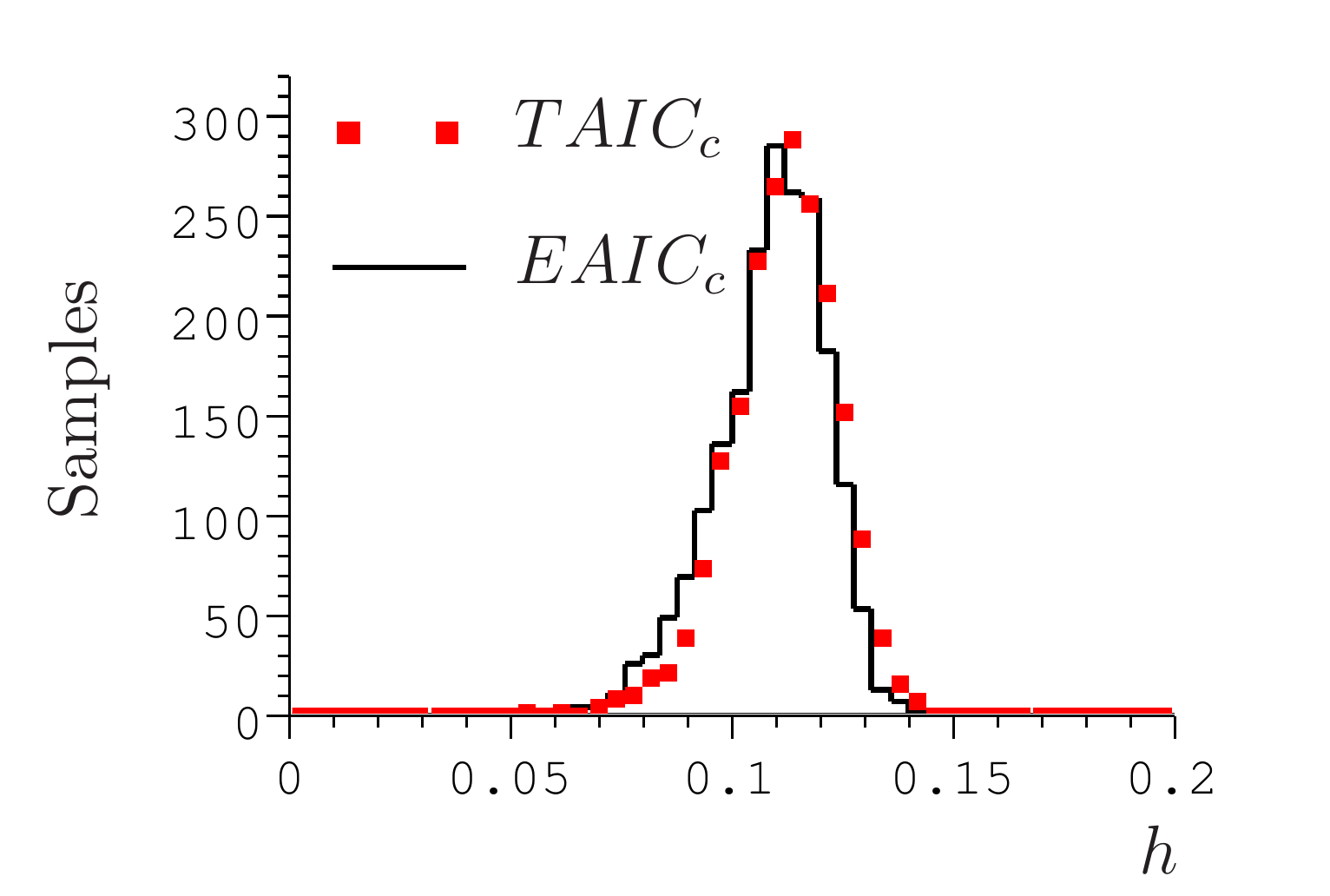}
\caption{Left: smoothing matrix 
        bandwidth selected by the $AIC_c$ criteria for the
        first example distribution. 2000 samples are used,
        with 1,000 points per sample on average.
        Right: bandwidth selected for the
        second example distribution. 2000 samples, 10,000 points per sample.
        It appears that for both distributions
        there is no substantial difference between bandwidth
        values selected by $EAIC_c$ and $TAIC_c$.}
\label{fig:bwdistro}
\end{center}
\end{figure}

\begin{figure}[h!]
\begin{center}
\includegraphics[width=.49\textwidth]{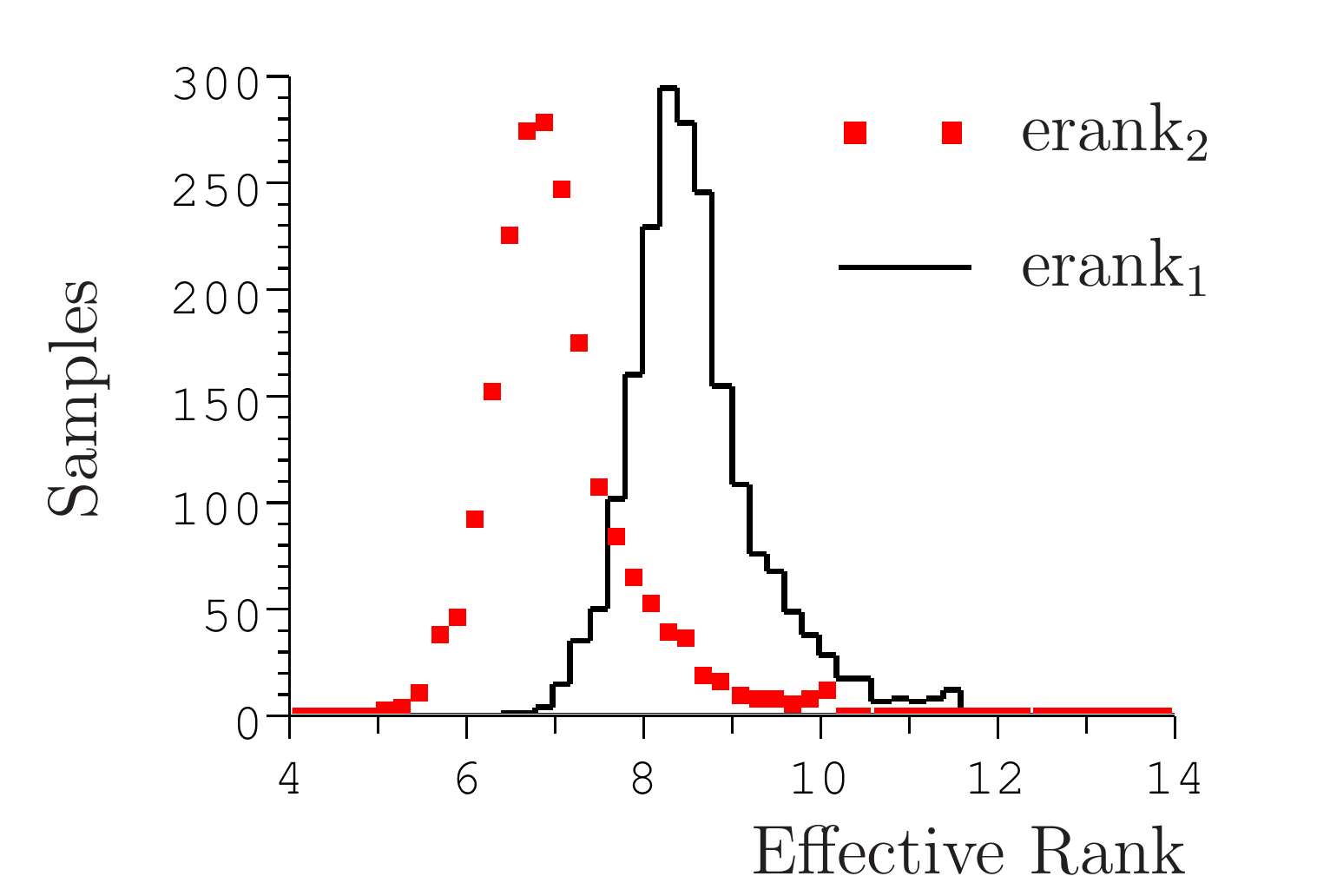} 
\includegraphics[width=.49\textwidth]{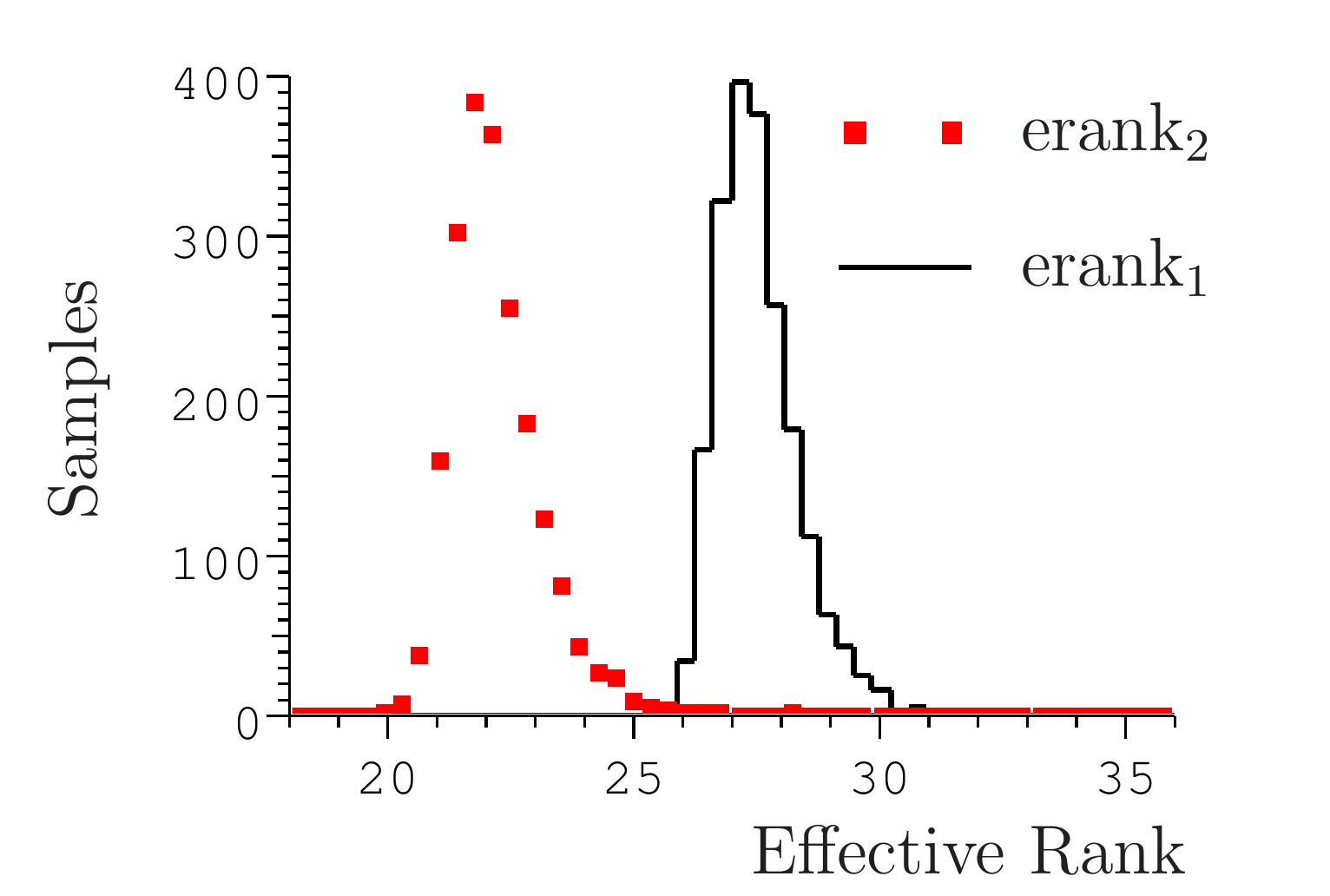}
\caption{Distributions of the effective numbers of model parameters
        for the fits to the simulated observations. The plot on the
        left corresponds to the first example density, and the plot
        on the right to the second. While distributions of the $\mbox{erank}_1$
        and $\mbox{erank}_2$ values (calculated, respectively,
        according to Eqs.~\ref{eq:erank1} and~\ref{eq:erank2})
        are substantially different, the
        corresponding $AIC_c$ criteria depend not on the values themselves
        but rather on their derivatives with respect to the bandwidth.}
\label{fig:effparams}
\end{center}
\end{figure}

The distributions of bandwidth values selected by the $AIC_c$ criteria
are shown in Fig.~\ref{fig:bwdistro}, and the corresponding
distributions of the effective number of 
model parameters used to fit simulated observations are
presented in Fig.~\ref{fig:effparams}. 
For the first example density (plots on the left),
the bandwidth value chosen by $AIC_c$ is, on average,
consistent with the bandwidth
that would be selected on the basis of MISE. For the second
example, bandwidth choice based on $AIC_c$ results in substantial
oversmoothing. Due to a significant fraction of empty bins in
the observation space (57.8\% on average for $N = 10,000$),
for this example the effective number of parameters in the
fit is overestimated. As this number is monotonously decreasing
as a function of bandwidth, its overestimation results in the shift
of the $AIC_c$ minimum towards increased bandwidth values.

\begin{figure}[h!]
\begin{center}
\includegraphics[width=.49\textwidth]{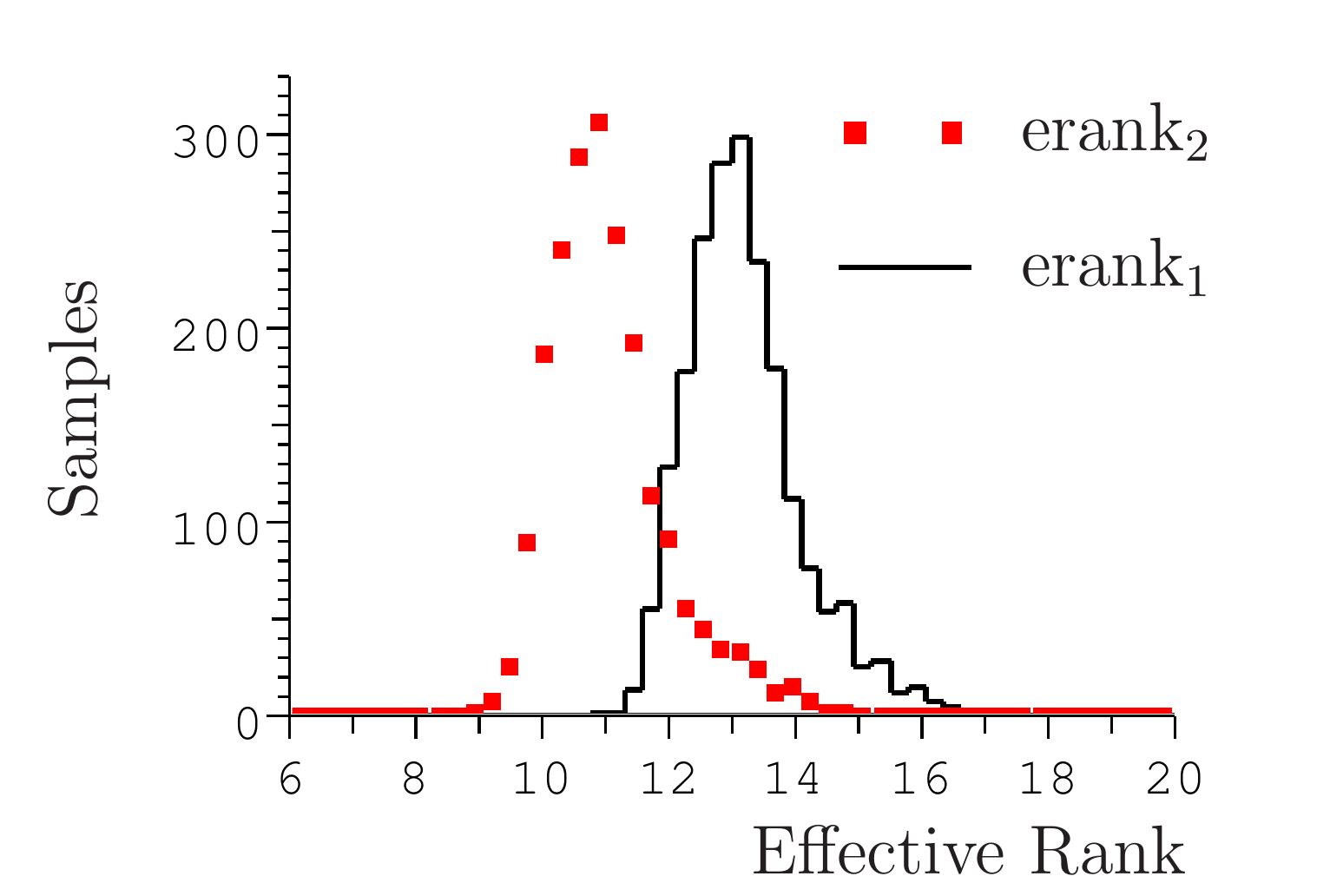} 
\includegraphics[width=.49\textwidth]{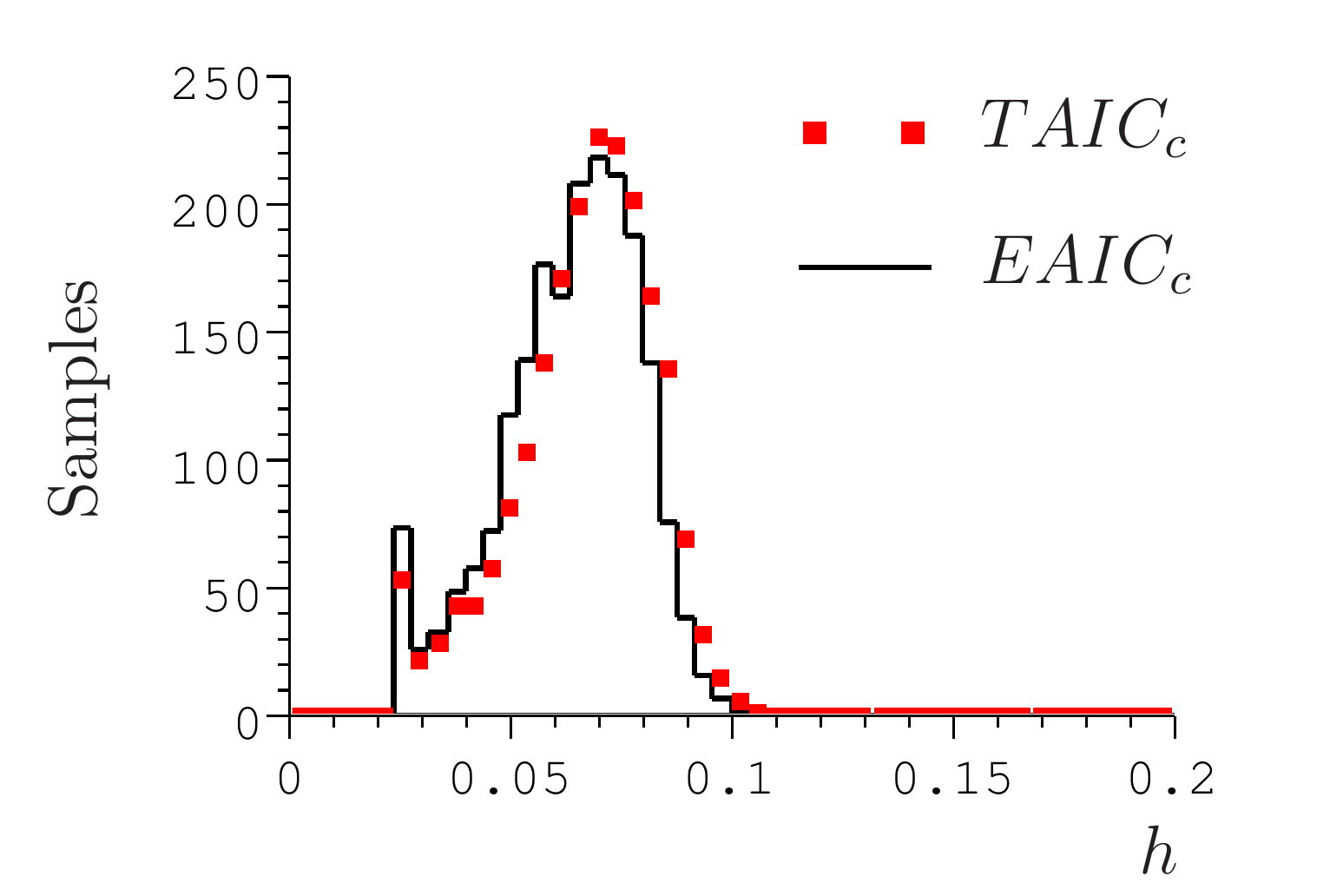}
\caption{Distributions of the effective numbers of model parameters (left)
        and of the smoothing matrix 
        bandwidth (right) selected by the $AIC_c$ criteria for the
        second example after correcting for sparsely populated bins.
        The starting pile-up of bandwidth values at $h = (\sqrt{1000} - \sqrt{50})/1000 \approx 0.026$ is due to the fact that the smallest bandwidth value
        considered was set to the width of one bin in the discretized
        physical process space.}
\label{fig:corrparams}
\end{center}
\end{figure}

A simple method of mitigating the effect of sparsely populated bins
consists in scaling the effective number of fitted parameters by the fraction
of populated bins in the observed 
data\footnote{The real intent of this method is to discard
  sections of the distribution support in the observation space which
  have very low probabilities that any random point ends up inside them for
  the expected sample size, $N$. For a known density, one can simply
  eliminate the lowest probability bins whose combined
  probability content does not exceed a~threshold of the order $1/N$.
  The optimal way to proceed without such an {\it a priori} knowledge is not
  obvious. For example, the simple approach proposed in the text breaks
  down in case the number of bins gets so large
  that the bin width becomes comparable with the typical distance between two
  neighboring sample points.}.
Distributions of bandwidth values
selected by the $AIC_c$ criteria with this scaling are presented
in Fig.~\ref{fig:corrparams} for the second example density.
This adjustment significantly improves the MISE-based bandwidth
selection performance of $AIC_c$.
MISE and a few other useful characteristics
of the EMS unfolding method
are summarized in Table~\ref{tab:effparam} for several sample sizes.

\begin{table}[h!]
\caption{Summary of the EMS unfolding performance for samples of average size $N$
drawn from the example densities.
For $N = 10^5$, the number of bins in the
discretization of both $x$ and $y$ spaces was increased.
$h$ is the smoothing matrix bandwidth selected by $EAIC_c$. 
$\mbox{erank}_1$ is the effective number of
model parameters calculated according to Eq.~\ref{eq:erank1}.
For the second example density, $\mbox{erank}_1$ is adjusted for sparsely populated bins.
$N_{eigen}$ is the number of principal
components of the result covariance matrix with frequentist
coverage above 61.4\% ({\it i.e.}, at least 0.9 of the expected 68.3\%),
determined with a bandwidth-dependent bias correction.
The principal component coverage is discussed further in Section~\ref{sec:resultpresent}.
Values given for each example in the first four rows
represent the medians of the corresponding distributions
and the ranges that cover 68.3\% of the samples.}
\label{tab:effparam}
\begin{center}
\begin{tabular}{c|l|ccc|}
\cline{2-5}
& $N$                                            & $10^3$ & $10^4$ & $10^5$ \\
\hline
\multicolumn{1}{|c|}{$\mbox{ }$} & $h$                & $0.188^{+0.055}_{-0.053}$ & $0.082^{+0.014}_{-0.022}$ & $0.041^{+0.005}_{-0.012}$ \\
\multicolumn{1}{|c|}{$\mbox{ }$} & $\mbox{erank}_1$ & $8.5^{+0.6}_{-0.5}$ & $10.0^{+0.5}_{-0.3}$ & $11.2^{+0.6}_{-0.2}$ \\
\multicolumn{1}{|c|}{First} & ISE    & $2.7^{+2.4}_{-1.3} \times 10^{-3}$ & $5.9^{+8.0}_{-3.0} \times 10^{-4}$ & $1.4^{+2.7}_{-0.7} \times 10^{-4}$ \\
\multicolumn{1}{|c|}{example} & SISE & $1.4^{+2.1}_{-0.7} \times 10^{-3}$ & $4.4^{+8.3}_{-2.3} \times 10^{-4}$ & $1.2^{+2.8}_{-0.6} \times 10^{-4}$ \\
\multicolumn{1}{|c|}{density} & MISE     & $3.7 \times 10^{-3}$ & $1.4 \times 10^{-3}$ & $3.0 \times 10^{-4}$ \\
\multicolumn{1}{|c|}{$\mbox{ }$} & MSISE & $2.6 \times 10^{-3}$ & $1.3 \times 10^{-3}$ & $2.9 \times 10^{-4}$ \\
\multicolumn{1}{|c|}{$\mbox{ }$} & $N_{eigen}$      &  11  &  14  &  17 \\
\hline 
\multicolumn{1}{|c|}{$\mbox{ }$} & $h$               & $0.120^{+0.035}_{-0.043}$ & $0.065^{+0.014}_{-0.019}$ & $0.031^{+0.006}_{-0.007}$ \\
\multicolumn{1}{|c|}{$\mbox{ }$} & $\mbox{erank}_1$ & $7.1^{+0.8}_{-0.6}$ & $13.1^{+0.9}_{-0.7}$ & $20.9^{+1.0}_{-0.8}$ \\
\multicolumn{1}{|c|}{Second} & ISE   & $8.5^{+7.8}_{-4.0} \times 10^{-3}$ & $3.9^{+2.5}_{-1.7} \times 10^{-3}$ & $1.9^{+0.9}_{-0.7} \times 10^{-3}$ \\
\multicolumn{1}{|c|}{example} & SISE & $1.7^{+4.3}_{-1.1} \times 10^{-3}$ & $4.3^{+9.8}_{-2.7} \times 10^{-4}$ & $1.3^{+2.3}_{-0.8} \times 10^{-4}$ \\
\multicolumn{1}{|c|}{density} & MISE     & $1.1 \times 10^{-2}$ & $4.3 \times 10^{-3}$ & $2.0 \times 10^{-3}$ \\
\multicolumn{1}{|c|}{$\mbox{ }$} & MSISE & $4.4 \times 10^{-3}$ & $9.5 \times 10^{-4}$ & $2.3 \times 10^{-4}$ \\
\multicolumn{1}{|c|}{$\mbox{ }$} & $N_{eigen}$      &  9  &  24  &  34 \\
\hline
\end{tabular}
\end{center}
\end{table}

\begin{figure}[h!]
\begin{center}
\includegraphics[width=.49\textwidth]{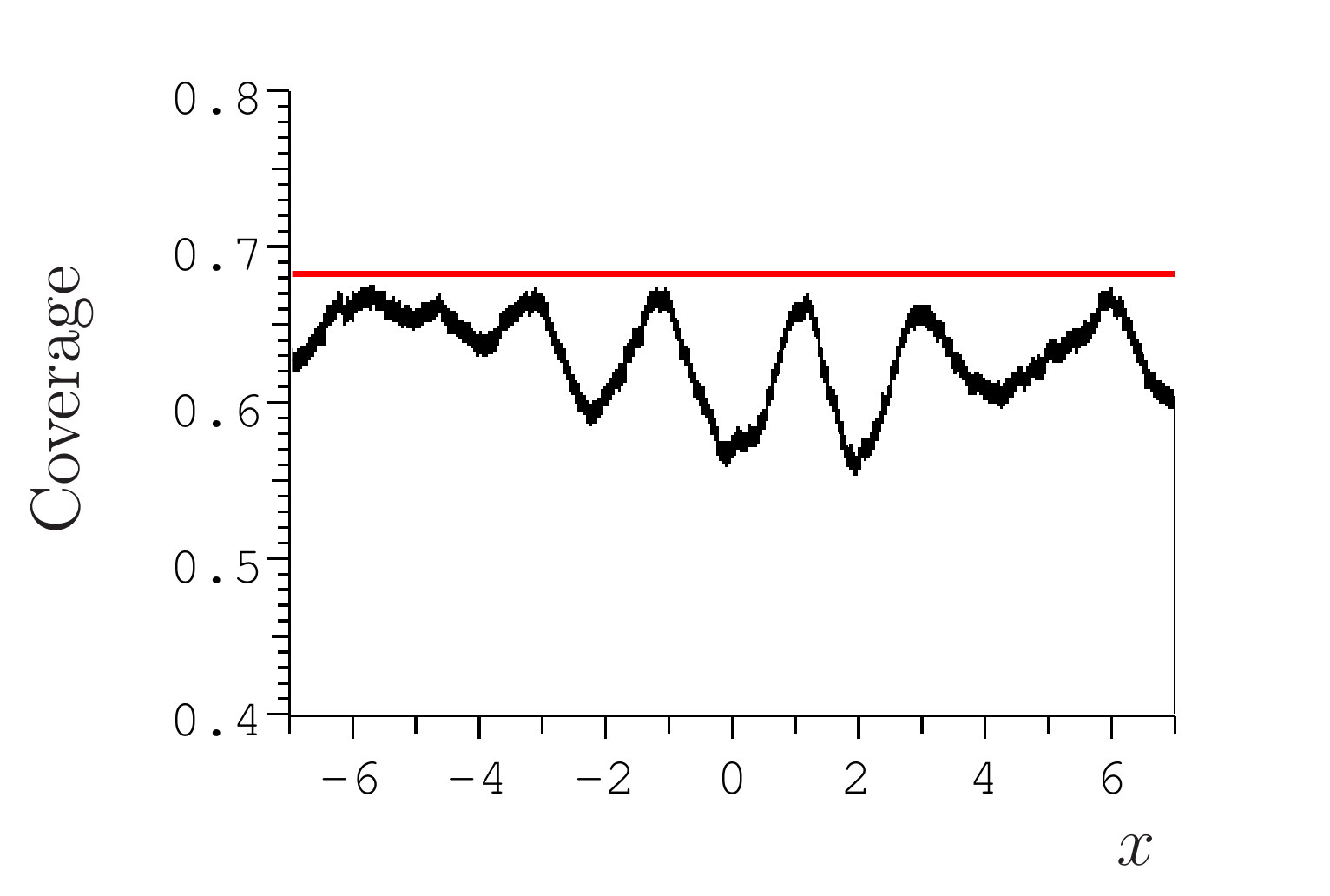} 
\includegraphics[width=.49\textwidth]{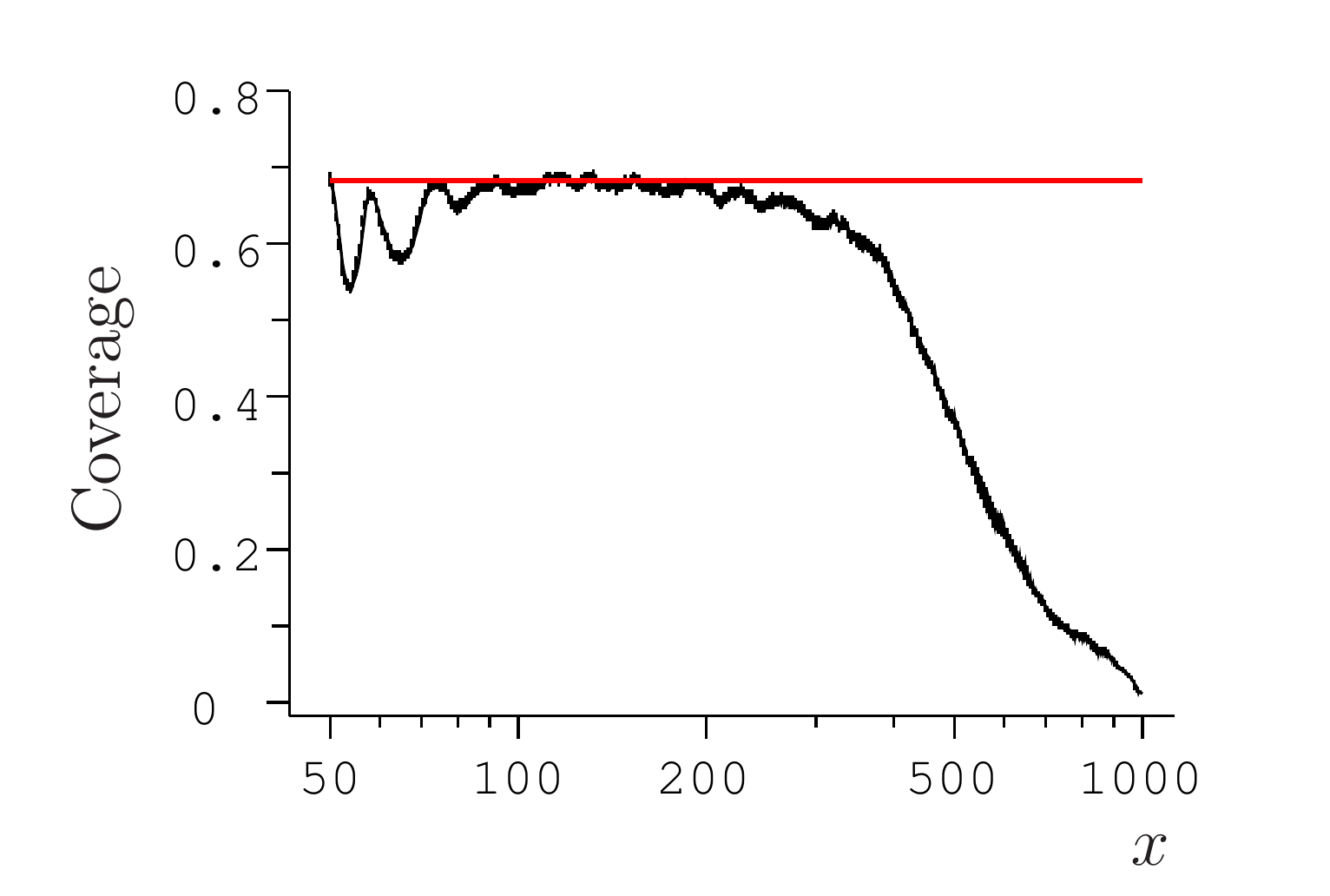}
\caption{Pointwise frequentist coverage for unfolded
         example densities with smoothing matrix bandwidth
         selection by the $EAIC_c$ (without adjustment
         for sparsely populated bins).
         The bias is corrected by subtracting a~single average curve
         from all unfolded results.
         On average, 1,000 points
         per sample are used for the first model (left plot) and
         10,000 points per sample are used for the second model (right plot).}
\label{fig:eaiccoverage}
\end{center}
\end{figure}

\begin{figure}[h!]
\begin{center}
\includegraphics[width=.49\textwidth]{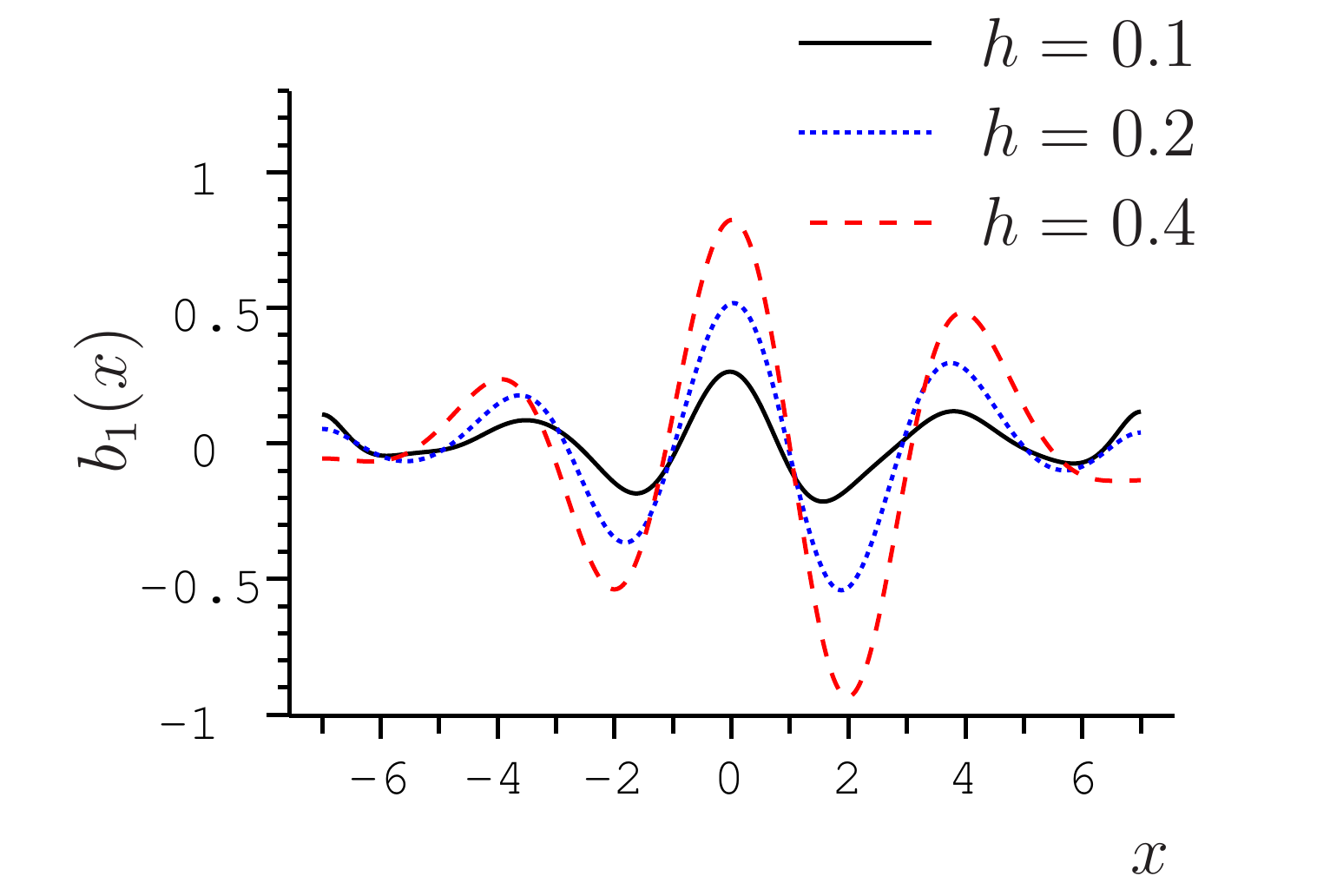} 
\includegraphics[width=.49\textwidth]{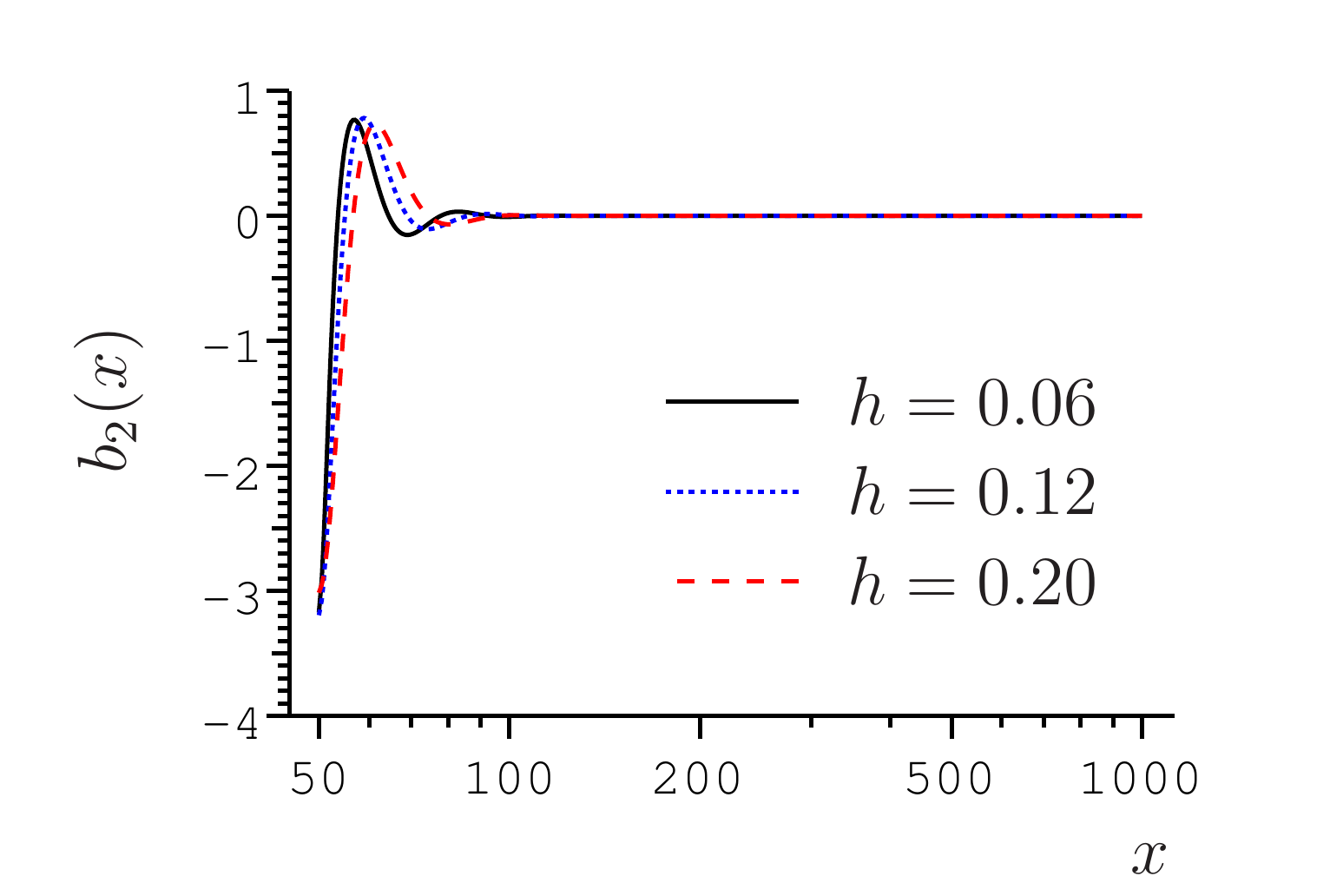}
\caption{Unfolding bias for three different values of the smoothing matrix
         bandwidth for the first example model with 1,000 points on average
         per sample (left)
         and the second model with 10,000 points on average (right).}
\label{fig:bwdependentbias}
\end{center}
\end{figure}

\begin{figure}[h!]
\begin{center}
\includegraphics[width=.49\textwidth]{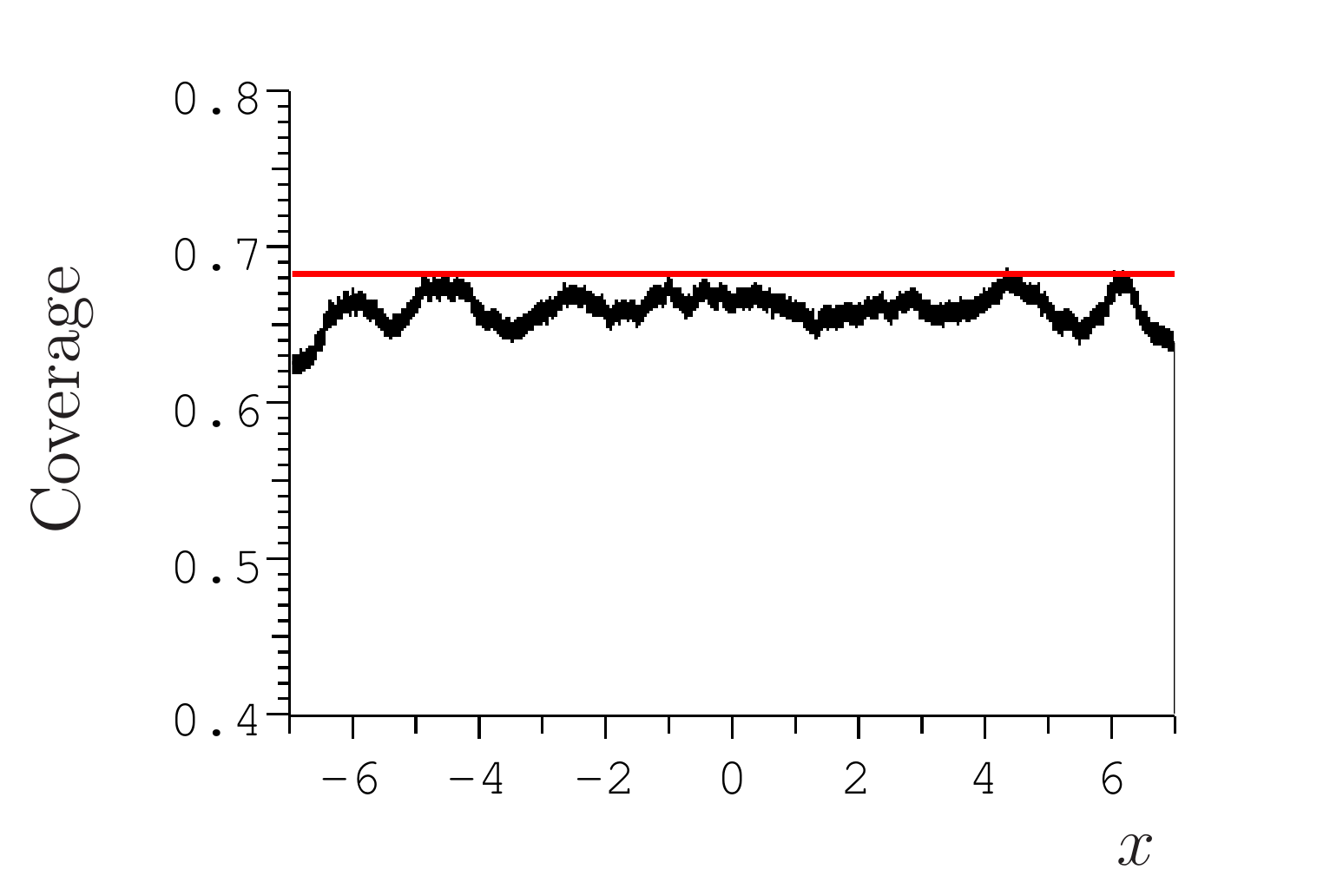} 
\includegraphics[width=.49\textwidth]{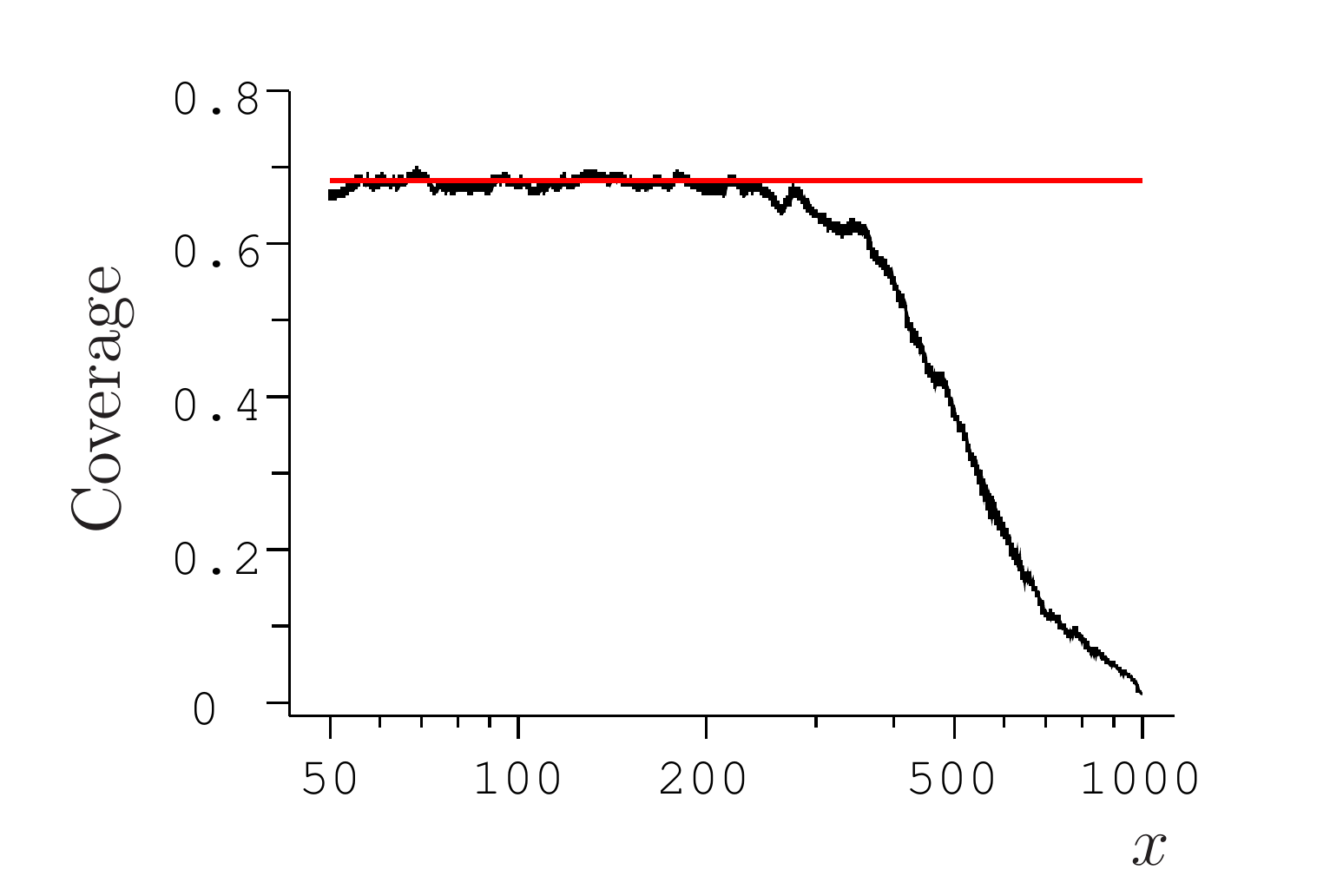}
\caption{Coverage for bias subtraction conditioned upon selected bandwidth.
         The first example distribution is on the left,
         and the second is on the right.
         A minor coverage deterioration is noticeable for the first
         example in comparison with the left plot in Fig.~\ref{fig:covdegrade}.
         This deterioration can be attributed to the assumption that
         ${\bf S}$ is fixed in the derivation of the error propagation
         formula ({i.e.}, Eq.~\ref{eq:errprop}). With the data-dependent
         choice of bandwidth,
         this assumption is no longer valid.}
\label{fig:eaicbwdependent}
\end{center}
\end{figure}

The bias-corrected pointwise frequentist coverage of the
uncertainty estimated according to Eq.~\ref{eq:covmat} is shown in
Fig.~\ref{fig:eaiccoverage}.
The effect of coverage reduction due to the data-dependent choice of
the regularization strength ({\it i.e.}, the bandwidth) is apparent.
The increase in the uncertainty can be attributed to the dependence of
the unfolding bias on the bandwidth. This dependence is illustrated in
Fig.~\ref{fig:bwdependentbias}. 
The coverage can be restored almost fully by
subtracting, for each simulated sample, 
the bias curve conditioned upon
the bandwidth value used to unfold the sample. The effect of this conditional
bias correction is shown in Fig.~\ref{fig:eaicbwdependent}.

In realistic data analysis scenarios, dependence of the unfolding bias on
the regularization strength is not known {\it a priori}, and the uncertainty
must be adjusted by other means. A promising but computationally intensive
approach consists in evaluating the uncertainty by
simulations~\cite{ref:cowling1996, ref:kuuselapreprint}. By analogy
with a similar statistical technique utilized without
deconvolution~\cite{ref:smoothingbootstrap}, its application
to EMS unfolding will
be called ``folded smoothed bootstrap''.
In this method, the estimate of the density of the observed data, $\hat{q}(y)$,
is derived from the unfolded result,  $\hat{p}(x)$, according to Eq.~\ref{eq:kernelaction}.
$\hat{q}(y)$ is then sampled repeatedly, and all these samples are in turn unfolded
so that a collection of secondary unfolded results, $\hat{\hat{p}}(x)$, is generated.
The result covariance matrix is estimated from this
collection\footnote{One can also
construct a hybrid estimate in which the variances are coming from simulations
while the correlation coefficients are determined by error propagation.},
possibly utilizing robust techniques~\cite{ref:robcov}.
It has also been suggested that the difference between the
$\hat{\hat{p}}(x)$ mean and $\hat{p}(x)$ could be used to
estimate the unfolding bias~\cite{ref:kuuselapreprint}.

\begin{figure}[h!]
\begin{center}
\includegraphics[width=.49\textwidth]{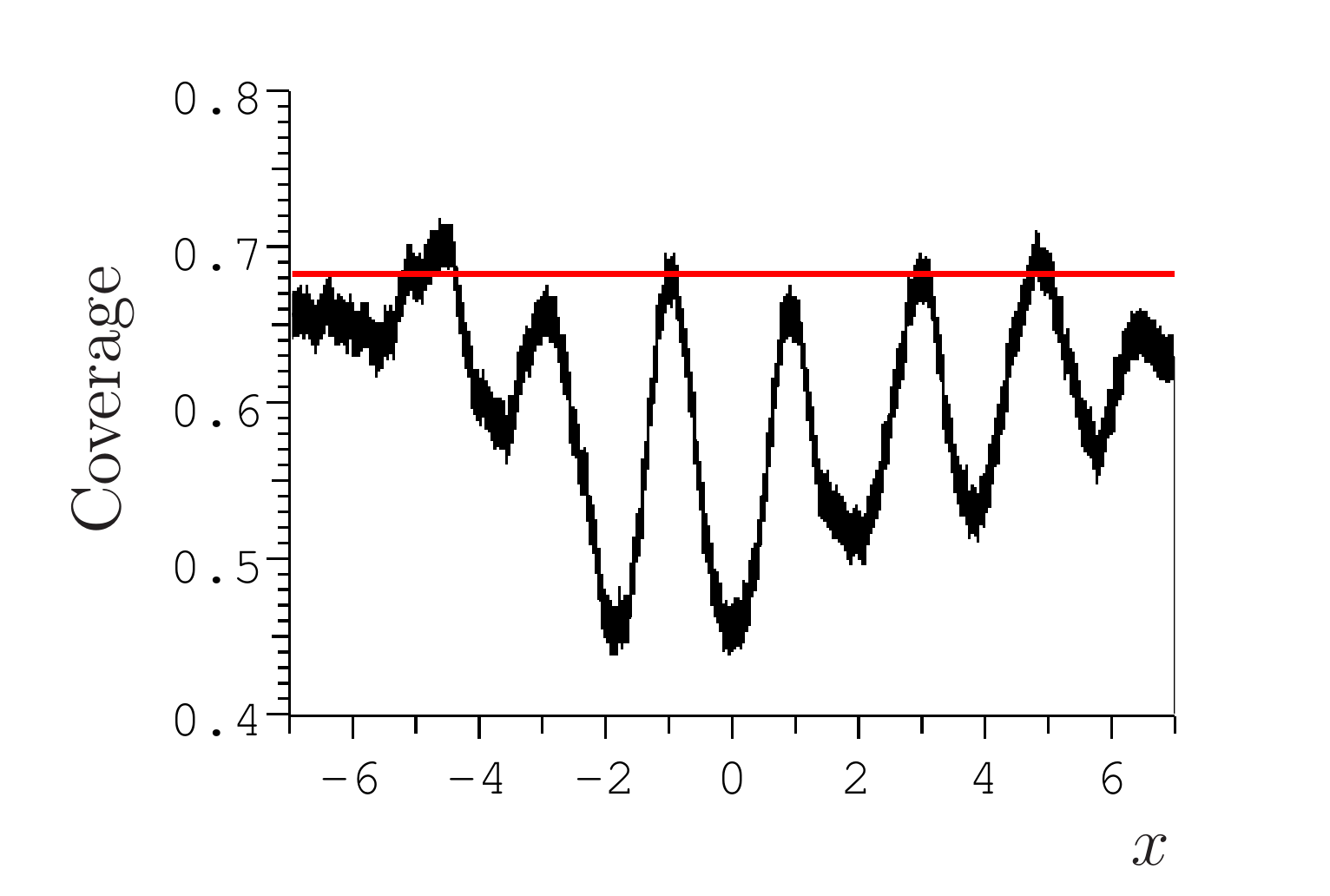} 
\includegraphics[width=.49\textwidth]{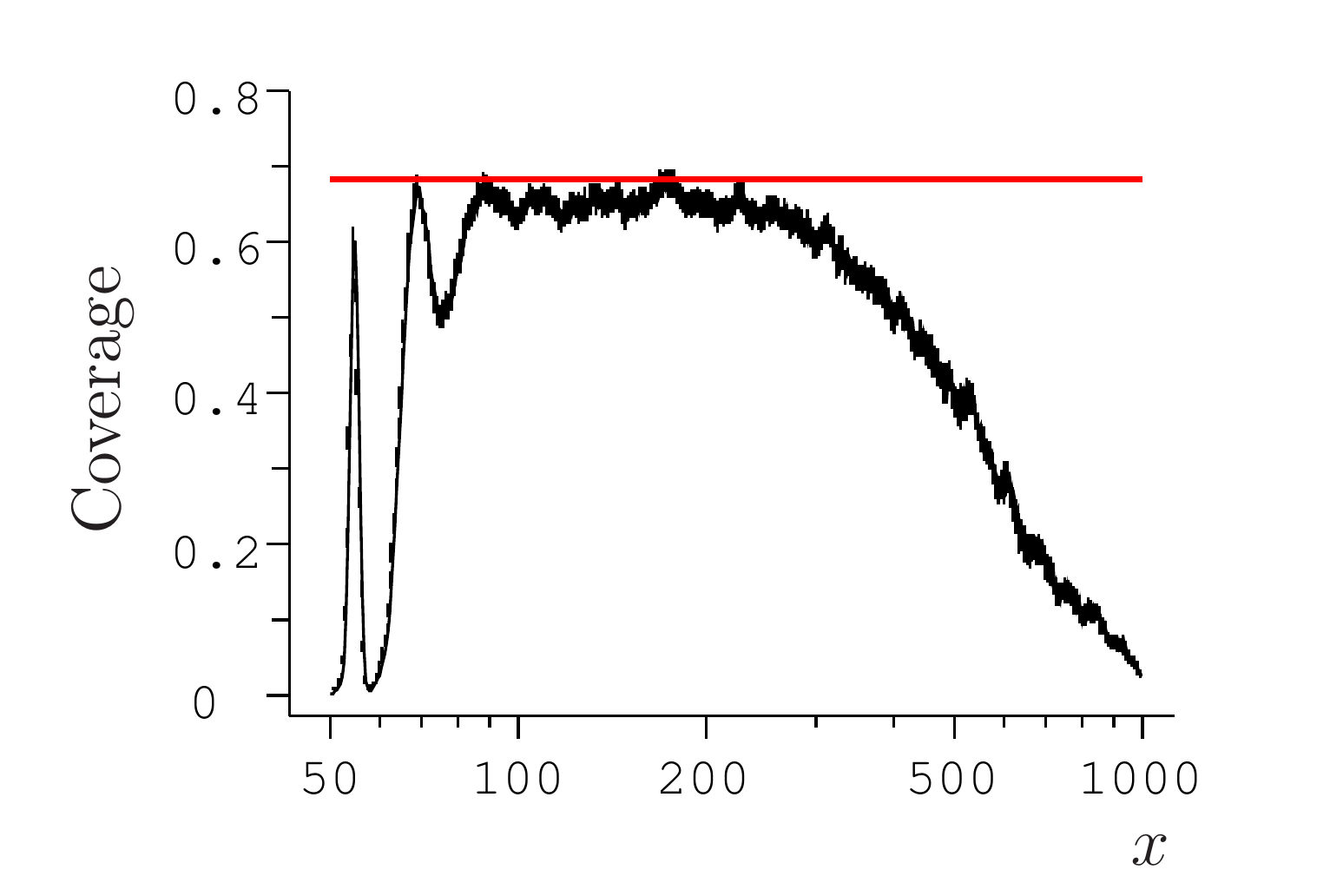}
\includegraphics[width=.49\textwidth]{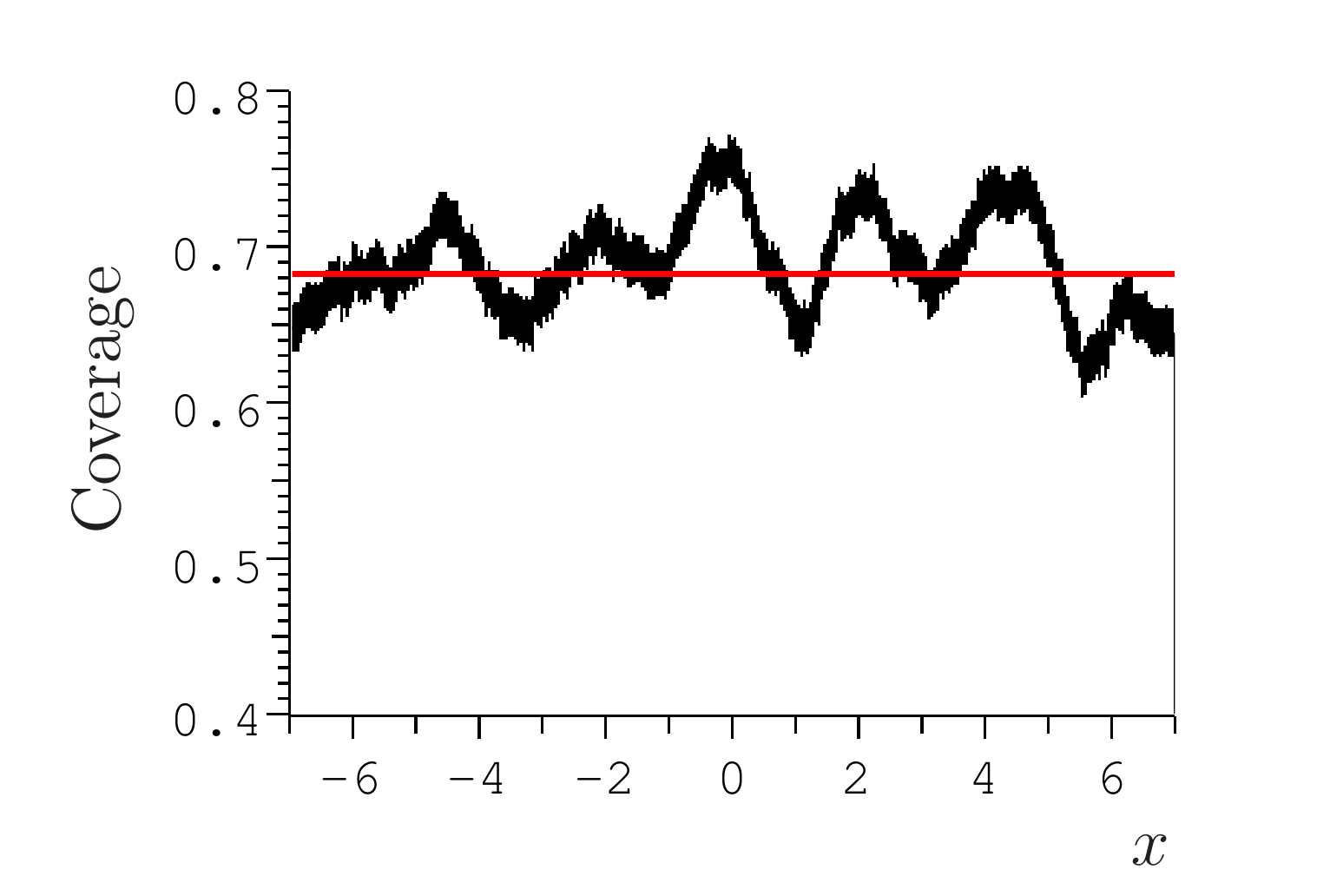} 
\includegraphics[width=.49\textwidth]{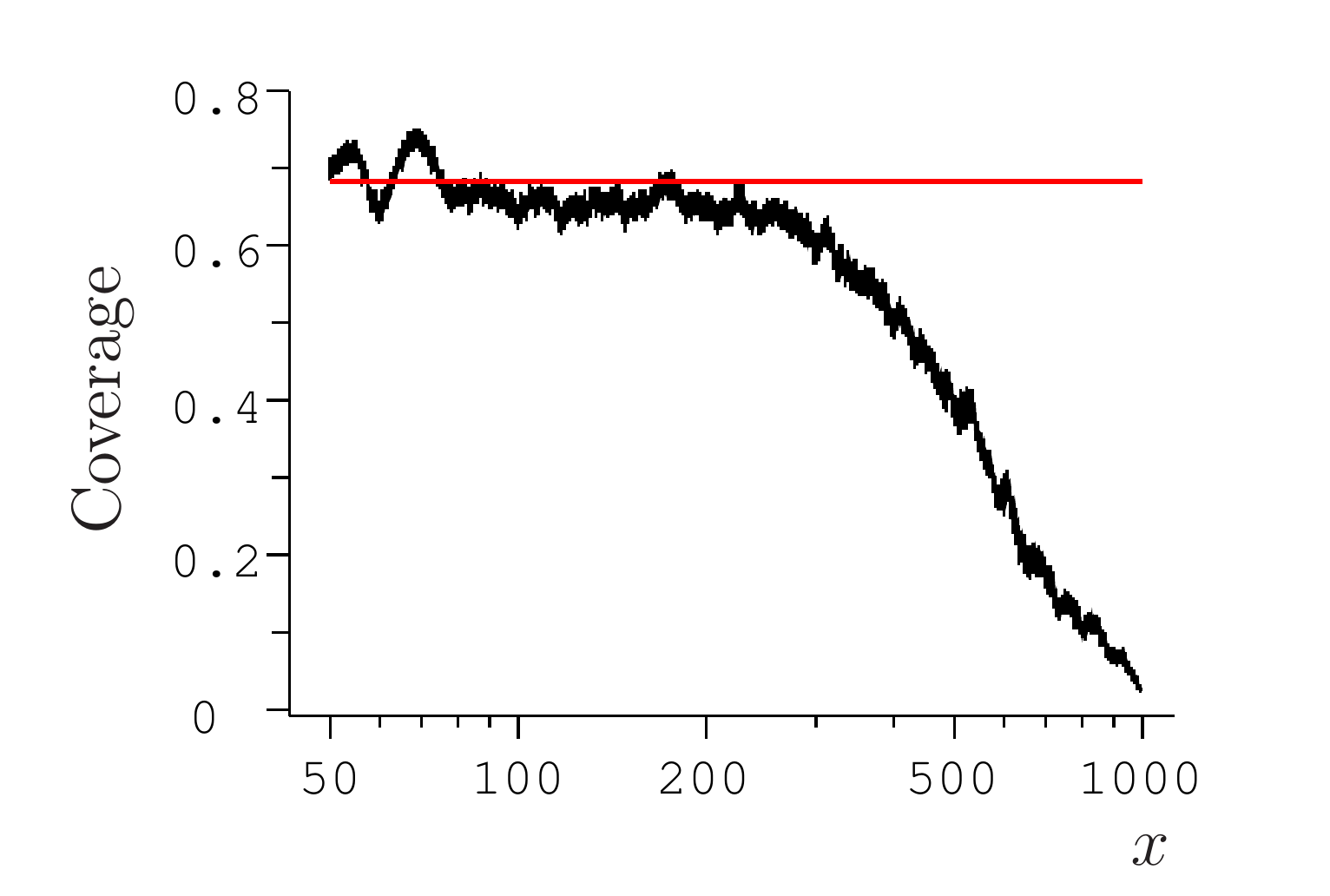}
\caption{Pointwise frequentist coverage for the folded smoothed bootstrap method.
         The first example distribution is on the left
         ($N = 1,000$ points per sample) and the second
         distribution is on the right ($N = 10,000$).
         1,000 bootstrapped replicas are generated for each unfolded result.
         For each bin of each sample, the uncertainty is defined as half of the difference
         between 84.13\supers{th} and 15.87\supers{th} percentiles
         of unfolded replica results in that bin (using asymmetric uncertainties
         makes virtually no difference).
         For the top plots, bias correction was not performed. The plots
         at the bottom were made by subtracting the known bandwidth-dependent bias,
         as in Fig.~\ref{fig:eaicbwdependent}, which leads to reasonable coverage
         properties.}
\label{fig:bootstrapcoverage1}
\end{center}
\end{figure}

\begin{figure}[h!]
\begin{center}
\includegraphics[width=.49\textwidth]{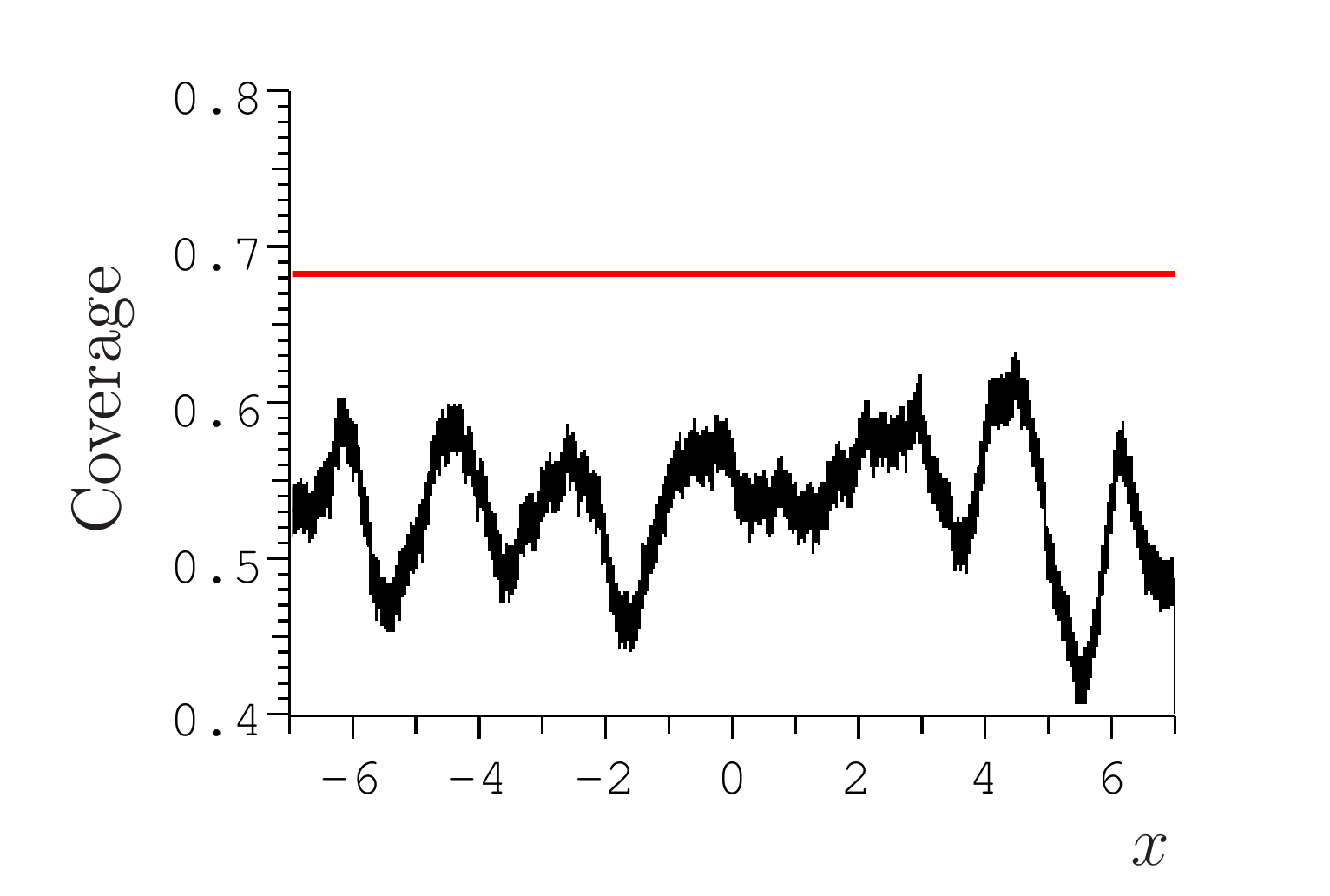} 
\includegraphics[width=.49\textwidth]{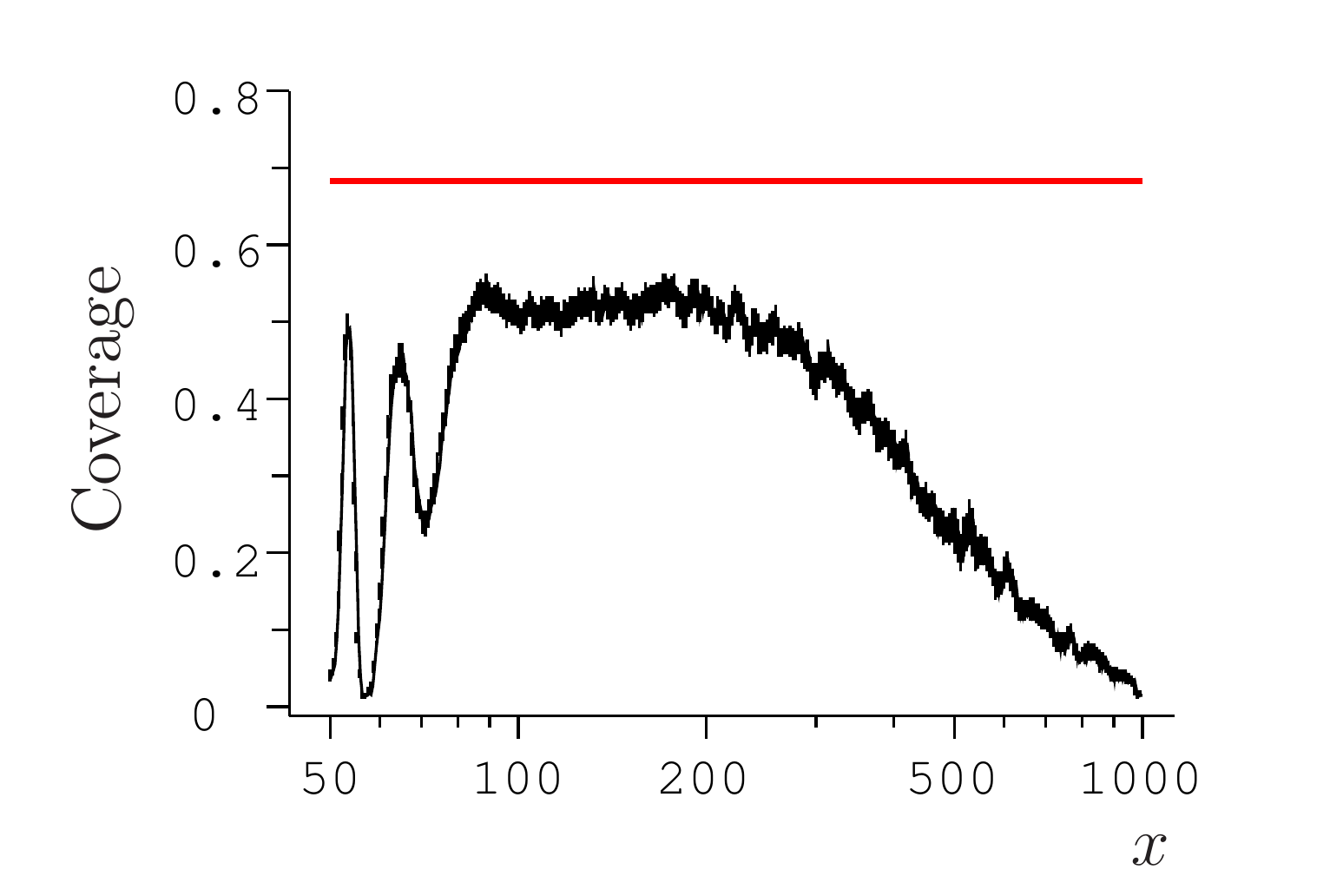}
\includegraphics[width=.49\textwidth]{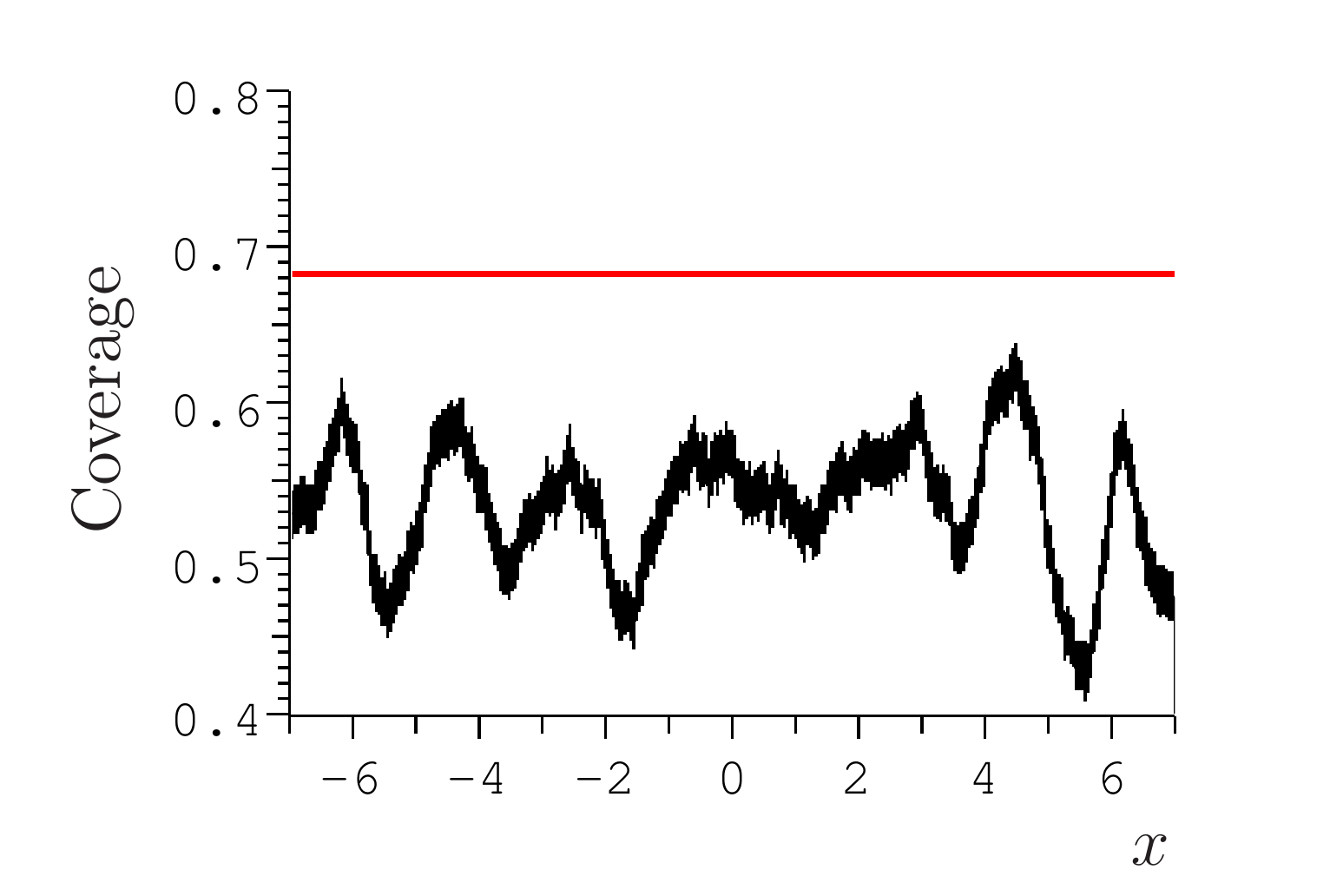} 
\includegraphics[width=.49\textwidth]{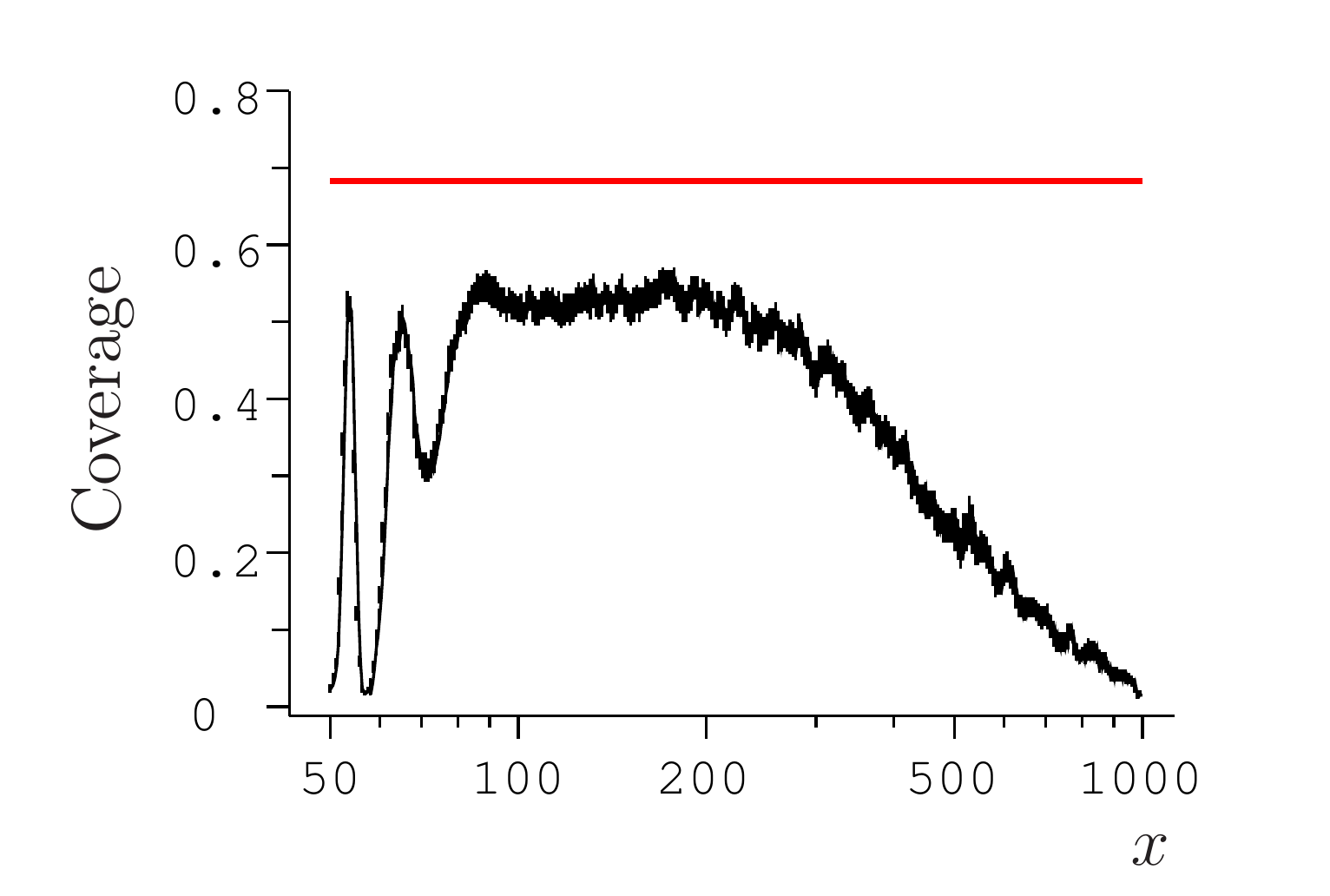}
\caption{Coverage for the folded smoothed bootstrap method with
         two bias correction schemes that can be realized in practice.
         For the top plots, the
         subtracted bias is defined by the difference between the
         $\hat{\hat{p}}(x)$ mean and $\hat{p}(x)$. For the plots
         at the bottom, the bias is conditioned upon the smoothing matrix bandwidth
         chosen by $EAIC_c$ for the primary sample.
         Additional unfolding is performed, with that bandwidth value only, for the
         same bootstrap replicas. The additional results are averaged and used
         in place of $\hat{\hat{p}}(x)$ to define the bias. The estimate of the
         variance still comes from the unfolding of the replicas with
         adaptive regularization strength.}
\label{fig:bootstrapcoverage2}
\end{center}
\end{figure}

The pointwise frequentist coverage of the uncertainties obtained for
the example distributions by the folded smoothed bootstrap method with
different bias correction schemes is presented in
Figs.~\ref{fig:bootstrapcoverage1} and~\ref{fig:bootstrapcoverage2}.
The regularization strength is chosen by $EAIC_c$. It appears that
the method provides an~acceptable estimate of the variance. However,
the utility of obtaining the bias correction from the bootstrap
remains questionable\footnote{Similar observations have been
made for bootstrap applications in nonparametric density estimation
without deconvolution~\cite{ref:hallbootstrap}.}.
In fact, such a bias correction introduces an additional
source of variability into the unfolding procedure and, for densities
under consideration, leads to a~noticeable overall reduction in
coverage.

\section{Presentation of Unfolded Results}
\label{sec:resultpresent}

The uncertainties of the unfolded results can be presented in
a~convenient form by utilizing the principal component
analysis~\cite{ref:principalcomponents} of the result covariance
matrix. Discretization of the physical process space with a large
number of bins avoids the dependence of the response function (and,
consequently, of the unfolded result) on the behavior of the physical
process density inside each bin. At the same time, information
content of large covariance matrices becomes an~important issue that
needs to be explicitly addressed.

\begin{figure}[h!]
\begin{center}
\includegraphics[width=.49\textwidth]{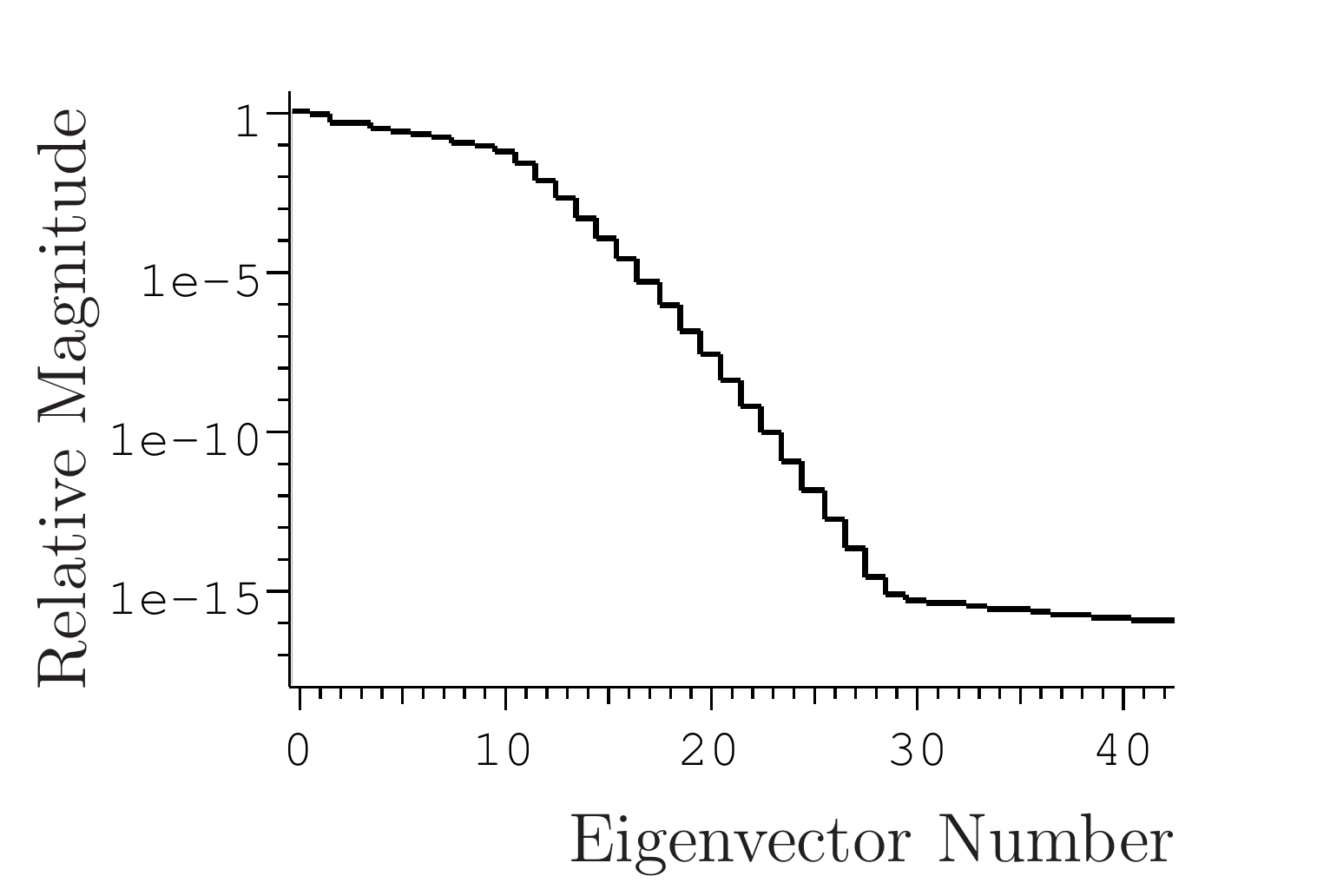} 
\includegraphics[width=.49\textwidth]{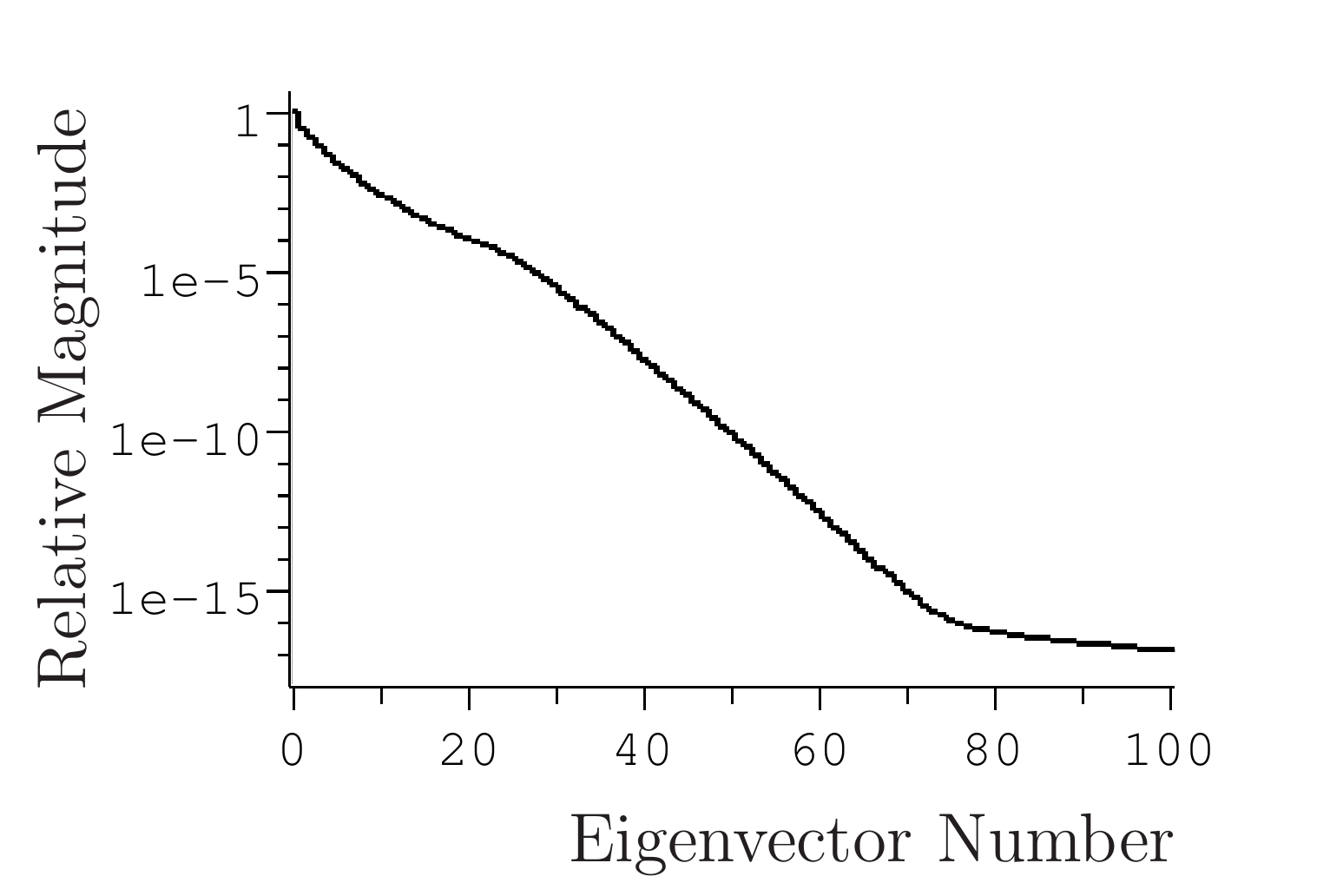}
\caption{Average relative eigenspectra of the covariance matrices of
         the unfolded results for the example densities considered in
         the previous section. The plot for the first example density is on the
         left, with sample size
         $N = 1,000$. The plot for the second example density is on the
         right, with $N = 10,000$. For both examples, the smoothing matrix bandwidth
         is chosen by $EAIC_c$ (same procedure as in Fig.~\ref{fig:eaicbwdependent}).
         Before averaging, eigenvalues are sorted in the decreasing
         order and divided by the largest one. For the left plot,
         covariance matrix dimensions are $420 \times 420$, so only about
         1/10\supers{th} of the complete eigenspectrum is shown. For the
         right plot, matrix dimensions are $1000 \times 1000$. Three parts
         can be visibly discerned in each spectrum: the structure of the
         largest eigenvalues (the left side of the curve) is dominated by the
         covariance matrix of the signal density, sampling
         noise determines the middle part of the spectrum, and the onset of the
         numerical round-off noise is apparent on the right.}
\label{fig:eigenspectrum}
\end{center}
\end{figure}

According to the Cramer–-Rao bound, for unbiased estimators
the largest amount
of information is associated with the smallest covariance matrix 
eigenvalues. The eigenvalue spectrum of the covariance matrices
of the unfolded results obtained with the simulations described
in the previous section is shown in Fig.~\ref{fig:eigenspectrum}.
The spectrum decays quickly, so the amount of information associated
with the covariance matrices appears to be disproportionately large.
This effect, however, is a~consequence of defining the covariance
matrix for the complete set of $m$ parameters $\lambda_j^{(\infty)}$
using a model with a substantially lower effective number of
parameters. Regularization (by smoothing, as in the EMS unfolding,
or by other techniques)
suppresses high frequency components of the unfolded result,
thus preventing unbiased estimation of these components.
Variance of a~biased estimator is no longer
subject to the Cramer–-Rao bound, while the lack of
coverage after bias correction can be attributed to the
imperfections of the linear error propagation approximation. On the
other hand, for sufficiently smooth densities, reasonable pointwise
bias-corrected coverage is attained because the combined
variance of high frequency components is substantially smaller than
the variance of low frequency components not affected by
regularization.

\begin{figure}[h!]
\begin{center}
\includegraphics[width=.49\textwidth]{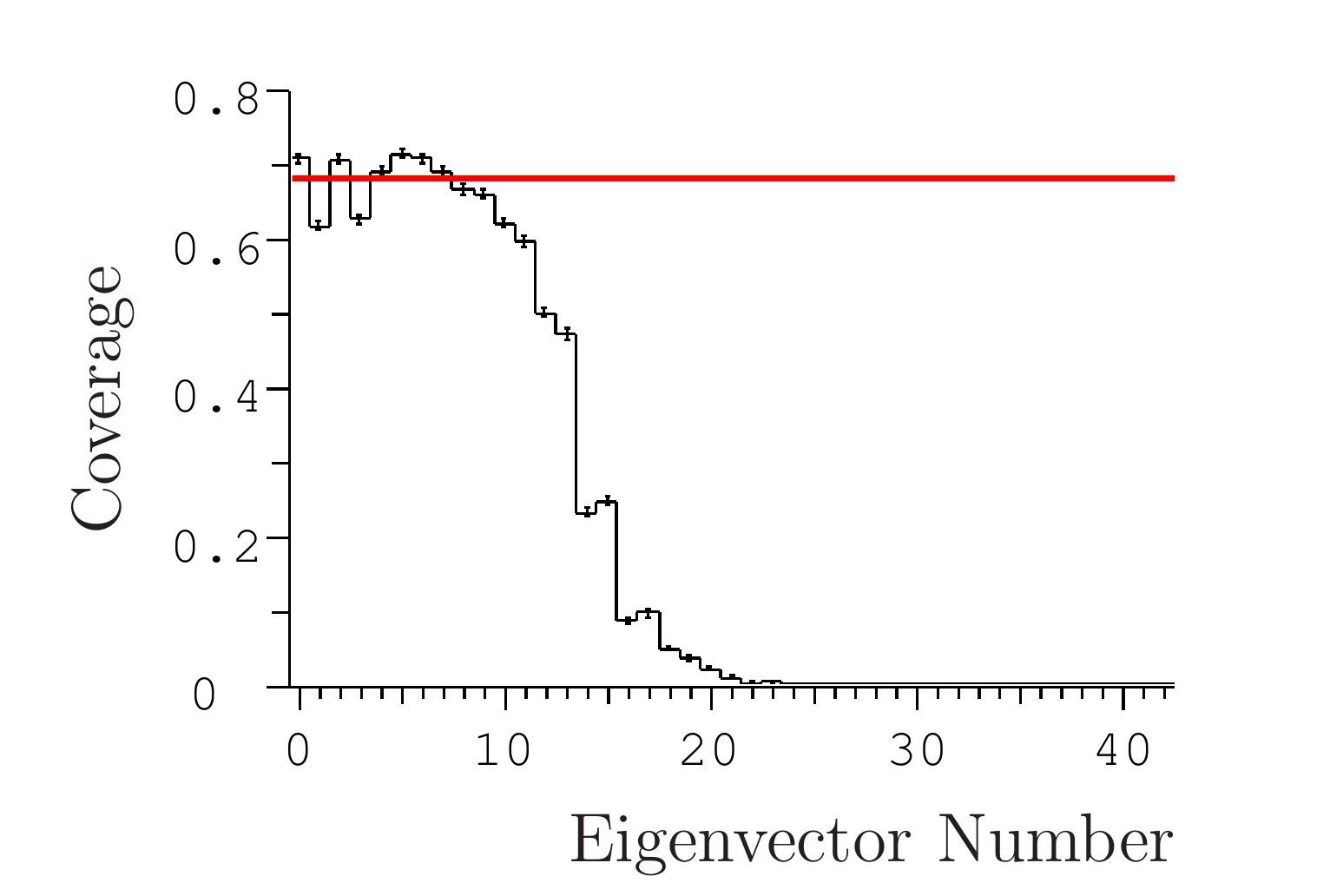} 
\includegraphics[width=.49\textwidth]{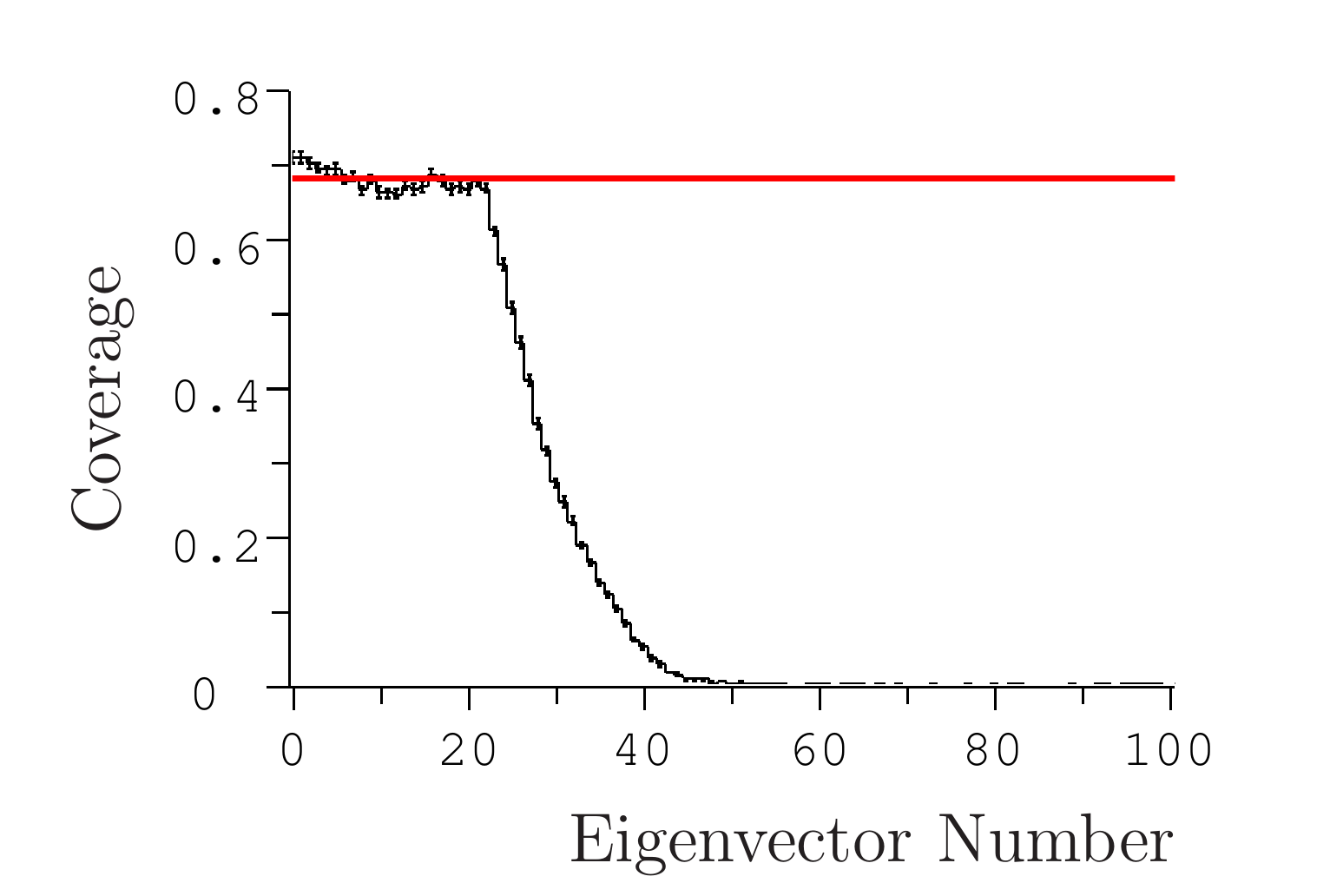}
\caption{Frequentist coverage of the principal components
         of the covariance matrices of unfolded results.
         The unfolding procedure and the simulated samples are
         the same as in Figs.~\ref{fig:eaicbwdependent}
         and~\ref{fig:eigenspectrum}.}
\label{fig:eigencoverage}
\end{center}
\end{figure}

\begin{figure}[h!]
\begin{center}
\includegraphics[width=.49\textwidth]{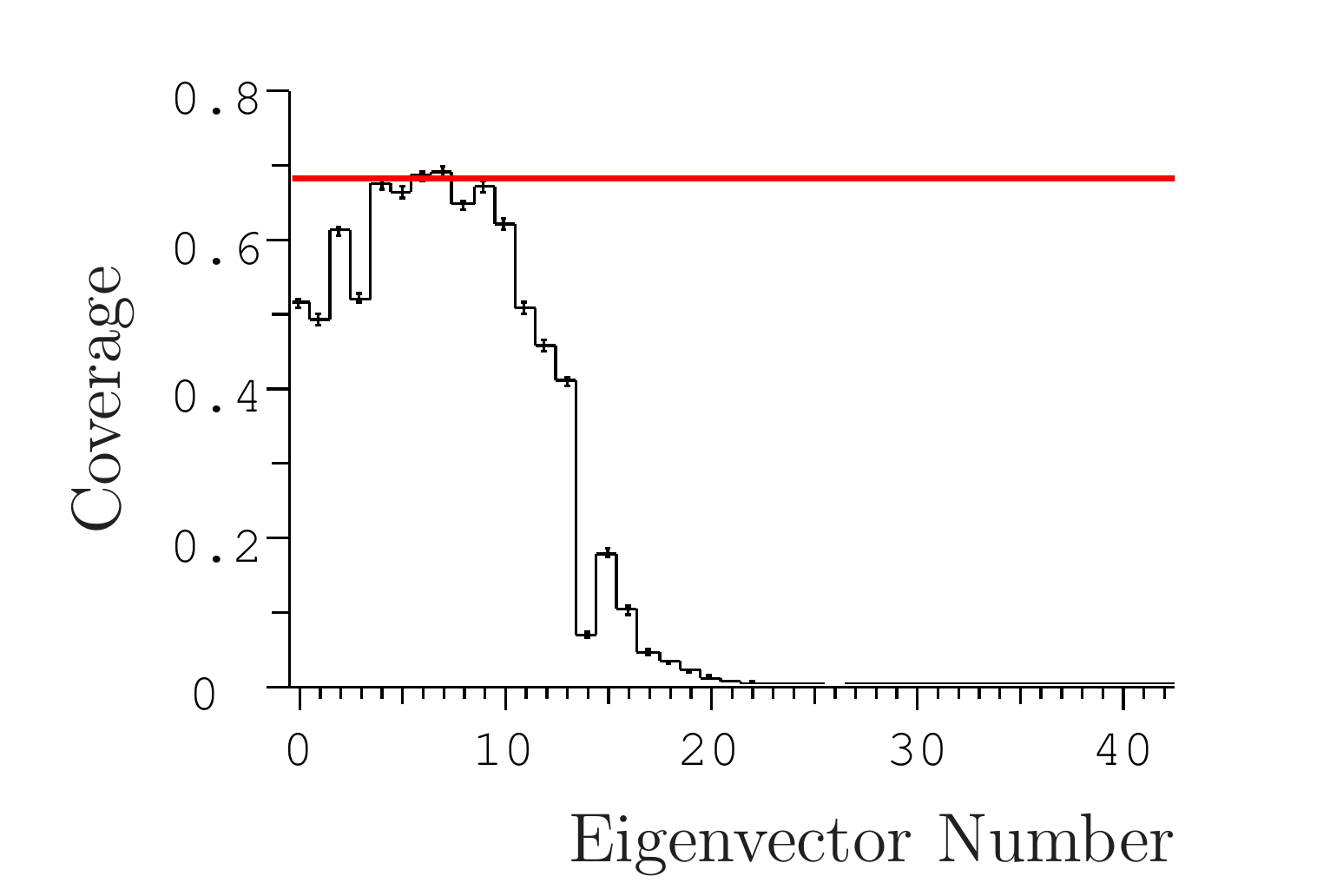} 
\includegraphics[width=.49\textwidth]{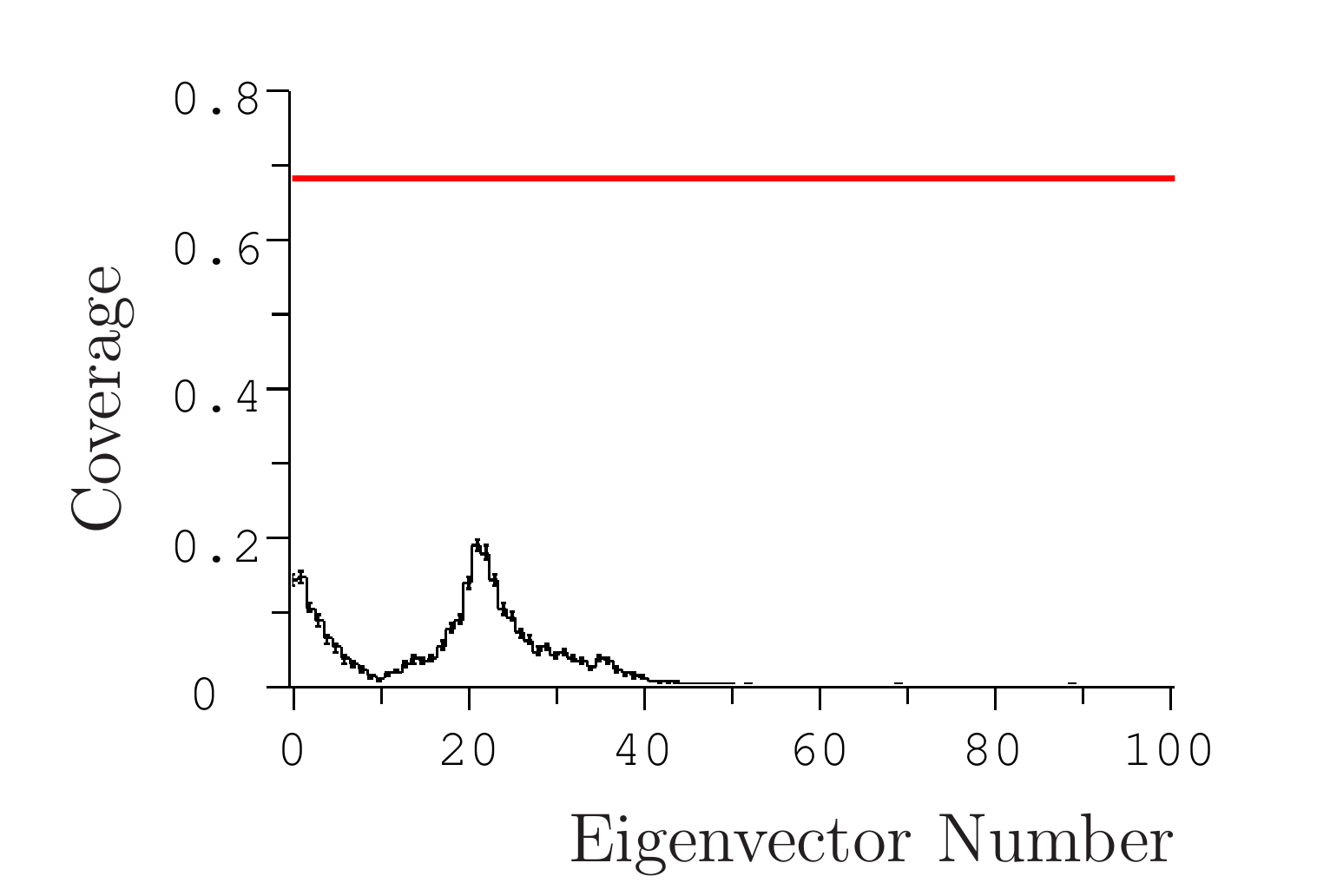}
\caption{This is what happens to the principal component coverage
         if the bias correction is not performed. The coverage breakdown
         for the second example density (on the right) is consistent
         with the left plot
         in Fig.~\ref{fig:coverage2}. There, the section with
         poor coverage coincides with the most populous region
         in the support of the density.}
\label{fig:biasedeigencoverage}
\end{center}
\end{figure}

For the example densities considered in the previous section,
frequentist coverage of the covariance matrix
principal components is illustrated in Figs.~\ref{fig:eigencoverage}
and~\ref{fig:biasedeigencoverage}. To construct these figures,
the principal components of the covariance matrix returned by
the unfolding procedure for each simulated sample
were arranged in the order of decreasing
eigenvalues. This ordering determined the ``eigenvector number'' represented
by the horizontal axes on the plots. The difference between the unfolded
result (bias-corrected in case of Fig.~\ref{fig:eigencoverage})
and the known ``true'' density processed by the smoothing
matrix was decomposed using the basis provided by
the normalized eigenvectors. The components of the difference in this basis
were divided by their estimated standard deviations ({\it i.e.}, square
roots of the corresponding eigenvalues). The ratios with the
same eigenvector number were grouped together, and the fraction
of the samples falling inside the $[-1, 1]$ interval was determined
for the resulting distributions.

The lack of coverage by the principal components of the
covariance matrix is also expected for the unfolding procedures
that utilize {\it ad hoc} binning.
The author, therefore, suggests that statistical uncertainties of
data analysis results obtained with unfolding procedures should
be presented by publishing, electronically,
the eigenvalues and the corresponding eigenvectors
of the result covariance matrix that are expected to possess proper
frequentist coverage in the assumption that the procedure
is unbiased\footnote{The bias has to be accounted
for by systematic uncertainties which can be presented in a~similar manner.}.
Table~\ref{tab:effparam} hints that, in addition to the eigenspectrum shape,
the effective number of model parameters 
corresponding to the $AIC_c$-selected bandwidth might
be instrumental in predicting
how many principal components of the result covariance matrix will have
appropriate coverage.
The rest of the statistical uncertainty
should be bundled together
into a~single ``overflow'' set encompassing the orthogonal complement
of these components. The formal variance of this set is simply the
difference between the trace of the result covariance matrix and the sum
of the eigenvalues for the principal components that have been specified
explicitly. However, proper frequentist coverage for this set
is not guaranteed.

Such a decomposition of experimental uncertainty
will provide specific diagnostic
information during comparisons of theoretical models with
unfolded results, will be instrumental in tuning model parameters,
and will avoid spurious rejection of models due to
inadequate coverage.

\section{Conclusions}
\label{sec:conclusions}

Due to a significant amount of arbitrariness associated with the
choice of regularization method and strength,
attaining correct frequentist coverage is difficult for the
unfolding procedures most commonly employed in the particle physics
data analysis practice. Use of wide bins in the physical
process space leads to a hard-to-quantify dependence
of the discretized response function on the assumed process density
inside each bin.

The automated regularization
optimization technique developed in this article addresses several
methodological issues in unfolding applications.
It permits a~fine-grained discretization
of the physical process space and enables the use
of precise response functions not affected by prior distribution
assumptions. The choice of the regularization strength is based
on a~well-established model selection criterion. The empirical
success of this regularization choice and the expected frequentist coverage
of the resulting unfolding procedure are validated by simulations.
The software package implementing this type of unfolding
is freely available~\cite{ref:npstat}.
The proposed presentation of
uncertainties is based on the expected coverage of the covariance
matrix principal components and offers a~considerable improvement
upon the common practice of providing just the diagonal elements
of the covariance matrix.

Accounting for the unfolding bias, and especially for the part
of the bias that belongs to the nullspace of the
response function, remains a difficult problem. It is unclear whether
this issue could be adequately resolved within the unfolding paradigm.
The ultimate solution will consist in developing a standard for
publishing experimental response
functions for particle detector systems,
so that theory predictions could be compared
with experimental spectra in the space of observations. 

\section{Acknowledgments}

The author thanks G\"{u}nter Zech
and the Statistics Committee of the CMS Collaboration
for comments and productive discussions. This work was supported in part
by the United States Department of Energy grant DE-FG02-12ER41840.


\begin{thebibliography}{99}

\bibitem{ref:phystat11}
H.B.~Prosper and L.~Lyons (eds.), ``Proceedings of the PHYSTAT 2011 Workshop
on Statistical Issues Related to Discovery Claims in Search Experiments
and Unfolding, CERN, Geneva, Switzerland 17-20 January 2011'',
\href{https://cdsweb.cern.ch/record/1306523/files/CERN-2011-006.pdf}{CERN-2011-006} (2011).

\bibitem{ref:reviewspano}
F.~Span\`o, ``Unfolding in particle physics:
a window on solving inverse problems'',
{\it EPJ Web of Conferences} {\bf 55}, 03002 (2013).

\bibitem{ref:deconvolutionbook}
A.~Meister, ``Deconvolution Problems in Nonparametric Statistics'',
Springer Lecture Notes in Statistics, Vol. 193 (2009).

\bibitem{ref:roounfold}
T. Adye, ``Unfolding algorithms and tests using RooUnfold'',
p. 313 in Ref.~\cite{ref:phystat11}.

\bibitem{ref:svdunfold}
A. H\"ocker and V. Kartvelishvili, ``SVD approach to data unfolding'',
{\it Nuclear Instruments and Methods in Physics Research A}
{\bf 372}, 469 (1996).

\bibitem{ref:dagounfold}
G. D'Agostini, ``A multidimensional unfolding method based on Bayes' theorem'', 
{\it Nuclear Instruments and Methods in Physics Research A}
{\bf 362}, 487 (1995).

\bibitem{ref:nychka}
D. Nychka, ``Some Properties of Adding a Smoothing Step to the EM Algorithm'',
{\it Statistics \& Probability Letters} {\bf 9}, 187 (1990).

\bibitem{ref:smoothedem}
B.W. Silverman \etal, ``A Smoothed EM Approach to Indirect Estimation Problems,
with Particular Reference to Stereology and Emission Tomography'',
{\it Journal of the Royal Statistical Society B} {\bf 52}, 271 (1990).

\bibitem{ref:numrecipes}
W.H.~Press, S.A.~Teukolsky, W.T.~Vetterling, and B.P.~Flannery,
``Numerical Recipes: the Art of Scientific Computing'', 3rd ed.,
Cambridge University Press (2007).

\bibitem{ref:richardson}
W.H. Richardson, ``Bayesian-Based Iterative Method of Image Restoration'',
{\it Journal of the Optical Society of America} {\bf 62}, 55 (1972).

\bibitem{ref:lucy}
L.B. Lucy, ``“An iterative technique for the rectification
of observed distributions'',
{\it The Astronomical Journal} {\bf 79}, 745 (1974).

\bibitem{ref:dempster}
A.P.~Dempster, N.M.~Laird and D.B.~Rubin, ``Maximum
likelihood from incomplete data via the EM algorithm'',
{\it Journal of the Royal Statistical Society B} {\bf 39}, l (1977).

\bibitem{ref:vardi}
Y.~Vardi, L.A.~Shepp and L.~Kaufman, ``A Statistical Model for
Positron Emission Tomography'',
{\it  Journal of the American Statistical Association}
{\bf 80}, No. 389, 8 (1985).

\bibitem{ref:lapack}
E.~Anderson \etal, ``LAPACK Users' Guide'', SIAM Series
on Software, Environments and Tools (Book 9), 1987.

\bibitem{ref:dagostini2}
G. D'Agostini, ``Improved Iterative Bayesian unfolding'',
\href{http://arxiv.org/abs/1010.0632}{arXiv:1010.0632 [physics.data-an]} (2010).

\bibitem{ref:lindemann}
L.~Lindemann and G. Zech, ``Unfolding by weighting Monte Carlo events'',
{\it Nuclear Instruments and Methods in Physics Research A}
{\bf 354}, 516 (1995).

\bibitem{ref:zech11}
G.~Zech, ``Regularization and error assignment to unfolded distributions'',
p. 252 in Ref.~\cite{ref:phystat11}.

\bibitem{ref:vonneumann}
John Von Neumann, ``Mathematical Foundations of Quantum Mechanics'',
Princeton University Press (1955).

\bibitem{ref:vershynin}
R. Vershynin, ``Introduction to the non-asymptotic analysis of random matrices'',
in Y.~Eldar and G.~Kutyniok (eds.), ``Compressed Sensing, Theory and
Applications'', Cambridge University Press (2012).

\bibitem{ref:aic}
H.~Akaike, ``A new look at the statistical model identification'',
{\it IEEE Transactions on Automatic Control} {\bf 19}, 716 (1974).

\bibitem{ref:modelsel}
K.P. Burnham and D.R. Anderson, ``Model Selection and Multi-Model Inference:
A~Practical Information-Theoretic Approach'', 2nd ed., Springer (2004).

\bibitem{ref:loader}
C. Loader, ``Local Regression and Likelihood'', Springer Series on
Statistics and Computing (1999).

\bibitem{ref:kuuselapreprint}
M.~Kuusela and V.M.~Panaretos, ``Empirical Bayes unfolding of
elementary particle spectra at the Large Hadron Collider'',
\href{http://arxiv.org/abs/1401.8274}{arXiv:1401.8274 [stat.AP]} (2014).

\bibitem{ref:duffy}
D.G.~Duffy, ``Green's Functions with Applications'',
Chapman \& Hall/CRC (2001).

\bibitem{ref:npstat}
\href{http://npstat.hepforge.org/}{http://npstat.hepforge.org/}

\bibitem{ref:kde}
M.P.~Wand and M.C.~Jones, ``Kernel Smoothing'',
Chapman \& Hall/CRC (1995).

\bibitem{ref:cowling1996}
A.~Cowling, P.~Hall and M.J.~Phillips, ``Confidence Regions
for the Intensity of a Poisson Point Process'', {\it Journal of
the American Statistical Association} {\bf 91}, 1516 (1996).

\bibitem{ref:smoothingbootstrap}
D. de Angelis and G.A.~Young, ``Smoothing the Bootstrap'',
{\it International Statistical Review} {\bf 60}, 45 (1992).

\bibitem{ref:hallbootstrap}
P.~Hall, ``Using the bootstrap to estimate mean squared error and select
smoothing parameter in nonparametric problems'',
{\it Journal of Multivariate Analysis} {\bf 32}, 177 (1990).

\bibitem{ref:robcov}
W.J. den Haan and A. Levin, ``A Practitioner's Guide to
Robust Covariance Matrix Estimation'', in
``Handbook of Statistics 15: Robust Inference'', G.S. Maddala
and C.R. Rao (eds.), Elsevier Science (1997).

\bibitem{ref:principalcomponents}
I.T. Jolliffe, ``Principal Component Analysis'', 2\supers{nd} ed.,
Springer Series on Statistics (2002).

\end{thebibliography}
\end{document}